\pdfoutput=1
\documentclass{IEEEtran} 
\usepackage[T1]{fontenc}
\usepackage[protrusion=true,expansion=true,final]{microtype} 

\usepackage[backend=biber, hyperref=true, style=ieee, isbn=false, doi=true]{biblatex}

\usepackage{amsmath,amsthm,amssymb,amsfonts,bm,nccmath,mathdots,mathtools,bigints}

\usepackage[dvipsnames]{xcolor}	
\usepackage{stfloats}			

\usepackage{hyperref}			
\usepackage{enumitem}           
\usepackage{parskip}           	
\usepackage{makecell}           
\usepackage{caption}           	

% ...

\theoremstyle{definition}

\hypersetup{citecolor=black, linkcolor=black, urlcolor=RoyalBlue, colorlinks=true}

\makeatletter
\def\subsubsection{%
  \@startsection
    {subsubsection}                 
    {3}                             
    {\parindent}                    
    {\parskip}  					
    {0ex}     						
    {\normalfont\normalsize\itshape}
}
\makeatother

\definecolor{stateColor}{RGB}{224,60,48}		
\definecolor{inputColor}{RGB}{73,158,255}		%
\definecolor{disturbanceColor}{RGB}{102,43,153} 
\definecolor{outputColor}{cmyk}{0.82,0,0.92,0}  
\definecolor{paramColor}{RGB}{224,148,48}
\definecolor{monokaiBG}{RGB}{24,24,24}
\definecolor{monokaiOrange}{RGB}{253,151,31}

\newcommand{\clP}[1]{#1}
\newcommand{\clX}[1]{#1}
\newcommand{\clU}[1]{#1}
\newcommand{\clY}[1]{#1}
\newcommand{\clW}[1]{#1}

\newcommand*{\tran}{{\mkern-1.5mu\mathsf{T}}}

\newcommand\hlineskips{\rule{0pt}{3ex}\rule[-1ex]{0pt}{0pt}}
\newcommand\doublehline{\hline\hline\hlineskips}

\begin{document}
\title{A model-based framework for controlling activated sludge plants}

\author{
	Otac\'ilio B. L. Neto\IEEEauthorrefmark{1}, Michela Mulas\IEEEauthorrefmark{2}, and Francesco Corona\IEEEauthorrefmark{1}
	\thanks{\IEEEauthorrefmark{1} School of Chemical Engineering, Aalto University, Finland. (e-mails: \texttt{\{otacilio.neto, francesco.corona\}@aalto.fi})}
	\thanks{\IEEEauthorrefmark{2} Department of Teleinformatics Engineering, Federal University of Cear\'a, Brazil. (e-mail: \texttt{\{michela.mulas\}@ufc.br})} 
}

\maketitle

\begin{abstract}
This work presents a general framework for the advanced control of a common class of activated sludge plants (ASPs). 
Based on a dynamic model of the process and plant sensors and actuators, we design and configure a highly customisable Output Model-Predictive Controller (Output MPC) for the flexible operation of ASPs as water resource recovery facilities. 
The controller consists of a \textit{i}) Moving-Horizon Estimator for determining the state of the process, from plant measurements, and \textit{ii}) a Model-Predictive Controller for determining the optimal actions to attain high-level operational goals. 
The Output MPC can be configured to satisfy the technological limits of the plant equipment, as well as operational desiderata defined by plant personnel. 
We consider exemplary problems and show that the framework is able to control ASPs for tasks of practical relevance, ranging from wastewater treatment subject to normative limits, to the production of an effluent with varying nitrogen content, and energy recovery.
\end{abstract}


\section{Introduction} \label{sec: Intro}

The conventional purpose of a municipal wastewater treatment plant (WWTP) is to depurate an influent sewage or wastewater stream, before it can be safely discharged into the environment. Central to conventional WWTPs is the biological treatment of wastewater through an activated sludge process and activated sludge plants (ASPs) have become a widely diffused technology, with clear societal importance. At the same time, the interest in recovering valuable resources existing in wastewater has become pervasive. To refer to the wealth technologies aiming at capturing such resources from otherwise unused wastewater streams, the notion of a WWTP as a water resource recovery facility (WRRF) has emerged. 

Chemicals containing nitrogen and phosphorus, abundant in wastewater, are main contributors to crop growth \cite{Ricart2019, Kundu2022}. Disposed sludge can be harvested for materials and used to generate electricity, potentially allowing treatment plants to be self-sufficient, if not producers of energy \cite{Sarpong2020}. Water reuse, the use of raw, or partially treated, wastewater for beneficial purposes, is a related practice which alleviates the need for freshwater \cite{WHO2006}. Current efforts have focused on the design of new, or the adaptation of existing, treatment processes to include resource recovery functionalities \cite{ValverdePerez2015, Khiewwijit2015, ValverdePerez2016, Wan2016, FernandezArevalo2017, Huang2020, JZonta2022}. Despite the remarkable results, these solutions are undeniably costly, because of the capital and maintenance investments needed to revamp and operate these modernised facilities. Conversely, too feeble is the effort to understand how the operation of the existing wastewater treatment infrastructure can be rendered flexible enough to sustain both changing quality standards and resource recovery objectives. 

The complexity of these bioprocesses is overwhelming. How to determine and and adapt in real-time the operational policies for activated sludge plants is a daunting task to be handled manually by plant operators and engineering. From this perspective, the arising paradigm of perceiving wastewater as a sustainable source of water of different grades, of nutrients, and energy, together with stricter regulations \cite{EUreuse2017, EU2022}, demands for soft and automatic technology aiming at an optimised operation of  existing treatment facilities. 

Automatic control provides the mathematical framework for the design of control policies capable of steering in real-time activated sludge plants toward desired objectives. Model predictive control (MPC), specifically, has been the technology of choice in many industrial applications \cite{Forbes2015}. In this model-based approach, optimal actions to the plant actuators are recursively computed based on measurements and a dynamical model representation of the plant \cite{Rawlings2020}. Aligned with the emerging Measurement-Analysis-Decision concept for wastewater treatment (M-A-D, \cite{Ingildsen2016}), this control methodology is a promising framework for the automation of wastewater treatment facilities of the future.

The availability of support tools \cite{WWTModels} that provide protocols for simulating activated sludge plants has initiated the computer-assisted design of automatic control solutions for efficiently operating real-world facilities. The most advanced strategies use dynamic process models as their core technology, starting from \cite{Rosen2002} and \cite{Alex2002} who both investigated model-based approaches for controlling high peaks in effluent ammonia. The classic technique of dynamic-matrix control was used by \cite{Corriou2004} to maintain effluent's quality within regulation-specified limits. For nitrogen removal, nonlinear MPC was shown to outperform traditional model-free controllers during peaks in the influent load \cite{Stare2007}. For the same task, \cite{Shen2008} and \cite{Shen2009} augmented the actuation layout and designed MPCs that improve effluent quality at the expense of increased energy costs. In \cite{Foscoliano2016}, a predictive controller is designed for a neural-network model of an activated sludge process with plant-wide actuation. Similarly, various approaches to predictive control were investigated in different types of activated sludge processes \cite{Sotomayor2002, Ekman2008, Ostace2011, Mulas2015}. A self-optimising procedure to select the best controlled variables to be used in optimal control strategies was proposed by \cite{Francisco2015}. The problem has also been tackled using economic model-predictive controllers: Several works report cost-effective control of ASPs under different technological assumptions \cite{Zeng2015, Zhang2019, Zhang2019b, Moliner2019, Kalogeropoulos2022}. Recently, the technological scope has been extended to the design of several centralised and distributed strategies in which predictive models are used not only for control, but also for estimation \cite{Yin2018, Yin2018b, Yin2019}. Importantly, while these contributions show the potential of model-based strategies, they are \textit{ad hoc} designs centred on a specific goal: Removal of effluent nitrogen to comply with normative limits. As such, they do not provide a unified control framework which is valid at system-level and general enough to incorporate resource recovery operations.

\begin{figure}[t!] \centering
	\includegraphics[width=\columnwidth]{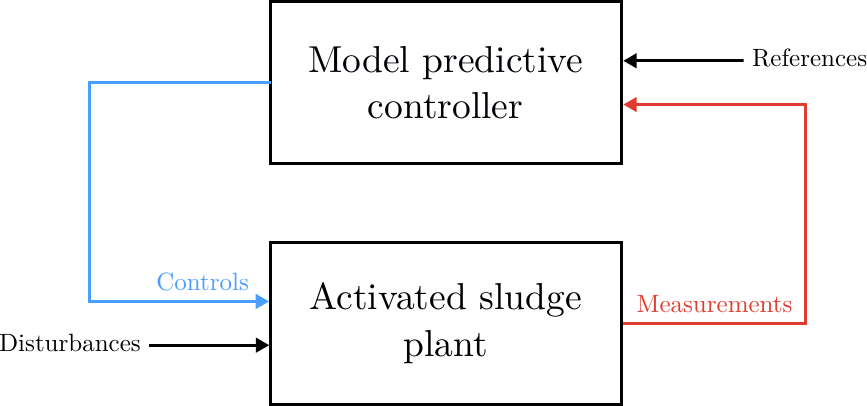}
    \caption{Model-based control of activated sludge plants.}
	\label{fig: ControlDiagram}
\end{figure}

In this study, we present a general framework for operating an important class of activated sludge plants with model predictive controllers (Figure \ref{fig: ControlDiagram}). Specifically, we
\begin{enumerate}[label={\roman*.}, nosep, itemsep=0.5ex]
    \item formulate an Output MPC for conventional ASPs and show how to configure it to operate a full-scale plant according to high-level and yet practical objectives;
    
    \item consider the objective of operating ASPs as WRRFs that produce an effluent water whose quality varies in time, in response to a downstream demand;
    
    \item demonstrate how the controller is able to dynamically operate an ASP to achieve these objectives, while satisfying legislative and technological limits, rejecting influent disturbances, and maximising energetic autonomy by modulating the production of sludge. 
\end{enumerate}
Due to its generality, our framework accommodates any control objective of operational, environmental, and/or economic nature. They are assumed to be planned by plant management and given to the controller as reference trajectories for the plant to follow. Their attainment is based on the recursive solution of an optimal control problem to compute the control actions which best track the reference over a future horizon. Importantly, the control actions are determined to satisfy any constraints that managers and engineers have requested to be enforced. The optimal control element (MPC) of the controller operates in conjunction with a moving-horizon state estimator (MHE), their functioning depends on a dynamic model of the process and their operation relies on the exchange of data with plants' actuators (the control actions) and measurements (the process sensors and laboratory analysis). 

Our study is presented as follows. Section \ref{sec: ASP} overviews a conventional activated sludge plant and the reference benchmark model commonly used to describe its main dynamics. The complete reference benchmark will be used as experimental ASP. In Section \ref{sec: OutputMPC}, the general framework of model-based predictive control is presented in detail. In Section \ref{sec: ControlFramework}, the controller is customised with linearised models of the activated sludge process and configured to operate the ASP towards the attainment of pre-assigned objectives. Section \ref{sec: Results} discusses the aforementioned applications designed to show how the controller can autonomously operate an activated sludge plant to produce water of varying quality. We discuss two high-level tasks of practical relevance: \textit{i}) production of effluent water whose quality satisfies conventional treatment limits and \textit{ii}) production of reuse effluent water with a varying nitrogen content. For completeness, models, additional experiments, and discussions are reported in the Supplementary Material.

\section{The activated sludge plant} \label{sec: ASP}

\begin{figure}[h!] \centering
	\includegraphics[width=\columnwidth]{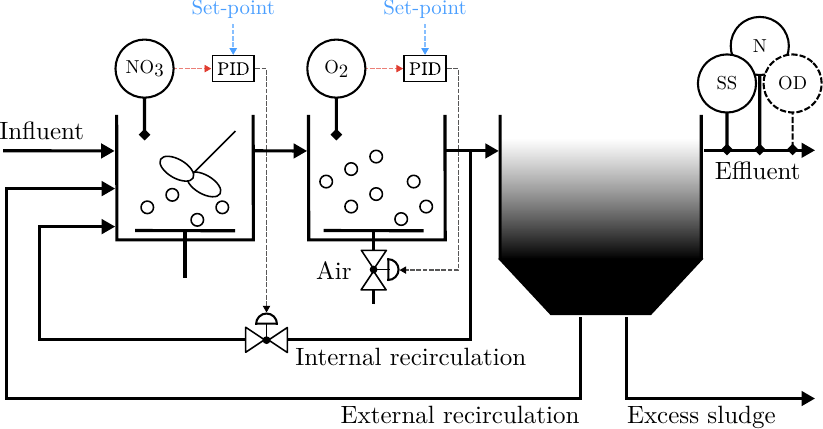}
    \caption{Activated sludge plant: Conventional process layout equipped with a basic setup for measurement and control.}
	\label{fig: CAS}
\end{figure}

We consider the activated sludge process in a conventional biological wastewater treatment plant \cite{Jenkins2014}. Based on denitrification-nitrification processes, micro-organisms are used to reduce the nitrogen present in the form of ammonia in the influent wastewater  into nitrate, which is then further reduced into nitrogen gas and released into the atmosphere. A typical process (Figure \ref{fig: CAS}) comprises two reacting sections and a settler. Each reacting section may consist of several tanks for the oxidation of organic matter. In the settler, the water clarified from suspended solids and flocculated particles before it is further processed or disposed.
 
The treatment begins in the first, anoxic, reacting section where influent wastewater from primary sedimentation, return sludge from secondary sedimentation, and internal recycle sludge are fed to the bioreactors where denitrification is performed. The outflow from the anoxic section is then fed to the second, aerated, reactors and, eventually, to the secondary settler. In the aerated section, ammonium nitrogen (NH$_4$-N) in the wastewater is oxidised into nitrate nitrogen (NO$_3$-N), which is in turn reduced into molecular nitrogen (N$_2$) in the anoxic section. This is achieved by recirculating mixed-liquor from the aerated section and recycle sludge from secondary sedimentation into the anoxic section. The oxygen used for oxidation is added by insufflating air to the bioreactors and a large part of the biodegradable organic matter is removed in the denitrification process. While supplementary carbon sources can be added to the reactors, no other major chemicals are used in the wastewater treatment process. Excess sludge from the settler is either removed from the plant or directed towards dedicated processes for sludge treatment. 

\begin{figure*}[htb!] \centering 
	\includegraphics[width=0.95\textwidth]{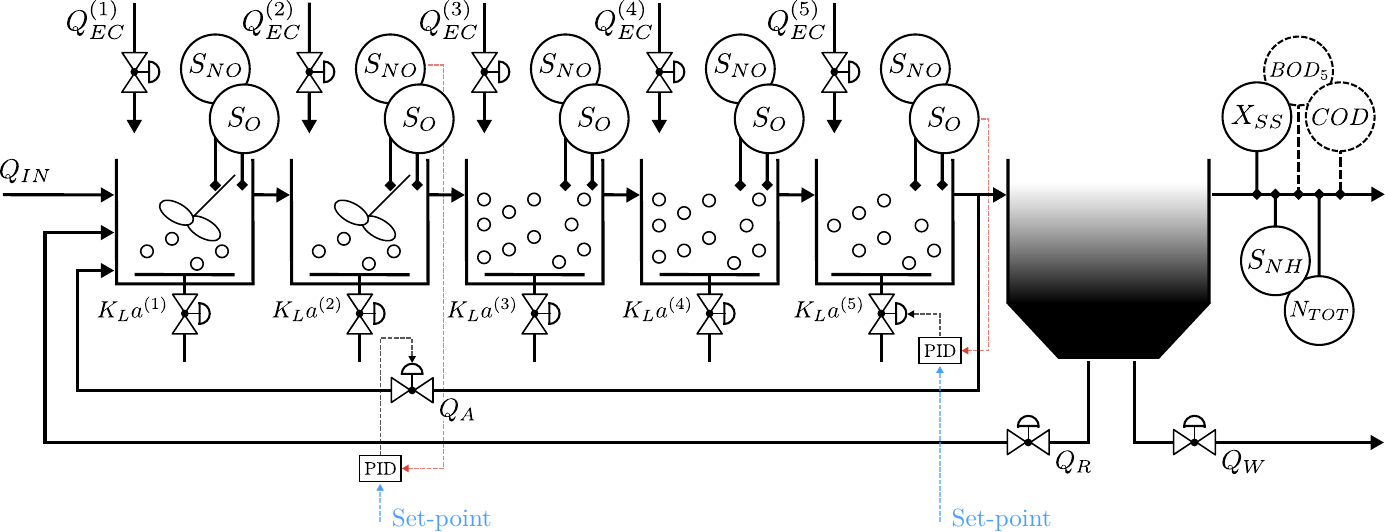} 

    \caption{Benchmark simulation model No. 1: Process layout equipped with a selection of measurement and actionable variables.}
	\label{fig: BSM1} 
\end{figure*}

The quality of the water from the top of the settler is routinely evaluated in terms of suspended solids (SS), nitrogen (N, in the form of ammonia/ammonium, and nitrates), and organic matter as quantified by oxygen demand (OD). Other characterisations, for instance in terms of phosphorous (P, mainly phosphates) and alkalinity (pH), are also common. Importantly, when no further processes exist downstream, water from settler must satisfy the disposal permits in force.

To produce an effluent water of consistent grade in spite of the variability of the influent, activated sludge plants use a combination of solutions from automatic control \cite{Olsson2013}. Simple  and yet effective strategies built upon two controllers for two key and readily measurable quantities:
\begin{itemize}
	\item[$\leadsto$] Nitrate nitrogen (NO$_3$-N) in the anoxic section is controlled by manipulating the internal recirculation;
	\item[$\leadsto$] Dissolved oxygen (O$_2$) in the aerated section is controlled by manipulating the air flow-rate.
\end{itemize}
Nitrates in the anoxic zone and dissolved oxygen in the aerated zone are regulated using low-level controllers (Figure \ref{fig: CAS}). Proportional-integral-derivative (PID) controllers, with set-points configured in such a way that the plant produces a water of desired quality, are the technology of choice. We refer to this scheme (Figure \ref{fig: CAS}) as the \textit{default control strategy}.

More advanced, though less common, strategies can and have been be developed. In Section \ref{sec: OutputMPC} and \ref{sec: ControlFramework}, we present a general framework for model-based control of activated sludge plants. We use the Benchmark Simulation Model no. 1 (BSM1, \cite{Gernaey2014}) as a reference process representative of a wide class of ASPs. The BSM1 is a well-established platform for simulation and control design \cite{Olsson2013, Yin2019, Zhang2019, Moliner2019}. The model and its parameters can be calibrated to represent specific real-world processes. The main elements of the BSM1 are reviewed in this section.

\subsection{Benchmark Simulation Model no. 1} \label{subsec: BSM1}

The Benchmark Simulation Model no. 1 describes a conventional activated sludge plant in which the anoxic and aerated sections consist of two and three bioreactors, respectively, and one settler with ten non-reacting layers. The dynamics of the bioreactors are described by the Activated Sludge Model no. 1 \cite{Henze2000}, while the model by Tak\'acs et~al. \cite{Takacs1991} is used for the settler. These are widely accepted models that were empirically validated in several studies \cite{Petersen2002,Petersen2003,Keskitalo2010,Ngo2021,Qiu2023}. Their dynamics and control properties are thoroughly studied by \cite{NMC_BSM1_Analysis}.

\begin{table*}[htb!] \centering
    \caption{Activated sludge plant: Process variables by location (`$A(r)$', $r=1,\dots,5$, in the $r$-th bioreactor, or `$IN$' in the influent wastewater; `$S(l)$', $l=1,\dots,10$, in the $l$-th settler layer) and type (`$D$', disturbance; `$S$', state variable; `$M$' measurement; and `$C$', control).}
    \begin{tabular} {l | l | l || c} 
        {Variable}										  	& {Description}						           	& {Units}				     & {Type}				\\ 
        \doublehline
        $S^{IN}_I$, $S^{A(r)}_I$, $S^{S(l)}_I$				& Soluble inert organic matter					& g COD m$^{-3}$			 & D, S, S			    \\[0.25ex]
        $S^{IN}_S$, $S^{A(r)}_S$, $S^{S(l)}_S$				& Readily biodegradable substrate				& g COD m$^{-3}$			 & D, S, S			    \\[0.25ex]
        $X^{IN}_I$, $X^{A(r)}_I$							& Particulate inert organic matter				& g COD m$^{-3}$			 & D, S				    \\[0.25ex]
        $X^{IN}_S$, $X^{A(r)}_S$							& Slowly biodegradable substrate				& g COD m$^{-3}$			 & D, S				    \\[0.25ex]
        $X^{IN}_{BH}$, $X^{A(r)}_{BH}$						& Active heterotrophic biomass					& g COD m$^{-3}$			 & D, S				    \\[0.25ex]
        $X^{IN}_{BA}$, $X^{A(r)}_{BA}$						& Active autotrophic biomass					& g COD m$^{-3}$			 & D, S				    \\[0.25ex]
        $X^{IN}_{P}$, $X^{A(r)}_{P}$						& Particulate products from biomass decay		& g COD m$^{-3}$			 & D, S				    \\[0.25ex]
        $S^{IN}_O$, $S^{A(r)}_O$, $S^{S(l)}_O$				& Dissolved oxygen								& g O$_2$ m$^{-3}$	    	 & D, S/M, S			\\[0.25ex]
        $S^{IN}_{NO}$, $S^{A(r)}_{NO}$, $S^{S(l)}_{NO}$		& Nitrate and nitrite nitrogen					& g N m$^{-3}$				 & D, S/M, S			\\[0.25ex]
        $S^{IN}_{NH}$, $S^{A(r)}_{NH}$, $S^{S(l)}_{NH}$ 	& Ammonium plus ammonia nitrogen				& g N m$^{-3}$				 & D, S, S/M($l=10$)	\\[0.25ex]
        $S^{IN}_{ND}$, $S^{A(r)}_{ND}$, $S^{S(l)}_{ND}$ 	& Soluble biodegradable organic nitrogen		& g N m$^{-3}$				 & D, S, S			    \\[0.25ex]
        $X^{IN}_{ND}$, $X^{A(r)}_{ND}$						& Particulate biodegradable organic nitrogen	& g N m$^{-3}$				 & D, S				    \\[0.25ex]
        $S^{IN}_{ALK}$, $S^{A(r)}_{ALK}$, $S^{S(l)}_{ALK}$	& Alkalinity									& mol HCO$_3^-$ m$^{-3}$     & D, S				    \\[0.25ex]
        $X^{S(l)}_{SS}$										& Total suspended solids						& g COD m$^{-3}$			 & S/M($l=10$)		    \\[0.25ex]
        $Q_{IN}$											& Influent flow-rate							& m$^{3}$ d$^{-1}$			 & D					\\[0.25ex]
        $Q_A$												& Internal recirculation flow-rate				& m$^{3}$ d$^{-1}$			 & C 				    \\[0.25ex]
        $Q_R$												& External recirculation flow-rate				& m$^{3}$ d$^{-1}$			 & C					\\[0.25ex]
        $Q_W$												& Wastage flow-rate 							& m$^{3}$ d$^{-1}$			 & C					\\[0.25ex]
        $Q_{EC}^{(r)}$										& External carbon source flow-rate				& m$^{3}$ d$^{-1}$			 & C					\\[0.25ex]
        $K_La^{(r)}$										& Oxygen transfer coefficient					& d$^{-1}$					 & C					\\[0.25ex]
        $BOD_5^{S(10)}$										& Biochemical oxygen demand						& g COD m$^{-3}$			 & M					\\[0.25ex]
        $COD^{S(10)}$										& Chemical oxygen demand						& g COD m$^{-3}$			 & M					\\[0.25ex]
        $N_{TOT}^{S(10)}$									& Total nitrogen								& g N m$^{-3}$               & M					\\[1ex]
        \hline 
    \end{tabular} 
    \label{tab: ASP_Notation}
\end{table*}

This section briefly overviews the main components of the BSM1. We present the set of BSM1's variables which are endowed with dynamics, the state variables, and highlight the set-up of measurement and control variables chosen for this study. The process layout is shown in Figure \ref{fig: BSM1} and the variables are overviewed in Table \ref{tab: ASP_Notation}. In this section, we also overview the stormy-weather scenario chosen for the influent wastewater used for the experiments. For completeness, the detailed model is reproduced in the Supplementary Material. 

\subsubsection{Model dynamics, measurements, and control} \label{subsubsec: BSM1-model}

The dynamics of each reactor $A{(r)}$, with $r=1,\dots,5$, are described by the evolution of the concentration of $13$ species
\begin{multline} \label{eq: Xa_vars}
	S_I^{A(r)}, S_S^{A(r)}, X_I^{A(r)}, X_S^{A(r)}, X_{BH}^{A(r)}, X_{BA}^{A(r)}, X_{P}^{A(r)}, \\ S_O^{A(r)}, S_{NO}^{A(r)}, S_{NH}^{A(r)}, S_{ND}^{A(r)}, X_{ND}^{A(r)}, S_{ALK}^{A(r)}. 
\end{multline}
The oxygen transfer coefficient $K_La^{(r)}$, a proxy quantity for characterising the degree of aeration of the reactor, and the flow-rate $Q_{EC}^{(r)}$ of external carbon are used as actionable quantities that can be manipulated to control the operating conditions of the $r$-th reactor. For each reactor, we assume that the concentration of dissolved oxygen ($S_O^{A(r)}$) and nitrate- and nitrite-nitrogen ($S_{NO}^{A(r)}$) can be measured with online sensors. 

As for the settler, the dynamics of each layer $S(l)$, with $l=1,\dots,10$, is described by concentration of eight species
\begin{multline} \label{eq: Xs_vars}
	X_{SS}^{S(l)}, S_{I}^{S(l)}, S_S^{S(l)}, S_O^{S(l)}, S_{NO}^{S(l)}, S_{NH}^{S(l)}, S_{ND}^{S(l)}, S_{ALK}^{S(l)}. 
\end{multline}
We assume that the quality of clarified water is measured with online analysers in terms of concentrations $X_{SS}^{S(10)}$, $S_{NH}^{S(10)}$, and $N_{TOT}^{S(10)}$, whereas oxygen demand ($BOD_{5}^{S(10)}$ and $COD^{S(10)}$) is only available from offline laboratory analysis. 

The internal and external recycle flow-rates ($Q_A$ and $Q_R$, respectively) and the wastage flow-rate ($Q_W$) are the other actionable quantities that can be used to manipulate the operating conditions of the treatment plant. As for the quality of these streams, we assume that concentrations in the internal recycle are equal to the concentrations in the fifth reactor $A(5)$, whereas the external recycle and wastage have properties equal to those at the bottom layer $S(1)$ of the settler.

\subsubsection{Characterisation of the influent wastewater} \label{subsubsec: BSM1-influent}

The BSM1 is accompanied by influent wastewater data corresponding to \textit{i)} dry-, \textit{ii)} rainy-, and \textit{iii)} stormy-weather scenarios. For each scenario, the wastewater from the primary settler entering the activated sludge plant is characterised in terms of flow-rate ($Q_{IN}$) and concentration of $13$ compounds,
\begin{multline} \label{eq: Xi_vars}
    S_I^{IN}, S_S^{IN}, X_I^{IN}, X_S^{IN}, X_{BH}^{IN}, X_{BA}^{IN}, X_{P}^{IN}, \\ S_O^{IN}, S_{NO}^{IN}, S_{NH}^{IN}, S_{ND}^{IN}, X_{ND}^{IN}, S_{ALK}^{IN}. 
\end{multline}
The influent is always of the municipal kind and non-actionable: Its properties are treated as process disturbances.

We focus on the problem of controlling ASPs subjected to a stormy weather. 
In the BSM1, this scenario lasts $T = 14$ days and it generates an incoming wastewater according to a model of urban activity which follows daily and weekly patterns of wastewater production, plus two high-intensity stormy events in week two \cite{Gernaey2014}. Due to the extreme events, this is the most challenging scenario from a control perspective. The data is based on observations from real plants and adapted to represent a 100~000 population-equivalent influent load, as described in \cite{Vanhooren1996}. The main characteristics of the influent wastewater, in the stormy-weather scenario, are in Table \ref{tab: Influent_Averages} in terms of average flow-rates and flow-weighted average compositions. Moreover, we have $X_{BA}^{IN} = X_{P}^{IN} = S_{O}^{IN} = S_{NO}^{IN} = 0$ g~m$^{-3}$ and $S_{ALK}^{IN} = 7$ mol~HCO$_3^-$~m$^{-3}$. 

\begin{table}[htb!] \centering 
    \caption{Influent wastewater: A selection of averaged conditions.}
    \begin{tabular} {c | c | c | l} 
    {Variable}		   & Week 1  & Week 2 & {Units}    \\ 
    \doublehline
    ${Q}_{IN} $        & 18446   & 21039   & m$^{3}$ d$^{-1}$\\[0.5ex]
    ${S}^{IN}_I $      & 30      &   26.30 & g COD m$^{-3}$	\\[0.25ex]	
    ${S}^{IN}_S $      & 69.5    &   60.93 & g COD m$^{-3}$	\\[0.25ex]	
    ${X}^{IN}_I $      & 51.2    &   52.54 & g COD m$^{-3}$	\\[0.25ex]	
    ${X}^{IN}_S $      & 202.32  &  185.45 & g COD m$^{-3}$	\\[0.25ex]	
    ${X}^{IN}_{BH} $   & 28.17   &   26.44 & g COD m$^{-3}$	\\[0.25ex]	
    ${S}^{IN}_{NH} $   & 31.56   &   27.66 & g N m$^{-3}$	\\[0.25ex]
    ${S}^{IN}_{ND} $   & 6.95    &    6.09 & g N m$^{-3}$	\\[0.25ex]	
    ${X}^{IN}_{ND} $   & 10.59   &    9.94 & g N m$^{-3}$	\\[1ex]
    \hline
    \end{tabular} 
    \label{tab: Influent_Averages}
\end{table}

\begin{figure}[ht!] \centering
	\includegraphics[width=\columnwidth]{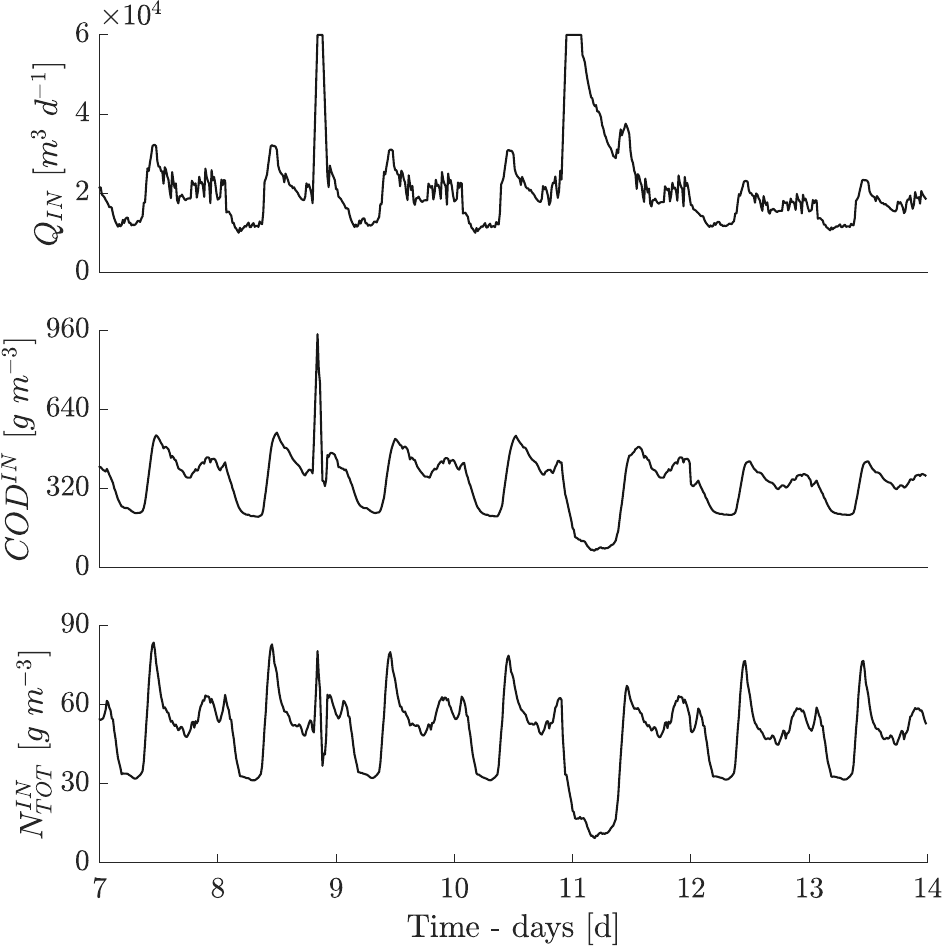}    
	\caption{Influent wastewater, week 2: Evolution of influent flow-rate $Q_{IN}$, oxygen demand $COD^{IN}$, and nitrogen $N_{TOT}^{IN}$.}
    \label{fig: W_Dryweather}
\end{figure}

In Figure \ref{fig: W_Dryweather}, we visualise the influent wastewater during the second week. To highlight the diurnal and weekly periodicities, we show flow-rate ($Q_{IN}$), chemical oxygen demand ($COD^{IN}$), and total nitrogen ($N_{TOT}^{IN}$). The stormy events occur around the $9$-th and $11$-th day, when the flow-rate peaks significantly and the concentrations of nitrogen and organic matter either increase or decrease noticeably due to dilution.

\section{Output model predictive control} \label{sec: OutputMPC}
Output model predictive control (Output MPC, \cite{Rawlings2020}) is a systematic approach to design and execute controllers to operate physical systems, like an activated sludge plant, according to practical objectives. The control actions deployed to the plant are computed as solution to an optimisation problem based on a predictive model of the system and the plant's actuators and measurements, in our case the BSM1. As the objectives can be set by the plant's management, they explicitly reflect certain technological, environmental, or economical targets. 

\begin{figure}[t!] \centering   
   	\includegraphics[width=\columnwidth]{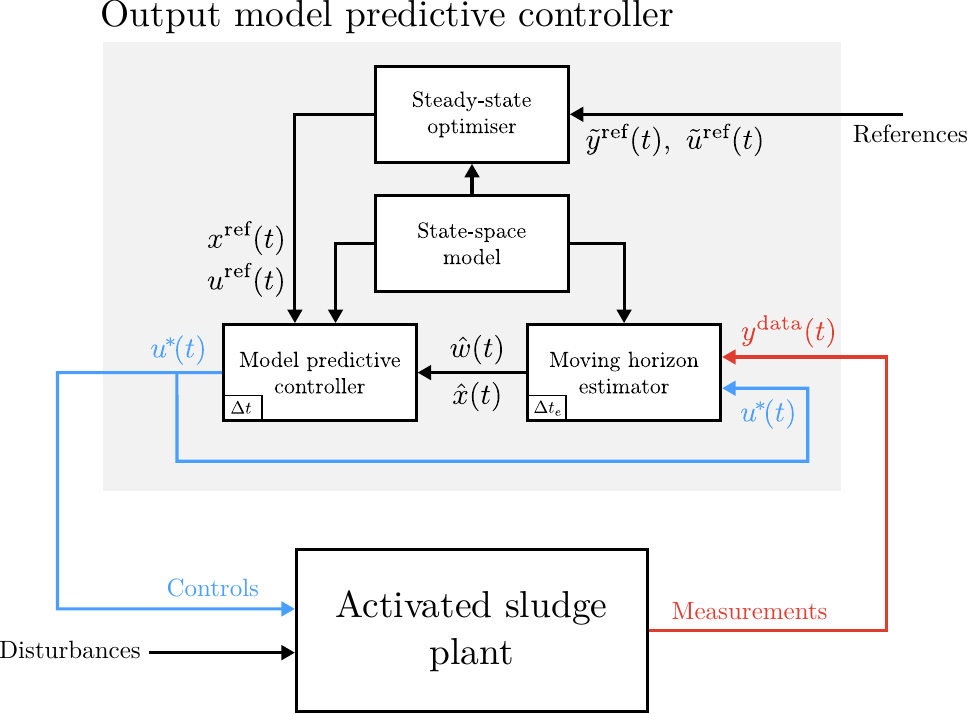}  
	
	\caption{Output MPC: Main components. To follow a reference, the controller uses a state-space model and measurement data (red) to determine the controls (blue) given to the plant.} 
    \label{fig: OutputMPC_Scheme} 
	\vspace*{-0.35cm}
\end{figure}

An Output MPC (Figure \ref{fig: OutputMPC_Scheme}) consists of two main components: \textit{i)} a model predictive controller (MPC) and \textit{ii)} a state estimator. The MPC computes the control actions that drive the plant to satisfy a reference trajectory as well as possible. MPC actions optimise the plant's evolution, as predicted by a dynamic model, from the knowledge of its current state. Since the state is not known and, possibly, not even measurable, the MPC relies on an estimator to reconstruct it from measurement data: In our framework, the moving-horizon estimator (MHE). MPC actions and MHE estimates are computed recursively as solutions to independent optimisation problems, subjected to technological constraints and operational desiderata explicitly set by the plant's personnel. Once computed, control actions are deployed to the plant's actuators as set-points to their low-level PIDs.

In the following, the Output MPC is overviewed for a general system for which a dynamic model is available: The configuration for activated sludge plants, like the BSM1, is presented in Section \ref{sec: ControlFramework}. Firstly, we review the predictive model (Section \ref{subsec: SS}) used by the Output MPC, then overview the controller (Section \ref{subsec: MPC}) and state estimator (Section \ref{subsec: MHE}). 

\subsection{The predictive model} \label{subsec: SS}
The Output MPC is based on a, possibly time-varying, representation in state-space form of the system to be controlled,
\vspace*{-0.40cm}
\begin{subequations}
\begin{align}
	x(t+dt)	& = x(t) + f_t(x(t), u(t), w(t) \mid \theta_x) dt	\label{eq: nonlinSS_x} \\
	y(t)	& = g_t(x(t) \mid \theta_y) + v(t).					\label{eq: nonlinSS_y}
\end{align} \label{eq: SS}%
\end{subequations} 
The state equation \eqref{eq: nonlinSS_x} models the evolution of the $N_x$ state variables, given \textit{i)} their value $x(t) \in \mathbb{R}^{N_x}$, \textit{ii)} the value $u(t) \in \mathbb{R}^{N_u}$ of $N_u$ actionable input or control variables, and \textit{iii)} $N_w$ disturbances $w(t) \in \mathbb{R}^{N_w}$, at time $t$. The output equation \eqref{eq: nonlinSS_y} models how the state $x(t)$ is emitted as $N_y$ noisy output variables $y(t) \in \mathbb{R}^{N_y}$. The vectors $\theta_{x}$ and $\theta_{y}$ collect the parameters in the dynamics $f_t$ and in the output function $g_t$.

The controller is understood as a device which computes, at each $t$, all the actions $u$ that evolve the state $x$ in such a way that the model output $y(x)$ follows, or tracks, a sequence of future reference values $y^{\text{ref}}$ provided by the plant's personnel. The assumption is that the physical plant evolves as the model predicts, thus model output $y$ and plant's measurements $y^{\text{data}}$ match to some degree. The calculation of the control actions must be amenable to a computer implementation: This can be achieved by adopting a \textit{discretise-then-optimise} strategy \cite{Betts2010}, according to which the continuous-time model (Eq. \eqref{eq: SS}) and the MPC and MHE optimisations (Section \ref{subsec: MPC} and \ref{subsec: MHE}), are discretised in time and then solved numerically. 

\begin{figure}[tb!] \centering   
    \includegraphics[width=0.95\columnwidth]{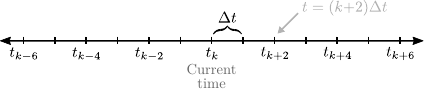}

    \caption{Discrete-time model: Partitioning of the time axis.}  
    \label{fig: Time_line_dynamics} 
	\vspace*{-0.18cm}
\end{figure}

In discrete-time, to evolve the model's state (Eq. \eqref{eq:  nonlinSS_x}), we partition the time axis in intervals of duration $\Delta{t}$ (Figure \ref{fig: Time_line_dynamics}). For state, controls, and disturbances at time $t_k = k\Delta{t}$, we thus have $x(t_k) = x(k\Delta{t}) = x_{k}$, $u(t_k) = u(k\Delta{t}) = u_k$, and $w(t_k) = w(k\Delta{t}) = w_k$. By keeping the inputs constant between $t_k$ and $t_{k+1}$ (the common zero-order-hold setting in which $u(t) = u_k$ and $w(t) = w_k$, for $t \in [t_k, t_{k+1})$), we have that the next state $x_{k+1} = x((k{+}1)\Delta{t}) = x(t_{k+1})$ is given by
\vspace*{-0.10cm}
\begin{equation} \label{eq: Transition_Function}
	x_{k+1} =  \underbrace{x_{k} + \int_{t_k}^{t_{k+1}}{f_t(x(t), u_k, w_k \mid \theta_x)dt}}_{f_{\Delta{t} | t_k}( x_k, u_k, w_k \mid \theta_x)}.
\end{equation}
The integral in Eq. \eqref{eq: Transition_Function} cannot always be evaluated in closed-form and the transition function $f_{\Delta{t}|t_k}$ need be approximated using numerics for integrating ordinary differential equations.

Similarly, in discrete-time, the output $y_k = y(k \Delta{t})$ at time $t_k$, when in state $x_k$, with noise $v_k = v(t_k)$, can be written as
\begin{equation} \label{eq: nonlinSS_yk}
y_k = g_{t_k}(x_k \mid \theta_y) + v_k.
\end{equation}

\subsection{The model predictive controller (MPC)} \label{subsec: MPC}

A model predictive controller plans the control actions which optimally evolve a system from its initial state $x(t_{k})$, at time $t_{k}$, for a period of time of duration $H_c$, the \textit{control-horizon}. The notion of optimality is general and problem-specific: In this work, it is defined in terms of \textit{closeness to a reference trajectory} and of \textit{magnitude of the control effort}. The planning is done under the assumption that \textit{i}) the actual system evolves as the model predicts and that \textit{ii}) future disturbances entering the system are known beforehand. To account for disturbance and model uncertainties, only the first planned action is actually deployed to the system, then the planning is repeated, at time $t_k + dt$, for the next horizon. After the first action is deployed, a new cycle is executed at time $t_k + 2dt$.

For a cycle starting at $t_k$, the sequence of optimal actions is the function (of time) $u^*: [t_{k},t_{k} + H_c] \to \mathbb{R}^{N_u}$ which solves
\begin{subequations}
    \begin{align}
        \min_{\scriptsize\begin{matrix} u(\cdot) \\ x(\cdot)\end{matrix}} 
        	\quad 
        	& \int_{t_{k}}^{t_k + H_c}{\hskip-0.50emL^{c}_{t}(x(t), u(t))dt} + L^{f}_{t_k} (x(t_k + H_c))	\label{eq: MPC_cost}		\\ 	
		\underset{\forall t \in [t_k,t_k+H_c]}{\text{s.t.}} 
			\ 
    	    & x(t + dt) = x(t) + f_t(x(t),u(t),\widehat{w}(t))dt,											\label{eq: MPC_dynamics}	\\
	        & x(t) \in \mathcal{X}^{c}(t), \ u(t) \in \mathcal{U}(t),											\label{eq: MPC_sets}		\\ 	
	        & x(t_{k}) = \widehat{x}(t_{k}).																\label{eq: MPC_init}
    \end{align} \label{eq: MPC}%
\end{subequations} 
$L^{c}_{t} : \mathbb{R}^{N_x} \times \mathbb{R}^{N_u} \to \mathbb{R}$ and $L^{f}_{t_k} : \mathbb{R}^{N_x} \to \mathbb{R}$ in Eq. \eqref{eq: MPC_cost} denote \textit{stage} and \textit{terminal cost functions}, respectively: In our case, they quantify how well the state evolution would follow a reference trajectory. By solving \eqref{eq: MPC}, we find the best sequence of control actions to keep the state as close as possible to the reference: Should this sequence $u^*$ be applied, the state evolution would be $x^*: [t_{k},t_{k} + H_c] \to \mathbb{R}^{N_x}$.

At each time $t  \in [t_{k},t_{k} + H_c]$, the solution is constrained to satisfy the dynamics (Eq. \eqref{eq: MPC_dynamics}) and to be inside the \textit{constraint sets} $\mathcal{X}^c$ and $\mathcal{U}$ (Eq. \eqref{eq: MPC_sets}). $\mathcal{X}^c$ and $\mathcal{U}$ are used to express technical conditions that states $x^*$ and controls $u^*$ must satisfy at each $t$. Often, constraint sets are stated as inequalities: $\mathcal{X}^c(t) = \{x(t): h^c_{x|t}(x(t)) \le 0\}$ and $\mathcal{U}(t) = \{u(t): h_{u|t}(u(t)) \le 0\}$. To account for arbitrary relations that state and control variables must satisfy, jointly across the control-horizon, the formulation can be generalised as $(x(\cdot),u(\cdot)) \in \mathcal{Z}_{t_k}$ such that $\mathcal{Z}_{t_k} = \{(u(\cdot),x(\cdot)) \mid h_{x,u | t_k}(x(\cdot),u(\cdot)) \le 0\}$.

Notice how, in problem \eqref{eq: MPC}, the initial and future disturbances and the initial state, unknown quantities at $t_k$, are replaced by estimates $\widehat{w}(t)$ and $\widehat{x}(t_k)$. Before each cycle, initial state and disturbance are reconstructed from data using the state estimator (Section \ref{subsec: MHE}), whereas future disturbances can be assumed to remain constant and equal to their initial value.

\begin{figure}[htb!] \centering   
    \includegraphics[width=0.95\columnwidth]{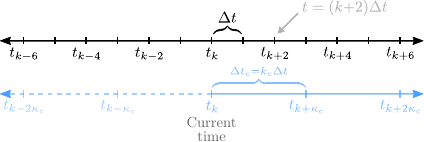}

    \caption{MPC: Partitioning of the time axis according to the interval $\Delta t_c$ at which the controller is operated, in blue.}  
    \label{fig: Time_line_mpc} 
\end{figure}

\begin{figure}[ht!] \centering
    \includegraphics[width=\columnwidth]{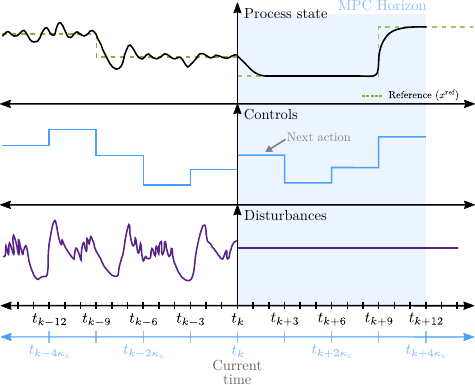}

    \caption{MPC: One control cycle, with known initial state $x_k$ and disturbance $w_k$ and constant future disturbances $\{w_k\}_{n_c}$.}
    \label{fig: MPC_Illustration}
\end{figure}

For computation, the continuous-time MPC \eqref{eq: MPC} is transcribed into discrete-time. This is done by firstly setting the interval $\Delta{t}_c$ at which the controller is operated (Figure \ref{fig: Time_line_mpc}). It is natural to cycle a MPC at $\kappa_c \in \{1,2,\dots\}$ times the model discretisation interval $\Delta{t}$ (Eq. \eqref{eq: Transition_Function}): That is, the controller is executed at times $\Delta{t}_c = \kappa_c\Delta{t}$. The control-horizon $H_c = N_c \times \Delta{t}_c$ is then partitioned accordingly in $N_c$ stages, corresponding to the number $N_c$ of $N_u$ actions to be computed in the cycle. In the continuous-time MPC \eqref{eq: MPC}, the integration bounds are then re-written  to get $t_k = k\Delta{t}$ and $t_k + H_c = k\Delta{t} + N_c(\kappa_c\Delta{t}) = (k + N_c\kappa_c)\Delta{t}$, the integral is approximated by a finite number of sums, and the constraints are adapted, accordingly. The resulting discrete-time controller takes the form of a constrained optimisation problem,
\begin{subequations}
\begin{align}
    \min_{\scriptsize\begin{matrix} \{u_{k + n_c\kappa_c} \}_{n_c=0}^{N_c-1} \\[1ex] \{x_{k + n_c\kappa_c}\}_{n_c=0}^{N_c} \end{matrix}}
        							& \sum_{n_c = 0}^{N_c-1}{L^{c}_{k + n_c\kappa_c}(x_{k + n_c\kappa_c}, u_{k + n_c\kappa_c})} 
									\nonumber\\[-4ex]
									& \hspace*{12em} + L^{f}_{k}(x_{k+N_c\kappa_c})	\label{eq: dMPC_cost}		\\
    	\underset{\forall n_c \in \{0,\dots,N_c-1\}}{\text{s.t.}}																								\nonumber 					\\
        x_{k + (n_c + 1)\kappa_c}	& = f_{\Delta{t}_c | k + n_c\kappa_c}(x_{k + n_c\kappa_c},u_{k + n_c\kappa_c},\widehat{w}_{k + n_c\kappa_c}),				\label{eq: dMPC_dynamics}	\\
        x_{k + n_c\kappa_c}			& \in \mathcal{X}^c_{k + n_c\kappa_c}, \ u_{k + n_c\kappa_c} \in \mathcal{U}_{k + n_c\kappa_c},	    						\label{eq: dMPC_sets}		\\ 	
        x_k							& = \widehat{x}_k.								        																	\label{eq: dMPC_init}
\end{align} \label{eq: MPCDiscr}%
\end{subequations} 
The sequence of optimal actions to be computed consists of the collection $(u^*_{k + n_c\kappa_c})_{n_c=0}^{N_c-1} = u^*_{k}, u^*_{k + \kappa_c}, \dots, \allowbreak u^*_{k + (N_c-1)\kappa_c}$ of $N_c$ controls to be applied at times $\{t_{k + n_c\kappa_c}\}_{n_c=0}^{N_c-1}$ and held constant between them. If applied, the resulting evolution of the state would be $(x^*_{k + n_c\kappa_c})_{n_c=0}^{N_c}$. However, only the first action $u^*_k$ is deployed at $t_k$ and held constant for a period $\Delta{t}_c$, after which all actions are re-computed in a new MPC cycle. The approach is shown in Figure \ref{fig: MPC_Illustration} for $\kappa_c = 3$ (for $\Delta{t}_c = 3\Delta{t}$), assuming that the disturbances remain constant over the control-horizon: That is, for $(\widehat{w}_{k+n_c\kappa_c} = \widehat{w}_{k})_{n_c=0}^{N_c-1}$. 

\subsubsection{Cost functions for reference tracking} \label{sec: Topt}  
The notion of being \textit{close to a reference trajectory} is encoded by the stage and terminal cost functions. For a $H_c$-long reference trajectory $(x_{k + n_c\kappa_c}^{\text{ref}})_{n_c=0}^{N_c}$, optimal tracking is achieved by minimising the (squared) mismatch between the state of the system ($x_{k + n_c\kappa_c}$, according to the model) and $x_{k + n_c\kappa_c}^{\text{ref}}$: That is,
\begin{subequations}
\begin{align} 
    L^{c}_{k + n_c\kappa_c}(\cdot, \cdot)	
        & = \| x_{k + n_c\kappa_c} - x_{k + n_c\kappa_c}^{\text{ref}} \|^2_{Q_{c|k + n_c\kappa_c}}	\nonumber\\ 
    	& \quad + \| u_{k + n_c\kappa_c} - u_{k + n_c\kappa_c}^{\text{ref}} \|^2_{R_{c | k + n_c\kappa_c}};		\label{eq: MPC_Quadratic_Costs_stage}	\\
    L^{f}_{k}(\cdot)
        & = \| x_{k+N_c\kappa_c} - x_{k+N_c\kappa_c}^{\text{ref}} \|^2_{Q_{f|k}}.					\label{eq: MPC_Quadratic_Costs_terminal}
\end{align} \label{eq: MPC_Quadratic_Costs}%
\end{subequations}
At each stage, we include an additional cost term quantifying the \textit{magnitude of the controls} ($u_{k + n_c\kappa_c}$) with respect to a reference value $u_{k + n_c\kappa_c}^{\text{ref}}$. This is a customary practice aiming at limiting the selection of large control actions. In Eqs. \eqref{eq: MPC_Quadratic_Costs}, the square matrices $Q_{c|k + n_c\kappa_c},Q_{f|k} \succeq 0$ and $R_{c|k + n_c\kappa_c} \succ 0$ are user-defined tuning parameters: At each stage, they quantify the relative relevance of individual state and control variables.

Reference trajectories of practical relevance are rarely formulated in terms of state-variables, but rather in terms of easy-to-measure or to compute quantities. From the model's perspective, this can be a selection $\tilde{y} \subseteq y$ of output variables for which an actual plant's measurement (or estimate) exists or, more generally, for some transformation $\tilde{y} = h(y)$ that models an existing instrument. To allow the MPC to track such references and thus effectively be an Output MPC, the $\tilde{y}^{\text{ref}}$-trajectory must be converted into an equivalent $(x^{\text{ref}},u^{\text{ref}})$-trajectory to be used in the cost functions in Eq. \eqref{eq: MPC} and \eqref{eq: MPC_Quadratic_Costs}.
\begin{itemize}
\item[$\leadsto$] Because the outputs are function of the state variables, Eq \eqref{eq: nonlinSS_yk}, the conversion can be done with a steady-state optimiser which determines the pair $(x^{\text{ref}},u^{\text{ref}})$ of state and control variables that stabilises the system at $\tilde{y}^{\text{ref}}$.
\end{itemize}
    
\subsubsection*{Steady-state optimiser} \label{sec: SSOpt}
Because $y_k$ is the collection of the $N_{y}$ outputs returned by the model (Eqs. \eqref{eq: nonlinSS_y} and \eqref{eq: nonlinSS_yk}) at time $t_k$, we let $\tilde{y}_k = h(g(x_k))$ be the $N_{\tilde{y}}$ key performance indicators for which a trajectory to be tracked is given. At time $t_k$, the state and control values that keep the model in equilibrium around the reference value $\tilde{y}_{k + n_c\kappa_c}^{\text{ref}}$ correspond to the pair $(x_{k+n_c\kappa_c}^{\text{ref}},u_{k+n_c\kappa_c}^{\text{ref}})$ which solves
\begin{subequations}\begin{align} 
    \min_{\scriptsize\begin{matrix} x^{\text{ref}}_{k+n_c\kappa_c} \\ u^{\text{ref}}_{k+n_c\kappa_c} \end{matrix}}
		& \quad \| h(g(x_{k+n_c\kappa_c}^{\text{ref}})) - \tilde{y}_{k+n_c\kappa_c}^{\text{ref}} \|^2_{W_{y|k+n_c\kappa_c}}  				\nonumber\\[-3ex]
    	& \quad \hspace*{6em} + \| u_{k+n_c\kappa_c}^{\text{ref}} - \tilde{u}_{k+n_c\kappa_c}^{\text{ref}} \|^2_{W_{u|k+n_c\kappa_c}}	\label{eq: SS_OPT_a}				\\ 
    \text{s.t.}
		& \quad 0 = f(x_{k+n_c\kappa_c}^{\text{ref}},u_{k+n_c\kappa_c}^{\text{ref}},w_{k+n_c\kappa_c}^{\text{ref}}|\theta_x),	\label{eq: SS_OPT_b}	\\
    	& \quad x_{k+n_c\kappa_c}^{\text{ref}} \in \mathcal{X}_{k+n_c\kappa_c}^{\text{ref}}, \ u_{k+n_c\kappa_c}^{\text{ref}} \in \mathcal{U}_{k+n_c\kappa_c}^{\text{ref}}.	\label{eq: SS_OPT_c} 
\end{align} \label{eq: SS_OPT}%
\end{subequations}    
The matrices $W_{y|k+n_c\kappa_c}, W_{u|k+n_c\kappa_c} \succeq 0$ are used to quantify the trade-off between matching the reference $\tilde{y}_{k+n_c\kappa_c}^{\text{ref}}$ and limiting the control effort $u_{k+n_c\kappa_c}^{\text{ref}}$: Setting $W_{u|k+n_c\kappa_c} = 0$ leads to solutions that only aim at matching $\tilde{y}_{k+n_c\kappa_c}^{\text{ref}}$, whereas $W_{u|k+n_c\kappa_c} \succ 0$ and $\tilde{u}_{k+n_c\kappa_c}^{\text{ref}} {=} 0$ leads to solutions of minimum control. $\mathcal{X}_{k+n_c\kappa_c}^{\text{ref}}$ and $\mathcal{U}_{k+n_c\kappa_c}^{\text{ref}}$ are sets of constraints: They can, but need not, be equal to $\mathcal{X}^c_{k+n_c\kappa_c}$ and $\mathcal{U}_{k+n_c\kappa_c}$ in Eq. \eqref{eq: MPCDiscr}. Steady-state conditions are enforced by an equality constraint that sets the model dynamics to be equal to zero, Eq. \eqref{eq: SS_OPT_b}.
   
\subsubsection{A constrained linear-quadratic MPC} \label{sec: lqMPC}
Solutions of the MPC in Eq. \eqref{eq: MPCDiscr} are computationally demanding. 
While considering quadratic costs as in Eq. \eqref{eq: MPC_Quadratic_Costs} sets the right path towards improvements, a substantial reduction of the load can be obtained when the static constraints (Eq. \eqref{eq: dMPC_sets}) are given as linear inequalities (for example, when they define minimum and maximum values that control and state variables can take) and the dynamic constraints (Eq. \eqref{eq: dMPC_dynamics}) are linear equalities, corresponding to linear-affine dynamics,
\begin{subequations}
\begin{align}
    x_{k + (n_c + 1)\kappa_c}	& = f_{\Delta{t}_c}(x_{k + n_c\kappa_c},u_{k + n_c\kappa_c},\widehat{w}_{k + n_c\kappa_c})							\nonumber \\
    							& \approx A_{\Delta{t}_c | k+n_c\kappa_c} x_{k+n_c\kappa_c} + B_{\Delta{t}_c | k+n_c\kappa_c} u_{k+n_c\kappa_c}	\nonumber \\
								& \hskip1.25em + G_{\Delta{t}_c | k+n_c\kappa_c} \widehat{w}_{k+n_c\kappa_c} + {z}_{\Delta{t}_c | k+n_c\kappa_c}.			
\end{align} \label{eq: MPC_DiscrSS}%
\end{subequations}
An MPC with quadratic costs, linear inequality constraints, and linear-affine dynamics, though still nonlinear, becomes a convex program and can be solved efficiently \cite{Boyd2004}.  

Convenient linearisations of the plant's dynamics can be obtained by evaluating the model $f$ and its Jacobian matrices ($\partial{f}/\partial{x}$, $\partial{f}/\partial{u}$ and $\partial{f}/\partial{w}$) at the reference points $p_{k+n_c\kappa_c} = (x_{k+n_c\kappa_c}^{\text{ref}}, u_{k+n_c\kappa_c}^{\text{ref}}, w_{k+n_c\kappa_c}^{\text{ref}}, y_{k+n_c\kappa_c}^{\text{ref}})$, with $n_c=0,\dots,N_c$. As a result, we get the offset vectors $z_{\Delta{t}_c} \in \mathbb{R}^{N_x}$ and the matrices $A_{\Delta{t}_c} \in \mathbb{R}^{N_x \times N_x}$, $B_{\Delta{t}_c} \in \mathbb{R}^{N_x \times N_u}$, and $G_{\Delta{t}_c} \in \mathbb{R}^{N_x \times N_w}$ for the linear-affine dynamics in Eq. \eqref{eq: MPC_DiscrSS}. Importantly, note that the Jacobian $\partial{g}/\partial{x}$ resulting from linearising the measurement equation at $p_{k+n_c\kappa_c}$ yields the output matrices $C_{k+n_c\kappa_c}\in \mathbb{R}^{N_y \times N_x}$. Such output matrices $C$ can be used to ensure the closed-loop stability of the resulting constrained linear-quadratic MPC \cite{Mayne2000}. 
\begin{itemize}
\item[$\leadsto$] Because of its computational efficiency, while retaining accurate dynamics, a constrained linear-quadratic MPC is the method of choice in our application (Section \ref{sec: ControlFramework}).
\end{itemize}

\subsection{The moving-horizon state estimator (MHE)}  \label{subsec: MHE}

The estimates $\widehat{x}(t_k)$ and $\widehat{w}(t_k)$ used in the continuous-time MPC \eqref{eq: MPC} can be obtained as terminal values of the state trajectory $\widehat{x} : [t_{k}-H_e,t_{k}] \to \mathbb{R}^{N_x}$ and the disturbance trajectory $\widehat{w} : [t_{k}-H_e,t_{k}] \to \mathbb{R}^{N_w}$ that solve an optimal state estimation problem, over a past period of duration $H_e$, the \textit{estimation-horizon}. Optimality is defined in terms of \textit{closeness between the past model outputs and plant's measurements}, and of the \textit{magnitude of the disturbances}, under the controls deployed to the plant over the estimation-horizon. 

For a controller's cycle starting at $t_k$, the associated moving-horizon estimator determines the  trajectories $\widehat{x}$ and $\widehat{w}$ from
\begin{subequations}
\begin{align}
        \min_{\scriptsize\begin{matrix} w(\cdot) \\ x(\cdot) \end{matrix}} \quad
        	& L^{i}_{t_{k}} (x(t_k - H_e)) + \int_{t_{k}-H_e}^{t_k}{\hskip-0.50emL^{e}_{t}({x}(t), {w}(t))dt}	\label{eq: MHE_cost}		\\ 	
		\underset{\forall t \in [t_{k}{-}H_e,t_{k}]}{\text{s.t.}} 
    	    & \ {x}(t + dt) = {x}(t) + f_t({x}(t),u^*(t),{w}(t))dt,										\label{eq: MHE_dynamics}	\\
	        & \ {x}(t) \in {\mathcal{X}^e}(t), \ {w}(t) \in {\mathcal{W}}(t).								\label{eq: MHE_sets}
\end{align} \label{eq: MHE}%
\end{subequations}
$L_{t_k}^{i} : \mathbb{R}^{N_x} \to \mathbb{R}$ and $L_{t}^{e} : \mathbb{R}^{N_x} \times \mathbb{R}^{N_u} \to \mathbb{R}$ denote \textit{initial} and \textit{stage cost functions}, Eq. \eqref{eq: MHE_cost}: They are used to quantify how well the estimated states and disturbances determine model outputs from their match to plant's data. Similarly to \eqref{eq: MPC}, at each time $t$, the solution is constrained to satisfy the dynamics (Eq. \eqref{eq: MHE_dynamics}) and the \textit{constraint sets} ${\mathcal{X}^e}(t) = \{{x}(t): h^e_{{x}|t}({x}(t)) \le 0\}$ and ${\mathcal{W}}(t) = \{{w}(t): h_{{w}|t}({w}(t)) \le 0\}$, which now encode restrictions on state and disturbances, Eq. \eqref{eq: MHE_sets}. Should the optimal estimates $(\widehat{w}(t))_{t=t_k-H_e}^{t_k}$ of the disturbances and the past optimal controls $(u^*(t))_{t=t_k-H_e}^{t_k}$ be used to evolve model \eqref{eq: SS}, the resulting state estimates $(\widehat{x}(t))_{t=t_k - H_e}^{t_k}$ would emit the sequence of output variables $(\widehat{y}(t) = g_t(\widehat{x}(t)))_{t=t_k-H_e}^{t_k}$ that best match the actual plant's data $(y^{\text{data}}(t))_{t=t_k - H_e}^{t_k}$ measured over the estimation-horizon.

\begin{figure}[b!] \centering   
    \includegraphics[width=0.95\columnwidth]{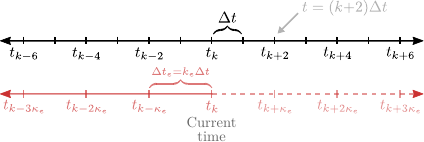}

    \caption{MHE: Partitioning of the time axis according to the interval $\Delta t_e$ at which data are collected, in red.}  
    \label{fig: Time_line_mhe} 
\end{figure}

\begin{figure}[htb!] \centering
    \includegraphics[width=\columnwidth]{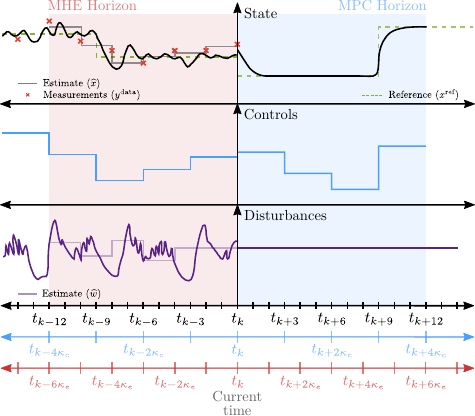}

    \caption{MHE: Illustration for a single estimation horizon.}
    \label{fig: MHE_Illustration}
\end{figure}

To account for plant's measurements that may be collected at different points in time, as well as for permitting practical computations, also the MHE \eqref{eq: MHE} is transcribed into a constrained optimisation problem. Following a \textit{discretise-then-optimise} strategy, firstly the interval $\Delta{t}_e = \kappa_e \Delta{t}$ at which data are collected is set and the estimation-horizon $H_e = N_e \times \Delta{t}_e$ is partitioned in $N_e$ intervals (Figure \ref{fig: Time_line_mhe}). Secondly, the integration limits in Eq. \eqref{eq: MHE} are re-written to get $t_k - H_e = k\Delta{t} - N_e(\kappa_e\Delta{t}) = (k - N_e\kappa_e)\Delta{t}$ and $t_k = k\Delta{t}$ and the constraint sets are adapted, accordingly. After approximating the integral, the discrete-time MHE is
\begin{subequations} 
\begin{align}
    \min_{\scriptsize\begin{matrix} \{{w}_{k - n_e\kappa_e} \}_{n_e=0}^{N_e} \\ \{{x}_{k - n_e\kappa_e}\}_{n_e=0}^{N_e} \end{matrix}}
        							& L^{i}_{k}({x}_{k - N_e\kappa_e}) \nonumber\\[-4ex]
									& \hspace*{4em} + \sum_{n_e = 0}^{N_e}{L^{e}_{k - n_e\kappa_e}({x}_{k - n_e\kappa_e}, {w}_{k - n_e\kappa_e})} \label{eq: dMHE_cost}\\
    	\underset{\forall n_e \in \{0,\dots,N_e\}}{\text{s.t.}}																																\nonumber 					\\
        {x}_{k - (n_e - 1)\kappa_e}	& = f_{\Delta{t}_e | k - n_e\kappa_e}({x}_{k - n_e\kappa_e},u^*_{k - n_e\kappa_e},{w}_{k - n_e\kappa_e}),							\label{eq: dMHE_dynamics}	\\
        {x}_{k - n_e\kappa_e}		& \in {\mathcal{X}^e}_{k - n_e\kappa_e}, \ {w}_{k - n_e\kappa_e} \in {\mathcal{W}}_{k - n_e\kappa_e}.	    					\label{eq: dMHE_sets}
\end{align} \label{eq: MHEDiscr}%
\end{subequations}
The sequence $(\widehat{w}_{k - n_e\kappa_e})_{n_e=0}^{N_e} = \widehat{w}_{k}, \widehat{w}_{k - \kappa_e}, \dots, \widehat{w}_{k - N_e\kappa_e}$ of $N_e{+}1$ estimated disturbances, together with the sequence of past controls $(u^*_{k-n_e\kappa_e})_{n_e=0}^{N_e}$ deployed to the plant at times $\{t_{k - n_e\kappa_e}\}_{n_e=0}^{N_e}$, result in a sequence $(\widehat{x}_{k - n_e\kappa_e})_{n_e=0}^{N_e}$ of state variables and associated output estimates $(\widehat{y}(\widehat{x}_{k - n_e\kappa_e}))_{n_e=0}^{N_e}$. Only the estimates $(\widehat{w}_k,\widehat{x}_k)$ at time $t_k$, at the end of the estimation-horizon, are used by the MPC \eqref{eq: MPCDiscr}. The computation is repeated anew at the next controller's cycle. The approach is illustrated in Figure \ref{fig: MHE_Illustration}, where we set $\kappa_e = 2$ and $\Delta{t}_e = 2\Delta{t}$.

\subsubsection{A constrained linear-quadratic MHE} \label{subsubsec: lqMHE}

To define meaningful criteria and computationally viable solutions also for estimation, we resort again to quadratic cost functions that aim at encoding a notion of closeness between model output estimates and measurement data. Thus, we set
\begin{subequations}\begin{align} 
    L^{e}_{k - n_e\kappa_e}(\cdot,\cdot)	
            & = \| g({x}_{k - n_e\kappa_e}){-}y^{\text{data}}_{k - n_e\kappa_e} \|^2_{Q_{v|k-n_e\kappa_e}^{-1}}	\nonumber\\
    		& \quad + \| {w}_{k - n_e\kappa_e} - \overline{w}_{k - n_e\kappa_e} \|^2_{R_{w|k-n_e\kappa_e}^{-1}};			\label{eq: MHE_Quadratic_Costs_stage}	\\
    L^i_{k}(\cdot)				& = \| {x}_{k-N_e\kappa_e} - \overline{x}_{k-N_e\kappa_e} \|^2_{Q_{x_{0|k}}^{-1}}.										 								\label{eq: MHE_Quadratic_Costs_terminal} 
\end{align}\end{subequations}
When adopted in \eqref{eq: MHEDiscr}, this choice of stage and initial costs corresponds to assuming that initial state, disturbances and measurements are Gaussian variables, {estimated as}
\begin{subequations}\begin{align}
    x_{k - N_e\kappa_e} & \sim	\mathcal{N}(\overline{x}_{k - N_e\kappa_e}, Q_{x_0 | k}); \\
    w_{k - n_e\kappa_e} & \sim	\mathcal{N}(\overline{w}_{k - n_e\kappa_e}, R_{w | k-n_e\kappa_e}); \\
    v_{k - n_e\kappa_e} & \sim	\mathcal{N}(0, Q_{v | k-n_e\kappa_e}).
\end{align} \label{eq: MHE_Quadratic_Costs}%
\end{subequations}
The estimates can be interpreted as \textit{maximum a posteriori} solutions to an equivalent inference problem \cite{Rao2002}. Mean vectors $\overline{x}_{k-N_e\kappa_e}$ and $(\overline{w}_{k-n_e\kappa_e})_{n_e=0}^{N_e}$ can be set recursively: $\overline{x}_{k-N_e\kappa_e} = \widehat{x}_{k-N_e\kappa_e}$ and $(\overline{w}_{k-n_e\kappa_e} = \widehat{w}_{k - n_e\kappa_e})_{n_e=0}^{N_e}$. The covariance matrices $Q_{x_0|k}, \allowbreak Q_{v|k - n_e\kappa_e}, R_{w|k - n_e\kappa_e} \succ 0$ can be set based on process knowledge and/or historic data.

If the static constraints (Eq. \eqref{eq: dMHE_sets}) are linear inequalities and the dynamic constraints (Eq. \eqref{eq: dMHE_dynamics}) are linear equalities, for example from an affine (approximation of the) dynamics,
\begin{subequations}
\begin{align}
    {x}_{k - (n_e - 1)\kappa_e}	
            & = f_{\Delta{t}_e}({x}_{k - n_e\kappa_e},u_{k - n_e\kappa_e}^*,{w}_{k - n_e\kappa_e})							\nonumber \\
    		& \approx A_{\Delta{t}_e | k-n_e\kappa_e} {x}_{k-n_e\kappa_e} + B_{\Delta{t}_e | k-n_e\kappa_e} u_{k-n_e\kappa_e}^*	\nonumber \\
			& \quad + G_{\Delta{t}_e | k-n_e\kappa_e} {w}_{k-n_e\kappa_e} + {z}_{f|\Delta{t}_e | k-n_e\kappa_e},
\end{align} \label{eq: MHE_DiscrSS}%
\end{subequations}
then also the MHE can be specialised into a convex program. Similarly, the output function in cost Eq. \eqref{eq: MHE_Quadratic_Costs_stage} can be approximated as $g({x}_{k-n_e\kappa_e}) \approx z_{g|\Delta{t}_e|k-n_e\kappa_e} + C_{\Delta{t}_e|k-n_e\kappa_e}{x}_{k-n_e\kappa_e}$.

Convenient linearisations of the dynamics and output equations can be obtained by evaluating the functions $(f,g)$ and their Jacobians ($\partial{f}/\partial{x}$, $\partial{f}/\partial{u}$, $\partial{f}/\partial{w}$, and $\partial{g}/\partial{x}$) at previous optimal points $\hat{p}_{k-n_e\kappa_e} = (\widehat{x}_{k-n_e\kappa_e}, u^*_{k-n_e\kappa_e}, \widehat{w}_{k-n_e\kappa_e}, \allowbreak \widehat{y}_{k-n_e\kappa_e})$. As a result, we get the offset vectors $z_{f|\Delta{t}_e} \in \mathbb{R}^{N_x}$ and $z_{g|\Delta{t}_e} \in \mathbb{R}^{N_y}$, and the system matrices $A_{\Delta{t}_e} \in \mathbb{R}^{N_x \times N_x}$, $B_{\Delta{t}_e} \in \mathbb{R}^{N_x \times N_u}$, $G_{\Delta{t}_e} \in \mathbb{R}^{N_x \times N_w}$, and $C_{\Delta{t}_e} \in \mathbb{R}^{N_y \times N_x}$. 

A constrained linear-quadratic MHE with affine dynamics is the method of choice used in the application (Section \ref{sec: ControlFramework}).

\section{Model-based control of ASPs} \label{sec: ControlFramework}

\begin{figure*}[htb!] \centering   
    \includegraphics[width=\textwidth]{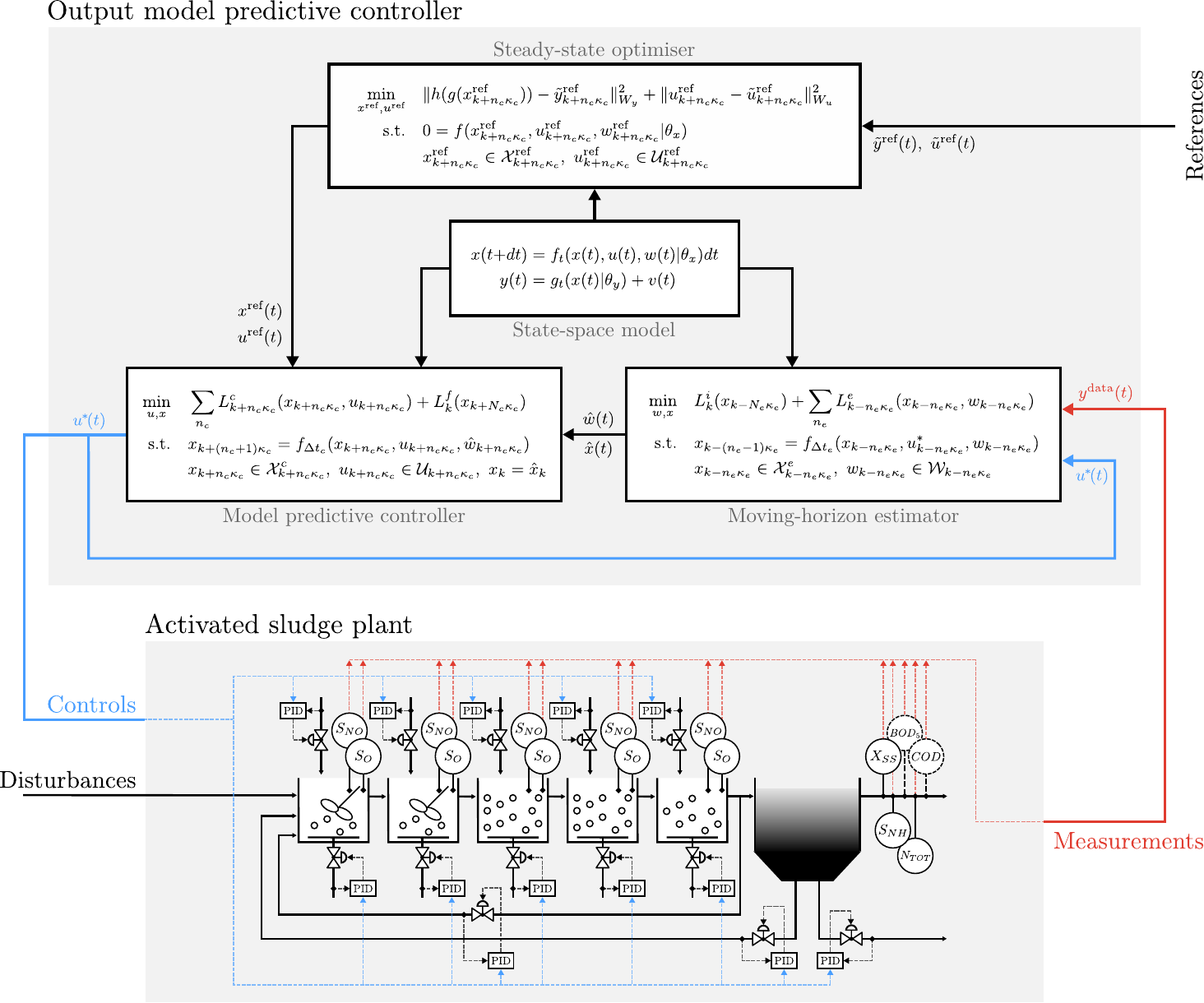}
    \caption{Model-based predictive control of an activated sludge plant: The controller receives \textit{i)} an external reference trajectory (black) and \textit{ii)} measurement data (red) from sensors/laboratory in the plant. These quantities, together with a state-space model of the plant dynamics and instruments, are used by a MPC and associated MHE to compute the control actions that best track the reference. The optimal actions (blue) are deployed to the plant actuators as set-points to their low-level (PID) controllers.} 
    \label{fig: ControlFramework_Scheme} 
\end{figure*}

This section shows how the Output MPC can be configured to autonomously operate conventional activated sludge plants: 
\begin{itemize}
\item We consider an ASP corresponding to the BSM1 and assume that all of its handles for measurement and control exist as actual instruments and actuators.
\item We assume the availability of a predictive model of the plant's dynamics and of its measuring equipment.
We consider the case in which only an approximation of the BSM1 dynamics is available: This simplified model is used by both the controller and state estimator.
\end{itemize}

We focus on a scenario in which the ASP must adapt its operations to produce an effluent whose quality varies according to a high-level demand.
The task is reminiscent of the contemporary objective of actuating the ASP as a resource-recovery facility.
The controller is asked to autonomously drive the plant to produce a water whose specifications change in time, while satisfying technical constraints, and in spite of the quantity and quality of the influent wastewater. 
We emphasise the task of tracking references that are of practical relevance:
\begin{itemize}
\item[$\leadsto$] We consider an exemplary reference trajectory for the effluent in terms of total nitrogen, ammonium and ammonia nitrogen, and suspended solids: These quantities can be readily measured and correspond to model outputs $N_{TOT}^{S(10)}$, $S_{NH}^{S(10)}$, and $X_{SS}^{S(10)}$, at the settler's top.
\end{itemize}
From an illustrative stand-point, this objective is rich enough and it is also challenging from a control perspective. 
Yet, our framework is general and other control objectives expressed in terms of measurable quantities could be defined, instead.

The architecture of the control framework is shown in Figure \ref{fig: ControlFramework_Scheme}. The control actions computed by the Output MPC are given as the model inputs $Q_A$, $Q_R$, and $Q_W$, $K_La^{(1)}, \dots, K_La^{(5)}$, and $Q_{EC}^{(1)}, \dots, Q_{EC}^{(5)}$.
They are deployed as recycle flow-rates, aeration intensities, and dosages of extra carbon. 
As in reality it is impossible to set process quantities directly, we assume that, for each of them, there exists an ideal PID controller whose set-point is changed in time to be equal to the control values computed by the Output MPC. 
\begin{itemize}
    \item[$\leadsto$] Thus, the Output MPC defines a supervisory control layer that operates above low-level regulatory PID controllers. 
By the same token, the reference trajectories are seen as generated by a planning layer that operates above the Output MPC and produces its `set-points'.
\end{itemize}
 
The Output MPC receives the sequence of reference values ($\tilde{y}^{\text{ref}} = N_{TOT}^{S(10)}, S_{NH}^{S(10)}, X_{SS}^{S(10)}$) to be tracked over the next control-horizon ($H_c = 1/2 \ \text{[days]}$) and it computes the control actions ($u^* = Q_A, Q_R, Q_W, K_La^{(1)}, \dots, K_La^{(5)}, Q_{EC}^{(1)}, \allowbreak \dots, Q_{EC}^{(5)}$) that produce the sequence of selected model outputs $(\tilde{y} = N_{TOT}^{S(10)}, \allowbreak S_{NH}^{S(10)}, X_{SS}^{S(10)})$ that best matches $\tilde{y}^{\text{ref}}$:
\begin{itemize}
    \item  To compute the controls $u^*$ with the MPC (Eqs. \eqref{eq: MPCDiscr} and \eqref{eq: MPC_Quadratic_Costs}), the output references $\tilde{y}^{\text{ref}}$ are firstly converted into an equivalent sequence of state and control pairs $(x^{\text{ref}},u^{\text{ref}})$ by the steady-state optimiser (Eq. \eqref{eq: SS_OPT}). 
    \item The current state and disturbances used by the MPC are estimated by the MHE (Eqs. \eqref{eq: MHEDiscr} and \eqref{eq: MHE_Quadratic_Costs}), from the sequence of plant's measurements ($y^{\text{data}} = S_{O}^{A(1)}, \dots, \allowbreak S_{O}^{A(5)}, S_{NO}^{A(1)}, \dots, S_{NO}^{A(5)}, X_{SS}^{S(10)}, S_{NH}^{S(10)}, N_{TOT}^{S(10)}$) from the past estimation-horizon ($H_e = 1/8 \ \text{[days]}$). 
\end{itemize}
The control actions $u^*$ are computed to satisfy various constraints. 
As essential requirement, here we always include constraints on the controls that guarantee compatibility with plant's equipment and satisfy its technological limits. 
Specifically, we require each control variable to be between a minimum and maximum value, at all times.
Again, the understanding is that the framework is general and we show that it can accommodate sophisticated formulations of the constraints.

In the following, the setup of the individual components of the Output MPC is discussed, starting from the predictive model (Section \ref{subsec: CTRL_SS}) in the MPC (Section \ref{subsec: CTRL_MPC}) and the MHE (Section \ref{subsec: CTRL_MHE}). 
We present the discretisation interval used for the dynamics ($\Delta{t}$), the operating periods for the controller ($\Delta{t}_c = \kappa_c\Delta{t}$), and for the estimator ($\Delta{t}_e = \kappa_e\Delta{t}$). 
The control- and estimation-horizon ($H_c = N_c\Delta{t}_c$ and $H_e = N_e\Delta{t}_e$) are then written, accordingly, in terms of number $N_c$ and $N_e$ of control actions and estimates at each cycle. 
The constraint sets for state ($\mathcal{X}^c$ and $\mathcal{X}^e$), control ($\mathcal{U}$), disturbance ($\mathcal{W}$), and joint variables ($\mathcal{Z}$) are also presented. 
Details about the tuning parameters of controller, steady-state optimiser, and estimator are given in the Supplementary Material.


\subsection{The model}\label{subsec: CTRL_SS}
We configure the Output MPC to use two linear-affine approximations of the plant, one for the MPC and one for the MHE.
Though formally equal, both approximations are derived from the BSM1, they are different because obtained at different points in time and about distinct process conditions
\begin{subequations} \label{eq: Model_DiscrSS}\begin{align}
    x_{k + \delta{t}}	& \approx A_{\delta{t} | k} x_{k} + B_{\delta{t} | k} u_{k}	+  G_{\delta{t} | k} {w}_{k} + {z}_{f | \delta{t} | k} \\
    y_{k}				& \approx C_{\delta{t} | k} x_{k} + {z}_{g | \delta{t} | k}
\end{align}
\end{subequations}
where $\delta{t}$ either equals $\Delta{t_c}$ (controller) or $\Delta{t_e}$ (state estimator).
Matrices $A_{\delta{t} | k}$, $B_{\delta{t} | k}$, $G_{\delta{t} | k}$, $C_{\delta{t} | k}$, and affine terms $z_{f | \delta{t} | k}$ and $z_{g |\delta{t} | k}$ in Eq. \eqref{eq: Model_DiscrSS} are re-evaluated before each controller's cycle, by linearising the discrete-time form of the BSM1:
\begin{subequations}
\begin{align}
	{x}_{k + \delta{t}}	& = f_{\delta{t}}(x_{k},u_{k},{w}_{k} | \theta_x)	\label{eq: ASP_x}; \\
	{y}_k				& = g(x_k | \theta_y )							\label{eq: ASP_y},
\end{align}\label{eq: ASP_SS}%
\end{subequations}
where $x(t) = (x^{A(1)}, \dots, x^{A(5)},x^{S(1)}, \dots, x^{S(10)}) \in \mathbb{R}_{\ge 0}^{N_x}$ are the $N_x = (13 \times 5) + (8 \times 10) = 145$ state variables with
\begin{align*}
	x^{A(r)}	& = \big( S_I^{A(r)}, S_S^{A(r)}, X_I^{A(r)}, X_S^{A(r)}, X_{BH}^{A(r)}, X_{BA}^{A(r)}, X_{P}^{A(r)}, 	\\
				& \qquad  S_O^{A(r)}, S_{NO}^{A(r)}, S_{NH}^{A(r)}, S_{ND}^{A(r)}, X_{ND}^{A(r)}, S_{ALK}^{A(r)} \big)	\\
				& \qquad (r=1,\dots,5); \\
	x^{S(l)}	& = \big( X_{SS}^{S(l)}, S_{I}^{S(l)}, S_S^{S(l)}, S_O^{S(l)},S_{NO}^{S(l)},S_{NH}^{S(l)}, S_{ND}^{S(l)}, S_{ALK}^{S(l)} \big) \\
				& \qquad (l=1,\dots,10).
\end{align*}
$u(t) = \big( Q_A, Q_R, Q_W, K_La^{(1)}, \dots, K_La^{(5)}, Q_{EC}^{(1)}, \dots, Q_{EC}^{(5)} \big)\allowbreak \in \mathbb{R}_{\ge 0}^{N_u}$, are the $N_u = 3 + (2 \times 5) = 13$ control variables, corresponding to the control handles of the BSM1, and $w(t) = (Q_{IN}, x^{A(IN)}) \in \mathbb{R}_{\ge 0}^{N_w}$ are the $N_w = 1 + 13 = 14$ disturbances including the properties of influent wastewater
\begin{align*}
	x^{A(IN)}	& = \big( S_I^{A(IN)}, S_S^{A(IN)}, X_I^{A(IN)}, X_S^{A(IN)}, X_{BH}^{A(IN)},				\\
				& \qquad   X_{BA}^{A(IN)}, X_{P}^{A(IN)}, S_O^{A(IN)}, S_{NO}^{A(IN)}, S_{NH}^{A(IN)},		\\
				& \qquad   S_{ND}^{A(IN)}, X_{ND}^{A(IN)}, S_{ALK}^{A(IN) }\big).
\end{align*}
All disturbances, expect for the flow-rate $Q_{IN}$ are not measured: That is, no actual instrument or laboratory analysis is available in the plant to measure them.
As for the $N_y = (2 \times 5) + 3 = 13$ output variables $y(t) \in \mathbb{R}_{\ge 0}^{N_y}$, in which 
\begin{multline*}
    y(t) = \big( S_{O}^{A(1)}, \dots, S_{O}^{A(5)}, S_{NO}^{A(1)}, \dots, S_{NO}^{A(5)}, \\ 
			\qquad X_{SS}^{S(10)}, S_{NH}^{S(10)}, X_{SS}^{S(10)}, S_{NH}^{S(10)}, N_{TOT}^{S(10)} \big),
\end{multline*}
we assume the existence of plant's sensors, corresponding to the measurement handles of the BSM1: That is, it assumed that dissolved oxygen and nitrate- and nitrite-nitrogen in the reactors, and total suspended solids and nitrogen at the top of the settler, the effluent, are available as measurements $y^{\text{data}}$.

The vectors $\theta_x$ and $\theta_y$ in the model equations \eqref{eq: ASP_SS} collect all the stoichiometric and kinetic parameters in the BSM1.

\subsection{The MPC} \label{subsec: CTRL_MPC}
The MPC is operated once every hour ($\Delta{t}_c = 1/24 \ \text{[days]}$) to plan over a half-day long control-horizon ($H_c = 1/2 \ \text{[days]}$).
\begin{itemize}
\item At each cycle, a sequence of $N_c = 12$ control actions is computed, for each of the $N_u = 13$ control variables
\begin{itemize}
\item $12 \times 13 = 156$ optimal actions $(u^*_{k + n_c\kappa_c})_{n_c = 0}^{N_c-1}$ are calculated, of which only the first ones ($13$ values $u^*_{k}$) are sent as set-points to the low-level PIDs;
\item The set-points are held constant during the cycle;
\end{itemize}
\item The rest ($143$ values) of the control actions is discarded;
\item After one hour, the control cycle is repeated anew. 
\end{itemize}

\subsubsection*{Tracking reference trajectories}

At each cycle, the MPC control actions track the reference $(x_{k+n_c\kappa_c}^{\text{ref}}, u_{k+n_c\kappa_c}^{\text{ref}})_{n_c=1}^{N_c}$, in which each pair $(x_{k+n_c\kappa_c}^{\text{ref}}, u_{k+n_c\kappa_c}^{\text{ref}})$ solves a steady-state optimisation problem \eqref{eq: SS_OPT}: That is,
\begin{itemize}
\item[$\leadsto$] each pair of the sequence corresponds to the stationary values of state and control variables that associate to the corresponding term in the reference output sequence $(\tilde{y}_{k+n_c\kappa_c}^{\text{ref}})_{n_c=1}^{N_c}$, the quality as $(X_{SS}^{S(10)},S_{NH}^{S(10)},N_{TOT}^{S(10)})$. 
\end{itemize}
The optimality of reference tracking is determined in terms of the quadratic functions $L^c(\cdot)$ and $L^f(\cdot)$ defined in Eq. \eqref{eq: MPC_Quadratic_Costs}.

\subsubsection*{Dynamic and static constraints and desiderata}
The MPC actions are computed to satisfy the dynamic constraints, the plant's dynamics approximated over the control-horizon by $\{(z_{f | \Delta{t}_c | k+n_c\kappa_c}, A_{\Delta{t}_c | k+n_c\kappa_c}, B_{\Delta{t}_c | k+n_c\kappa_c}, G_{\Delta{t}_c | k+n_c\kappa_c},\allowbreak C_{\Delta{t}_c | k+n_c\kappa_c})\}_{n_c=0}^{N_c}$ and evaluated in time at reference points $p_{k+n_c\kappa_c} = (x^{\text{ref}}_{k+n_c\kappa_c}, u^{\text{ref}}_{k+n_c\kappa_c}, \widehat{w}_{k+n_c\kappa_c}, y^{\text{ref}}_{k+n_c\kappa_c})$.
No other constraints are imposed to the state variables, thus 
$$
\mathcal{X}^c = \mathbb{R}^{N_x}.
$$
On the other hand, the actions are also constrained to take values within the limits of the BSM1 actuators (Table \ref{tab: Control_Limits}), 
\begin{multline}\label{eq: Box_Constraint_Set}
    \mathcal{U} = \{ u \in \mathbb{R}^{N_u} : 
        u_{1} \in [0, 92230],\ 
        u_{2} \in [0, 36892],\ 			\\ 
        \ \qquad u_{3} \in [0, 1844.6],\ 
        u_{4,\cdots,8} \in [0, 360],\ 
        u_{9,\cdots,13} \in [0, 5] \}.
\end{multline}

\begin{table}[htb!] \centering
    \caption{Activated sludge plant: Actuator limits.}
    \begin{tabular}{l | c | c c | l}
        Variable        			&						& Lower limit   & Upper limit   & Unit              \\
        \doublehline
        $Q_A$           			& $u_{1}$				&   0           & 92230         & m$^{3}$ d$^{-1}$  \\
        $Q_R$           			& $u_{2}$				&   0           & 36892         & m$^{3}$ d$^{-1}$  \\
        $Q_W$           			& $u_{3}$				&   0           & 1844.6        & m$^{3}$ d$^{-1}$  \\
        $K_La^{(1 \leadsto 5)}$		& $u_{4 \leadsto 8}$	&   0           & 360           & d$^{-1}$			\\
        $Q_{EC}^{(1 \leadsto 5)}$	& $u_{9 \leadsto 13}$	&   0           & 5             & m$^{3}$ d$^{-1}$  \\[0.5ex]
        \hline
    \end{tabular}
    \label{tab: Control_Limits}
\end{table}

To demonstrate the flexibility in the definition of the constraints, we formulate the requirement that a portion $\eta \in [0,1]$ of the energy demand of the process must be recovered from the energy produced by the digestion of wastage sludge:
\begin{itemize}
\item[$\leadsto$] Such operational desiderata can be explicitly stated as a constraint that regards both state and control variables
\end{itemize} 

By letting the Operational Cost Index ($\text{OCI}_{\text{kWh}}$) express the energy demand and $\text{RE}$ denote the amount of energy generated from wastage sludge, we define the energy constraint
\begin{equation} \label{eq: Energy_Constraint_Set} 
    \mathcal{Z} = \left\{ (x_n,u_n)_{n=0}^{N-1} :  \eta \text{OCI}_{\text{kWh}}(u_n)  \leq \text{RE}(x_n, u_n) \right\}
\end{equation}

Under the assumption that wastage sludge is anaerobically digested into methane gas which is in turn used to generate electricity, we quantify energy generation from sludge
\begin{equation} 
    \text{RE} = \eta_D \underbrace{Q_W}_{u_3} \Big(\underbrace{\dfrac{1}{0.75}\underbrace{X_{SS}^{S(1)}}_{x_{66}}{+}\underbrace{S_{I}^{S(1)}}_{x_{76}}{+}\underbrace{S_{S}^{S(1)}}_{x_{77}}}_{COD^{S(1)}} \Big).
\end{equation}
The generation efficiency $\eta_D = 0.35 \times (4.865/3600)$ kWh/g corresponds to $35\%$ of excess sludge being converted to methane gas, from which $4.865$ kJ/g can be generated \cite{Wan2016}. 
Note that the $\text{RE}$ only considers wastage sludge as sole source of energy and it ignores sludge fluxes from the primary settler.
For compactness, the quantification of the $\text{OCI}_{\text{kWh}}$ is fully detailed in the Supplementary Material.

\subsection{The MHE} \label{subsec: CTRL_MHE}

At each cycle, the MPC requires the current state $x_{t_k} = \widehat{x}_{t_k}$ and plans its actions assuming that the disturbances $\widehat{w}_{t_k}$ will remain constant and equal to their current value. 
These quantities are unknown and are estimated by an MHE operating at $15$-minute ($\Delta{t}_e = 1/96 \ \text{[days]}$) intervals, over a $3$-hour estimation-horizon ($H_e = 1/8 \ \text{[days]}$) during which $N_e = 12$ plant measurements were acquired. Thus, at each MHE cycle
\begin{itemize}
    \item  $13 \times (145 + 14) = 2067$ past values of state and disturbance variables $(\widehat{x}_{t_k-n_e\kappa_e},\widehat{w}_{t_k-n_e\kappa_e})_{n_e=0}^{N_e}$ are calculated. The last (most recent) estimates are used by the MPC;
    \item The rest ($1749$) of the estimates is discarded.
\end{itemize}

\subsubsection*{Fitting plant data}

Over the estimation-horizon, the MHE estimates the state and disturbance variables that would correspond to model outputs that best fit plant data $(y_{k-n_e\kappa_e}^{\text{data}})_{n_e = 0}^{N_e}$. 
Estimation accuracy is quantified by the quadratic costs $L^e(\cdot)$ and $L^i(\cdot)$, in Eq. \eqref{eq: MHE_Quadratic_Costs}.

\subsubsection*{Dynamic and static constraints}

The MHE estimates of the state satisfy the dynamic constraints, the linear-affine approximations $\{(z_{g | \Delta{t}_e | k-n_e\kappa_e}, A_{\Delta{t}_e | k-n_e\kappa_e}, B_{\Delta{t}_e | k-n_e\kappa_e}, G_{\Delta{t}_e | k-n_e\kappa_e},\allowbreak C_{\Delta{t}_e | k-n_e\kappa_e})\}_{n_e=0}^{N_e}$ of the dynamics around the fixed-points $\widehat{p}_{k-n_e\kappa_e} = (\widehat{x}_{k-n_e\kappa_e}, u_{k-n_e\kappa_e}^*, \widehat{w}_{k-n_e\kappa_e}, \widehat{y}_{k-n_e\kappa_e})$. 
The estimates of the disturbances are constrained to take on non-negative values ($\mathcal{W} = \mathbb{R}^{N_w}_{\geq 0}$), as they refer to positive quantities: influent's flow-rate and composition. 
Moreover, 
\begin{itemize}
    \item the concentrations $X_{BA}^{IN} = X_{P}^{IN} = S_{O}^{IN} = S_{NO}^{IN} = 0$ g m$^{-3}$ and $S_{ALK}^{IN} = 7$ mol HCO$_3^-$ m$^{-3}$ are constrained to remain constant over the chosen influent scenario;
    \item the influent flow-rate $Q_{IN}$ is measured at each $\Delta t_{e}$ . 
\end{itemize}

\section{Case-studies}\label{sec: Results}

We demonstrate the potential of the control framework for activated sludge plants on two different operational tasks:
\begin{enumerate}[label={\arabic*.},nosep]
    \item \textit{Conventional treatment of wastewater} (Section \ref{subsec: Treatment_Results}) - The ASP is requested to produce effluent water whose quality satisfies normative constraints;
    \item \textit{Nitrogen on-demand} (Section \ref{subsec: Tracking_Results}) - The ASP is requested to produce a water whose nitrogen content varies according to an external demand. 
\end{enumerate} 
We discuss the results obtained by the Output MPC (Figure \ref{fig: ControlFramework_Scheme}) under the two-week scenario of \textit{stormy weather} (Section \ref{subsubsec: BSM1-influent}).
For reference, the performances are compared to the default control strategy (Figure \ref{fig: BSM1}) consisting of two PIDs:
\begin{itemize}
	\item[$\leadsto$] Nitrate and nitrite nitrogen in the second reactor, $S_{NO}^{A(2)}$, is controlled by manipulating the internal recycle $Q_A$; 
	\item[$\leadsto$] Dissolved oxygen in the fifth reactor, $S_O^{A(5)}$, is controlled by manipulating the oxygen mass transfer coefficient $K_La^{(5)}$, a proxy variable to the air flow-rate.
\end{itemize}
The performance of the Output MPC is also compared to open-loop operations in which the ASP is run with no regulation.
This is interesting, as this mode highlights the treatment potential of the ASP as such, without automatic control.

\subsection{Case-study 1: Wastewater treatment}\label{subsec: Treatment_Results}

To operate the activated sludge plant to meet standard disposal regulations, the output MPC must track a constant-in-time trajectory corresponding to the effluent restrictions (Table \ref{tab: Treat_References}, for the BSM1 \cite{Gernaey2014}).
This is achieved by continuously determining the best actions and use them dynamically as set-point values for the $13$ PIDs (Figure \ref{fig: ControlFramework_Scheme}).

\begin{table}[htb!] \centering
    \caption{Case-study I: Quality limits and corresponding reference values.}
    \begin{tabular}{l | c c | l}
            Variable  & Reference & Limits & Units \\
            \doublehline
            $X_{SS}^{S(10)}$    & 12.5   & 30    & g COD m$^{-3}$ 	\\[0.25ex]
            $S_{NH}^{S(10)}$    &  1.7   & 4     & g N m$^{-3}$   	\\[0.25ex]
            $N_{TOT}^{S(10)}$   & 14.0   & 18    & g N m$^{-3}$
    \end{tabular}
    \label{tab: Treat_References}
\end{table}
 
We show the treatment performance with two configurations:
\begin{enumerate}[label={1\alph*.}, nosep]
    \item With technological constraints only, Eq. \eqref{eq: Box_Constraint_Set}; 
    \item With extra constraints on recovered energy, Eq. \eqref{eq: Energy_Constraint_Set}.
\end{enumerate}
When the controller is not asked to recover energy to sustain operations ($\eta=0$, in Eq. \eqref{eq: Energy_Constraint_Set}), the configurations are equal.

\subsubsection*{Wastewater treatment} \label{sec: Treatment_Eta0}

\begin{figure}[htb!] \centering 
    \includegraphics[width=\columnwidth]{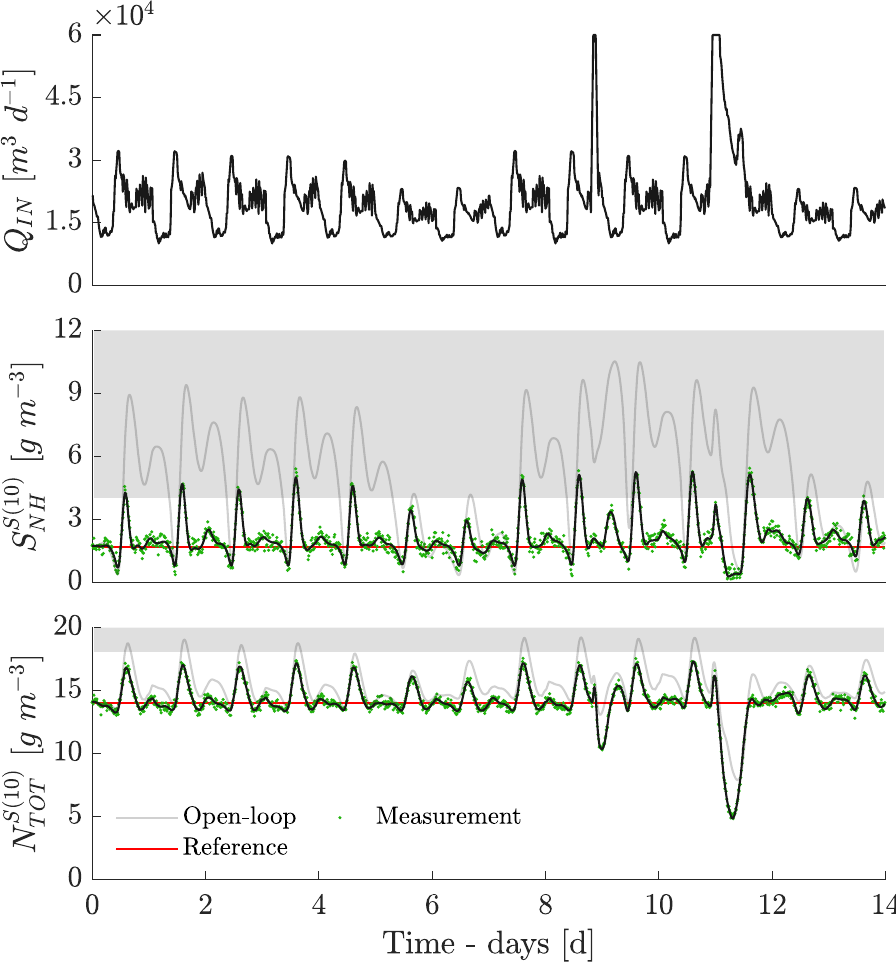}
    
    \caption{Case-study 1: Influent flow-rate, $Q_{IN}$ (top panel) and  concentration of $S_{NH}^{S(10)}$ and $N_{TOT}^{S(10)}$ in the effluent (bottom panels). The grey shade denotes regions above treatment limits.} 
    \label{fig: OutMPC_Treatment_Outputs} 
\end{figure}

\begin{figure}[htb!] \centering 
    \includegraphics[width=\columnwidth]{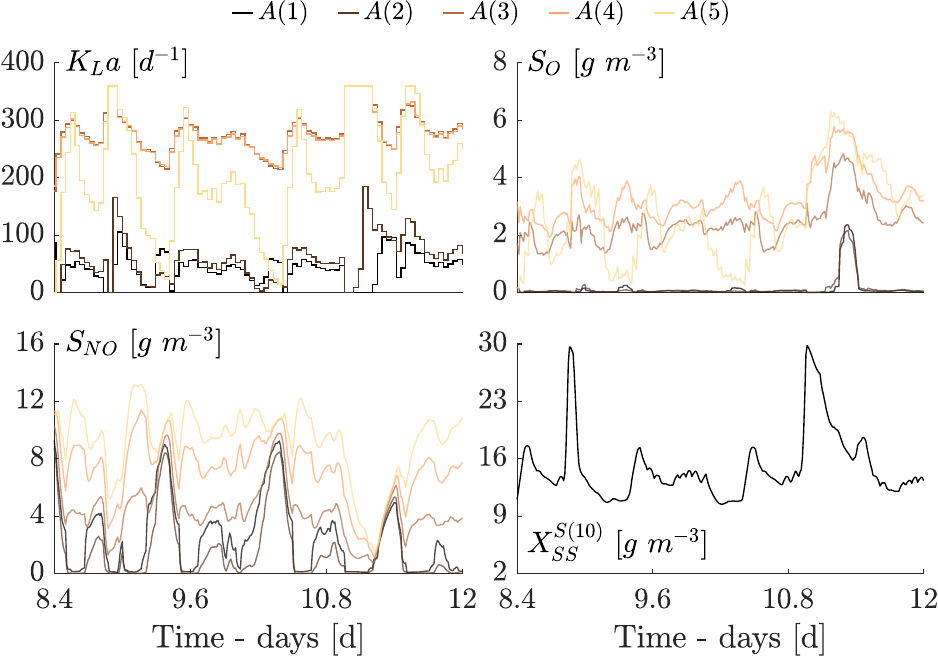} 

    \caption{Case-study 1, $t\in [8.4,12]$: Air flow-rate $K_La^{(1\leadsto 5)}$ and dissolved oxygen $S_{O}^{A(1\leadsto 5)}$ (top panels) and nitrogen form $S_{NO}^{A(1\leadsto 5)}$ and effluent suspended solids $X_{SS}^{S(10)}$ (bottom panels).} 
    \label{fig: OutMPC_Treatment_Summary} 
\end{figure}

Firstly, we analyse the treatment performance when the Output MPC operates the ASP only with actions that respect its equipment's limits. We present results (Figure \ref{fig: OutMPC_Treatment_Outputs}) in terms of quality of effluent water and violations of the specifications. Then, we look at the actions commanded by the controller and at the overall plant's behaviour (Figure \ref{fig: OutMPC_Treatment_Summary} and Table \ref{tab: Treatment_Results_Comparison}).

When treating the stormy-weather wastewater (its flow-rate $Q_{IN}$ is shown in Figure \ref{fig: OutMPC_Treatment_Outputs}, top panel), the Output MPC autonomously operates the plant to produce an effluent whose concentrations $S_{NH}^{S(10)}$ and $N_{TOT}^{S(10)}$ (middle and bottom panels in Figure \ref{fig: OutMPC_Treatment_Outputs}: the black lines, as state variables, and green dots, as plant measurements) closely follow the reference values (Table \ref{tab: Treatment_Results_Comparison} and red lines in Figure \ref{fig: OutMPC_Treatment_Outputs}). 
By rejecting the variations in typical municipal influents, as well as the upsets induced by the two storm events, the controller almost completely eliminates \textit{i}) violations of the specifications (Table \ref{tab: Treatment_Results_Comparison} and the grey-shaded area in Figure \ref{fig: OutMPC_Treatment_Outputs}) and \textit{ii}) the time in off-specification operations.
In the figure, this can be appreciated when comparing the performance of the Output MPC to the open-loop setup (grey lines).
Note that Figure \ref{fig: OutMPC_Treatment_Outputs} does not show the evolution of $X_{SS}^{S(10)}$, as its limits are never violated.

To understand how the tracking is achieved, in Figure \ref{fig: OutMPC_Treatment_Summary} we look closely at a selection of control actions deployed by the Output MPC.
For clarity, we zoom into a shorter period of time ($t \in [8.4,12]$ days) which includes the storm events, and we analyse how the controller operates the denitrification-nitrification process across reactors $A(1{\leadsto}5)$, and the settler.
Complete results are given in the Supplementary Material.

The tracking of the treatment reference is achieved by automatically adjusting the aeration intensities (via $K_La^{(1\leadsto 5)}$) and the sludge flow-rates ($Q_{A}$, $Q_{R}$ and $Q_{W}$, Figure \ref{fig: OutMPC_Treatment_Eta0_Summary} of the Supplementary Material).
To vary the aerated volume by changing the number aerated zones in a systematic way is a consolidated strategy based on modifying the usual \textit{two-anoxic and three-aerated} configuration of the bioreactors (as in Figure \ref{fig: BSM1}).
The Output MPC, on the other hand, utilises aeration of reactor $A(5)$ ($K_La^{(5)}$) as its primary control lever by switching between aerated and non-aerated modes, in response to the variations in influent load.
Moreover, we have
\begin{itemize}
    \item Aeration to the second most downstream reactors ($A(4)$ and $A(3)$) is kept strong and it is further increased whenever influent flow-rates increase, to favour nitrification, whereas less intense aeration is enforced whenever the flow-rates decrease, to favour denitrification;
    \item In the same fashion, the most upstream reactors ($A(1)$ and $A(2)$) are kept in virtually anoxic conditions and aeration slightly adjusted, to meet the incoming loads.
\end{itemize}
In addition to aeration control, the other actions applied the bioreactors are associated to the addition of carbon, via $Q_{EC}^{(1 \leadsto 5)}$ (Figure \ref{fig: OutMPC_Treatment_Eta0_Summary}, in the Supplementary Material). 
As overall effect, the concentrations of $S_{NO}^{A(1\leadsto 5)}$ (and $S_{NH}^{A(1\leadsto 5)}$, Figure \ref{fig: OutMPC_Treatment_Eta0_Summary} in the Supplementary Material) are smoothed out and follow the periodicities in the influent nitrogen load ($N_{TOT}^{IN}$ in Figure 4 and $S_{NH}^{A(IN)}$ in Figure \ref{fig: MHE_Treatment_Eta0_Disturbances_Summary} of the Supplementary Material).

In the secondary settler, the changes in feed characteristics are reflected by the changes in the effluent concentrations $S_{NH}^{S(10)}$ and $S_{NO}^{S(10)}$, and $N_{TOT}^{S{(10)}}$, as well as in the spatial distribution of suspended solids $X_{SS}^{(1 \leadsto 10)}$ and the high of the sludge blanket. 
\begin{itemize}
\item[$\leadsto$] It is natural to associate the changes in these variables mainly to the adjustment actions computed by the Output MPC for external recycle $Q_R$ and wastage $Q_W$ fluxes and to a lesser extent, to the internal recycle $Q_A$.
\end{itemize}

The controller reacts to the storm-events (at day $t \approx 8.8$ and $t \approx 11$) by increasing aeration in all reactors, thus raising the oxygen levels $S_{O}^{(1\leadsto 5)}$ and favouring nitrification throughout. 

The performance of the control framework on this case-study is in Table \ref{tab: Treatment_Results_Comparison} with respect to the conventional evaluation criteria used to assess the effluent quality \cite{Gernaey2014}. 
\begin{itemize}
\item In comparison with the open-loop operation (respectively, the default PID control strategy), our control strategy resulted in roughly 18\% (roughly 11\%) improvement in the effluent quality index (EQI). 
\item Moreover, the controller was able to significantly decrease the percentage of time in which effluent ammonium ($S_{NH}^{S(10)}$) and total ($N_{TOT}^{S(10)}$) nitrogen are in violation of their respective quality limits. 
\end{itemize}

\begin{table}[t!] \centering
	\caption{Case-study 1: Performance of each control strategies in terms of effluent quality index (EQI, kg PU d$^{-1}$), average operational cost index (OCI$_{\text{kWh}}$, kWh d$^{-1}$), number of quality limit crossings, and percentage of time in violation of the effluent restrictions (\#Crossings and \%Violation, respectively).}
	{\small\begin{tabular}{c | c c c }
	& Open-loop & PID & Out-MPC \\     
	\doublehline
	{EQI}	 			
	& 7246.1 & 6636.0 & 5915.2 \\[0.5ex] 
	{(avg.) OCI$_{\text{kWh}}$}
	& 3969.6 & 4202.1 & 4558.3 \\[1ex]
	\makecell{\#Crossings ($S_{NH}$)}
	& 14 & 12 & 10 \\[0.5ex] 
	\makecell{\#Crossings ($N_{TOT}$)}
	& 8 & 13 & 0 \\[1ex] 
	\makecell{\%Violation ($S_{NH}$)}
	& 62\% & 22\% & 6\% \\[0.5ex] 
	\makecell{\%Violation ($N_{TOT}$)}
	& 7\% & 16\% & 0\% \\[1.5ex] 
	\hline
	\end{tabular}}

	\label{tab: Treatment_Results_Comparison}
\end{table}

As expected, this improvement follow an increase in the average operational cost index (OCI$_{\text{kWh}}$) associated to the control actions. However, the process operation based on our framework still results in an effluent quality with almost no violations of regulatory constraints, as opposed to the results obtained by the control strategy proposed in the benchmark.

In conclusion, this case-study demonstrates that the controller is able to operate the process to comply with quality restrictions by tracking a constant reference profile for the effluent. 

\subsubsection*{Wastewater treatment, with energy recovery} \label{sec: Treatment_EtaNot0}

We expand upon the treatment results by studying the case in which the Output MPC is asked not only to satisfy the effluent regulations, but also to operate the ASP in a such way that partial or total energy recovery from sludge disposal is enforced (that is, for cases in which $\eta \in (0,1]$ in Eq. \eqref{eq: Energy_Constraint_Set}).

\begin{figure}[ht!] \centering
	\includegraphics[width=\columnwidth]{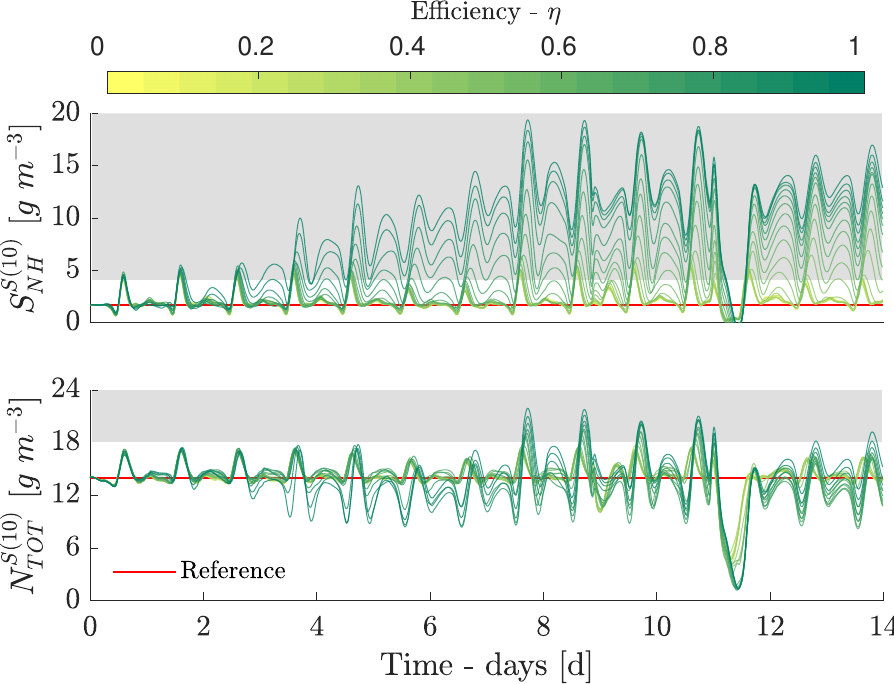}

	\caption{Case-study 1: Treatment of effluent nitrogen $S_{NH}^{S(10)}$ and $N_{TOT}^{S(10)}$ for different values of energy efficiency $\eta$.} 
	\label{fig: OutMPC_Sustainable_Treatment} 
\end{figure}

For the task, we analysed the treatment performance of the controller under different levels of energy recovery, individually: $\eta \in \{0.05,0.1,0.15,\dots,0.90,0.95,1\}$. As expected, the treatment quality, shown in Figure \ref{fig: OutMPC_Sustainable_Treatment}, worsens as larger portions of energetic demand are asked to be recovered:
\begin{itemize}
    \item Specifically, tracking effluent ammonium nitrogen $S_{NH}^{S(10)}$ becomes unfeasible when operations are also constrained to recover, from wastage sludge, $60\%$ or more of the energetic demand;
    \item Conversely, the performance when tracking the effluent $N_{TOT}^{S(10)}$ reference is not significantly affected as the recovery constraints are enforced.
\end{itemize}

\begin{figure}[t!] \centering
    \includegraphics[width=\columnwidth]{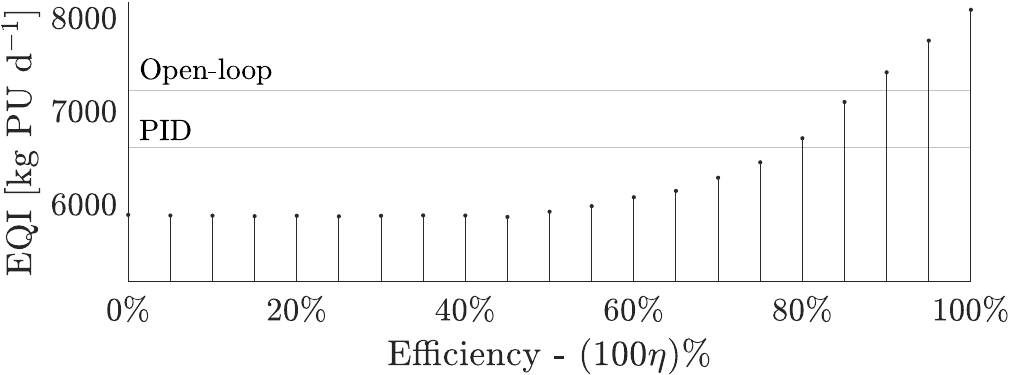}

    \caption{Case-study 1: Effluent quality index (EQI) for different energy-recovery levels $\eta \in [0,1]$. The gray lines refer to the EQI from the open-loop and default PID control operations.} 
    \label{fig: OutMPC_Treatment_Sustainable_EQI_Metrics}
\end{figure}

In Figure \ref{fig: OutMPC_Treatment_Sustainable_EQI_Metrics}, we show a comparison between the effluent quality index (EQI), under different energy-recovery levels, and against open-loop and PID operations. 
As mentioned, the Output MPC is capable to operate the ASP to produce a consistently excellent effluent profile, while recovering up to $60\%$ of its energetic demand. 
The effluent quality degrades as higher recovery levels are requested ($\eta \ge 65\%)$. 
Moreover, 
\begin{itemize}
    \item the Output MPC is able to recover $80\%$ of the energy costs, while still producing effluents whose quality matches those obtained by the default PID strategy. 
    \item effluents whose quality is to those obtained from the open-loop operation can still be produced, while operating the ASP with $90\%$ energy recovery efficiency. 
\end{itemize}

\begin{figure}[htb!] \centering
    \includegraphics[width=\columnwidth]{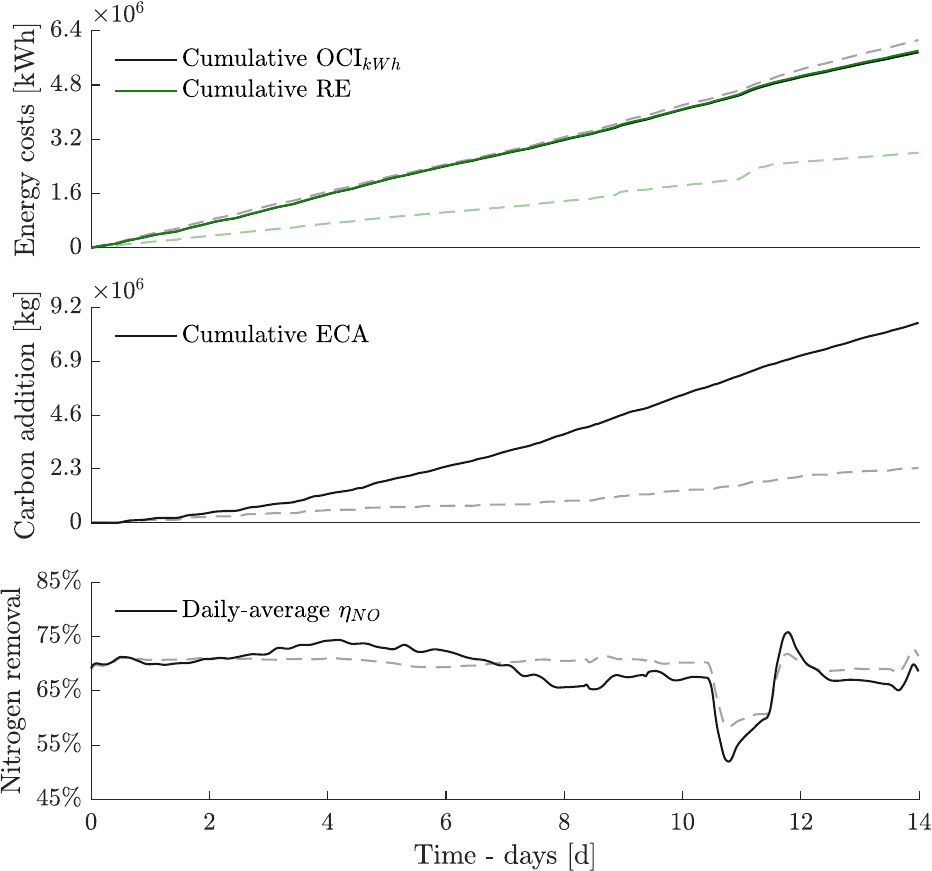}

    \caption{Case-study 1, $\eta = 1$: Control performance in terms of operational energy cost index, OCI$_{\text{kWh}}$, external carbon addition, ECA, and daily-averaged nitrogen removal efficiency, $\eta_{NO}$. Dashed lines refer to the results with efficiency $\eta = 0$.} 
    \label{fig: OutMPC_Sustainable_Control_Metrics} 
\end{figure}

The cumulative operational cost index (OCI$_{kWh}$) and external carbon (ECA) required for implementing zero-recovery ($\eta = 0$) and full-recovery ($\eta = 1$) treatment operations are illustrated in Figure \ref{fig: OutMPC_Sustainable_Control_Metrics}, for the two-week period under study. 
The daily-average of the resulting nitrogen removal efficiency (that is, $\eta_{NO} = (N_{TOT}^{IN} - N_{TOT}^{S(10)})/N_{TOT}^{IN}$) is also reported. 
\begin{itemize}
    \item The Output MPC can operate the plant without energy-related constraints ($\eta = 0$) and achieve a cumulative potential production of electricity (RE) equivalent to roughly $45\%$ of its operational costs. 
    \item The Output MPC can operate the plant with full energy-related constraints ($\eta = 1$) and achieve a cumulative potential energy production of electricity capable to satisfy its total energy demands. 
\end{itemize} 
The results also show an increased need for external carbon to implement the optimal actions from this controller.
Interestingly, both controllers obtain similar nitrogen removal performance. 
This reflects the already mentioned fact that the Output MPC operating under full-recovery constraints still provides good tracking accuracy for effluent $N_{TOT}^{S(10)}$, despite failing to generate the desired effluent $S_{NH}^{S(10)}$ concentrations.

We conclude that our model-based control framework can operate the activated sludge plant to comply with quality restrictions while ensuring that at least $60\%$ of its operational costs can be recovered from treating the wastage sludge. While the controller is able to enforce an operation which fully recovers the energy costs from actuators, the resulting control strategy leads to undesirable effluent qualities. Nevertheless, we remind that the results assume an anaerobic digester which is fed only by the sludge from the secondary settler: In full-scale treatment facilities, excess sludge from a primary clarifier is an additional raw material available for biogas production.

\subsection{Case-study 2: Nitrogen on-demand}\label{subsec: Tracking_Results}

In this section, we present the results when the Output MPC is configured to operate the activated sludge plant to perform tasks of a water resource recovery facility. As an exemplary problem, we consider tracking a reference trajectory for the total nitrogen $N^{S(10)}_{TOT}$ in the effluent. Specifically, we have
\begin{equation} \label{eq: NTOT_Reference} 
     N_{TOT}^{ref}(t) = 
    \begin{cases} 
        (5/3) \ N_{TOT}^{SS} , & t \in [2.8, 5.6)\text{ d} \\ 
        (2/3) \ N_{TOT}^{SS} , & t \in [8.4, 11.2)\text{ d} \\
        N_{TOT}^{SS} , & \text{otherwise}
    \end{cases}
    , 
\end{equation}
with $N_{TOT}^{SS} \approx 14$ g m$^{-3}$ being the benchmark's steady-state concentration. 
Tracking the reference in Eq. \eqref{eq: NTOT_Reference} equals to 
\begin{itemize}
\item operate the ASP to produce a water rich in nitrogen (for example, reusable for fertigation), $t \in [2.8, 5.6)$; 
\item operate the ASP to produce a water with low nitrogen (for example, due to stricter limits), $t \in [8.4, 11.2)$; 
\item operate the ASP to keep $N^{S(10)}_{TOT}$ constant at $N_{TOT}^{SS}$ in the other intervals ($t \in [0, 2.8) \cup [5.6,8.4) \cup [11.2, 14]$).
\end{itemize}
During the period, the Output MPC is also asked to maintain $(X_{SS}^{S(10)},S_{NH}^{S(10)})$ at the reference treatment values in Table \ref{tab: Treat_References}.

We firstly analyse the results when tracking the effluent trajectory without energy constraints ($\eta = 0$), then we extend the analysis when we enforce increasingly larger levels of energy recovery ($\eta \in \{0.05, 0.1, 0.15, \allowbreak \dots, 0.90, 0.95, 1\}$). 
Note that the usual technological constraints (Eq. \eqref{eq: Box_Constraint_Set}) must always be satisfied in both scenarios. 
We present two critical reference changes and refer to Section \ref{sec: Tracking_Appendix} of the Supplementary Material for a complete analysis of this case-study.

\subsubsection*{Nitrogen on-demand} \label{sec: Results_Reuse}

\begin{figure}[htb!] \centering 
    \includegraphics[width=\columnwidth]{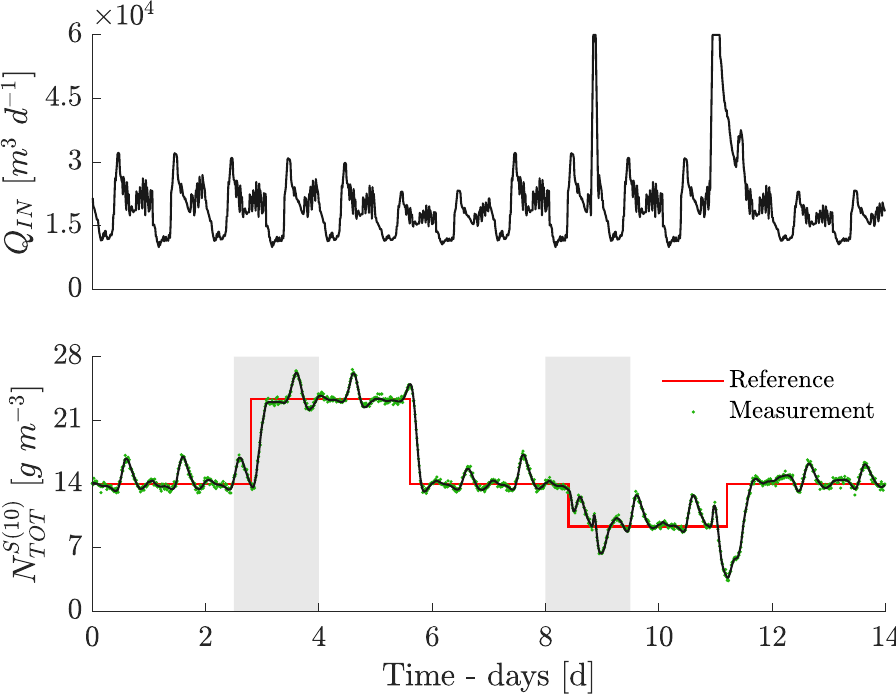}

    \caption{Case-study 2: Influent flow-rate $Q_{IN}$, top, and reference tracking of effluent total nitrogen $N_{TOT}^{S(10)}$, bottom. The shaded periods are analysed in detail (Figure \ref{fig: OutMPC_Reuse_Summary_SP1} and \ref{fig: OutMPC_Reuse_Summary_SP3}).} 
    \label{fig: OutMPC_Reuse_Outputs} 
\end{figure}

The results (Figure \ref{fig: OutMPC_Reuse_Outputs}) show how the Output MPC can operate the plant to track the reference trajectory in effluent nitrogen $N_{TOT}^{S(10)}$ (bottom panel), while rejecting the disturbances of the influent scenario (top panel). 
The tracking performance is consistently very good and it is only slightly degraded about the last change ($t = 11.2$ days), highlighting the challenges associated to requesting a reconfiguration of a large-scale facility whenever an extreme storm is occurring.

\begin{figure}[htb!] \centering 
    \includegraphics[width=\columnwidth]{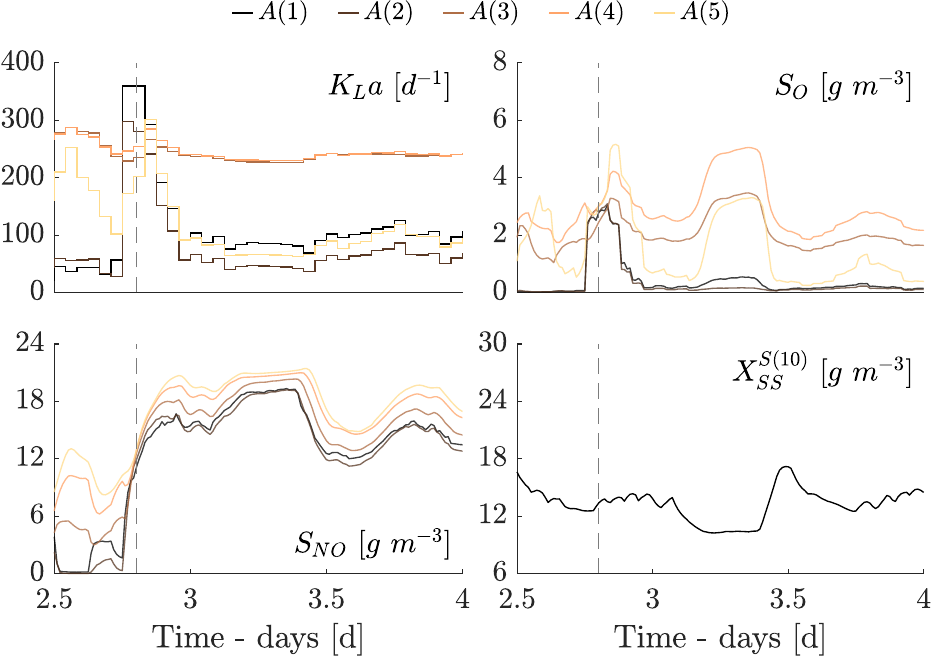} 

    \caption{Case-study 2, $t\in [2.5,4]$: Oxygen transfer coefficients $K_La^{(1\leadsto 5)}$ and dissolved oxygen $S_{O}^{A(1\leadsto 5)}$ (top panels) and nitrogen $S_{NO}^{A(1\leadsto 5)}$ and effluent suspended solids $X_{SS}^{S(10)}$ (bottom panels). The vertical dashed line indicates a reference change.}
    \label{fig: OutMPC_Reuse_Summary_SP1} 
\end{figure}

\begin{figure}[htb!] \centering  
    \includegraphics[width=\columnwidth]{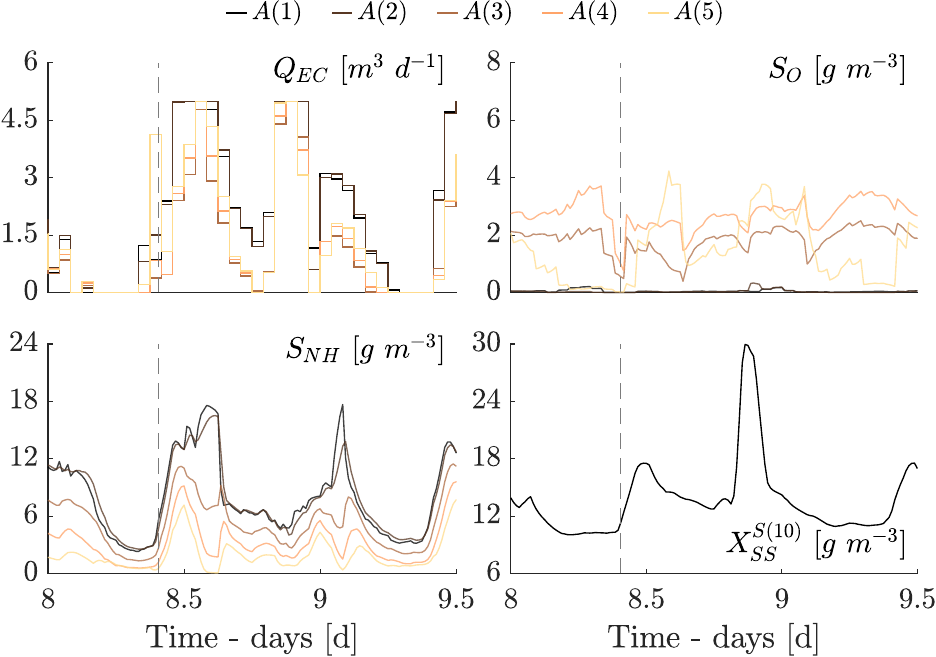}

    \caption{Case-study 2, $t\in [8,9.5]$: Extra carbon flow-rates $Q_{EC}^{(1\leadsto 5)}$and dissolved oxygen $S_{O}^{A(1\leadsto 5)}$ (top panels) and nitrogen $S_{NH}^{A(1\leadsto 5)}$ and effluent suspended solids $X_{SS}^{S(10)}$ (bottom panels). The vertical dashed line indicates the time of a reference change.}
    \label{fig: OutMPC_Reuse_Summary_SP3} 
\end{figure}

We analyse a selection of actions/responses that the Output MPC computes to perform the tracking (Figure \ref{fig: OutMPC_Reuse_Summary_SP1} and \ref{fig: OutMPC_Reuse_Summary_SP3}).

At the first reference change ($ t = 2.8$ days), the ASP serves the requested effluent total nitrogen, $N_{TOT}^{S(10)}$, by mainly producing $S_{NO}^{S(10)}$ nitrogen (Figure \ref{fig: OutMPC_Reuse_Summary_SP1}). This is done by increasing aeration in all reactors via $K_La^{(1\leadsto 5)}$,  favouring nitrification. As a result, the concentrations $S_{NO}^{A(1\leadsto 5)}$ quickly increase, too.
    \begin{itemize}
    \item[$\leadsto$] The reference is then maintained by instating a standard nitrification-denitrification layout, where reactors $A(1,2)$ are kept anoxic by reducing aeration ($K_La^{(1,2)}$), whereas reactors $A(3{\leadsto} 5)$ are kept aerated ($K_La^{(3\leadsto 5)}$). 
    \end{itemize}
In the settler, changes in $S_{NO}^{A(5)}$ in the stream from the bioreactors are reflected in the effluent $S_{NO}^{S(10)}$ and thus $N_{TOT}^{S(10)}$. Moreover, the Output MPC closely tracks the reference for suspended solids (Figure \ref{fig: OutMPC_Track_Eta0_Summary} of the Supplementary Material).

At the third reference change ($t = 8.4$ days), the aeration to the reactors $A(1{\leadsto} 5)$ is maintained in standard nitrification-denitrification layout via $K_La^{(1\leadsto 5)}$  (Figure \ref{fig: OutMPC_Reuse_Summary_SP3}).
The addition of external carbon is increased in all reactors via $Q_{EC}^{(1\leadsto 5)}$. 
\begin{itemize}
\item[$\leadsto$] The concentrations $S_{NH}^{A(1\leadsto 5)}$ are kept at desirable levels while decreasing the concentration of $S_{NO}^{A(1\leadsto 5)}$. As a result, the total nitrogen across the process is decreased. 
\end{itemize}
The changes in the feed are reflected in the settler by effluent $S_{NO}^{S(10)}$ and $S_{NH}^{S(10)}$, and $N_{TOT}^{S(10)}$. 
We point out that, while the Output MPC cannot closely track suspended solids, the limits are never violated (Figure \ref{fig: OutMPC_Track_Eta0_Summary} of the Supplementary Material).

\begin{figure}[h!] \centering
    \includegraphics[width=\columnwidth]{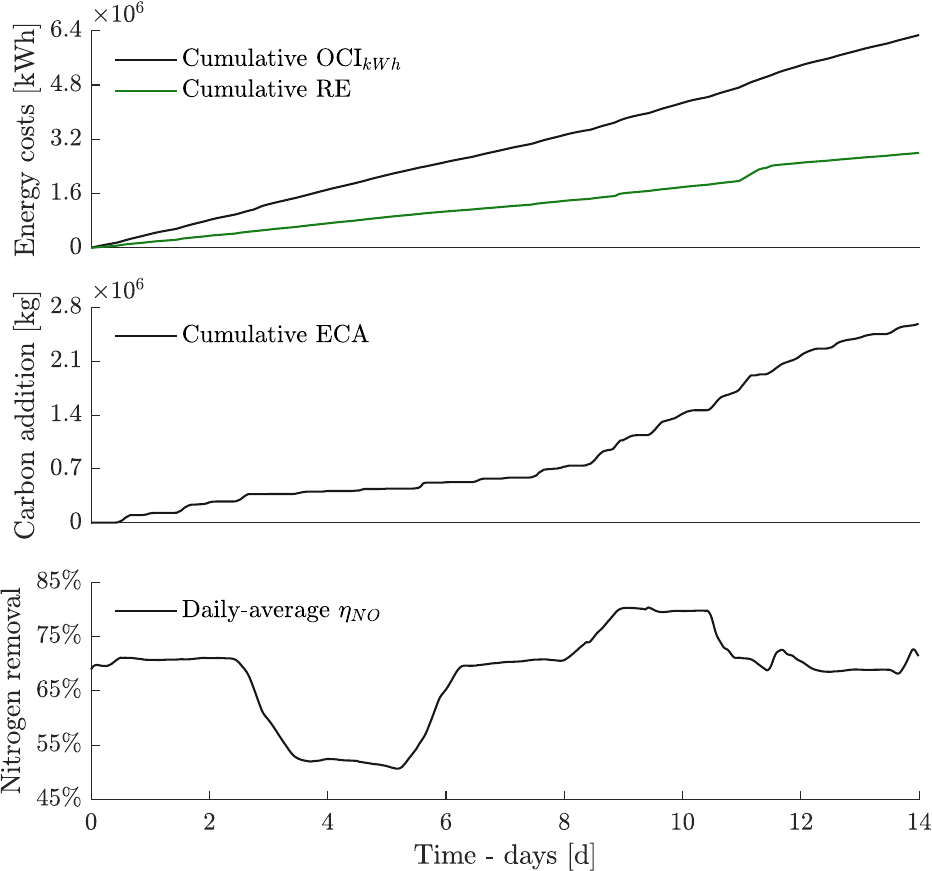}
 
    \caption{Case-study 2: Control performance in terms of operational cost index, OCI$_{\text{kWh}}$, external carbon addition, ECA, and daily-averaged nitrogen removal efficiency, $\eta_{NO}$.}  
    \label{fig: OutMPC_Reuse_Control_Metrics} 
\end{figure}

The performance of the Output MPC is summarised in Figure \ref{fig: OutMPC_Reuse_Control_Metrics} in terms of the cumulative operational cost index (OCI$_{\text{kWh}}$) and external carbon addition (ECA). 
The resulting nitrogen removal efficiency ($\eta_{NO}$, a daily-average) is reported, as well.
\begin{itemize}
\item The metrics indicate that the operational cost OCI$_{\text{kWh}}$ is not affect significantly during period, except for the first ($t = 2.8$) and last ($t=11.2$) reference change. 
\item The potential production of electricity RE indicates that approximately $44\%$ of the operational costs could be recycled even when the controller is not explicitly restricted to recover its energy demands from the waste. 
\end{itemize}

The results also show that a large quantity of external carbon is required to implement the actions obtained by the optimal controller. 
This is mainly due to the control strategy taken during the stricter nitrogen removal task, when the external carbon source flow-rate, $Q_{EC}^{(r)}$, is increased in all reactors. 
\begin{itemize}
\item[$\leadsto$] The efficiency in nitrogen removal reflects the expected performance for each task: Around $70\%$ of influent total nitrogen is removed during the conventional treatment, whereas $53\%$ and $77\%$ efficiencies are achieved for the reuse and nitrogen removal tasks, respectively.
\end{itemize} 

In conclusion, the Output MPC can operate the activated sludge plant to produce distinct nitrogen profiles, while satisfying the technological constraints of the equipment and keeping the other effluent concentrations at reference values. 

\subsubsection*{Nitrogen on-demand, with energy recovery} \label{sec: Results_Sustainable}
We expand upon the previous results by considering an Output MPC which, in addition to track the reference trajectory, must operate the ASP to recover energy from disposed sludge (that is, for cases with $\eta \in (0,1]$ in Eq. \eqref{eq: Energy_Constraint_Set}). We analyse the performance for increasing levels of energetic self-sufficiency for $\eta \in \{0.05,0.1,0.15,\dots,0.90,0.95,1\}$, individually.

\begin{figure}[htb!] \centering
	\includegraphics[width=\columnwidth]{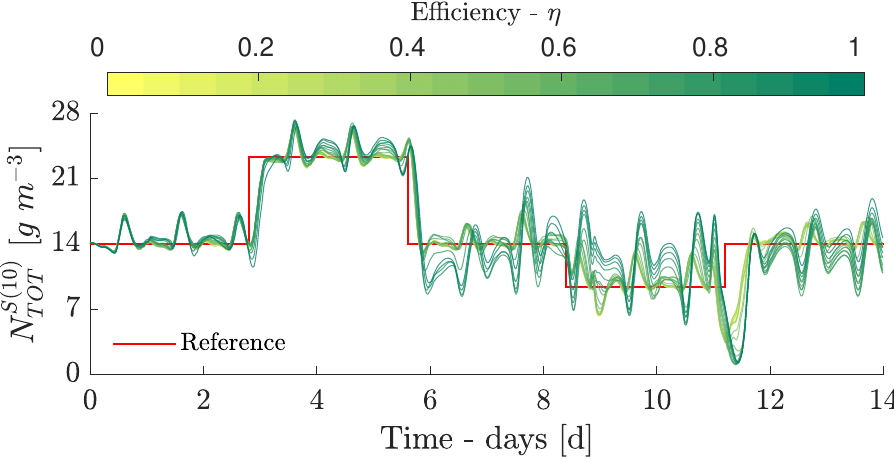}

	\caption{Case-study 2: Reference tracking of effluent total nitrogen $N_{TOT}^{S(10)}$ for different values of energy efficiency $\eta$.} 
	\label{fig: OutMPC_Sustainable_Reuse}
\end{figure}

The results, in Figure \ref{fig: OutMPC_Sustainable_Reuse}, show that the tracking accuracy slowly degrades as the Output MPC imposes more restrictive is requested to recover increasingly larger portions of its energetic demand from the wastage sludge. 
The tracking performance degrades when operating the ASP to perform extreme nitrogen removal ($t \in [8.4,11.2)$) while still attempting to recover $60\%$, or more, of the total energy needs. 
\begin{itemize}
    \item[$\leadsto$] Remarkably, excellent tracking is achieved before the occurrence of the extreme events ($t < 6$), almost regardless of the degree of energetic self-sustenance. 
    \item[$\leadsto$] This is an important result, as it highlights the flexibility potential of ASPs when efficiently controlled.
\end{itemize} 

\begin{figure}[htb!] \centering
    \includegraphics[width=\columnwidth]{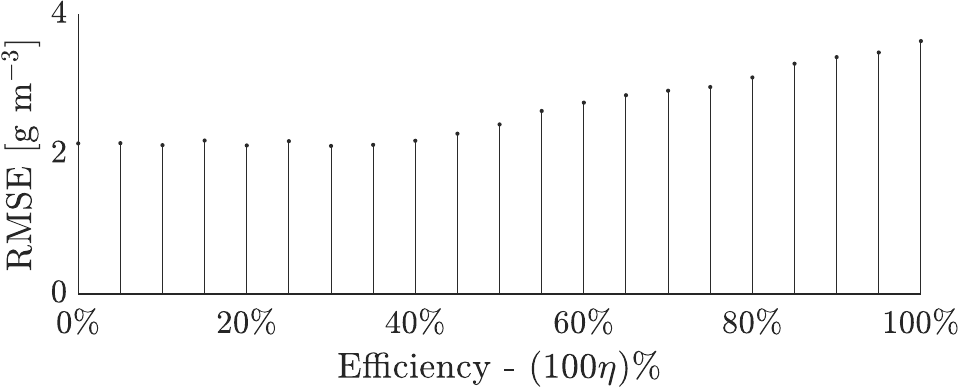}

    \caption{Case-study 2: Accuracy of reference tracking (RMSE), under different energy-recovery levels $\eta \in [0,1]$.} 
    \label{fig: OutMPC_Sustainable_RMSE_Metrics} 
\end{figure}

The controller performance in reference tracking is presented in Figure \ref{fig: OutMPC_Sustainable_RMSE_Metrics} in terms of the root-mean-squared-error (RMSE),
$$
    \text{RMSE} = \sqrt{ \frac{1}{14} \int_{0}^{14}\hspace*{-0.3em} \left( N_{TOT}^{ref}(t) - N_{TOT}^{S(10)}(t) \right)^2\hspace*{-0.2em} dt }.
$$
The results indicate that the controller is able to recover up to $60\%$ of the plant's energy demand, while still providing satisfactory tracking accuracy. The prevailing control strategy computed by the Output MPC leads to lower the aeration levels (to reduce operational costs) and to increase the wastage flow-rate (to biogas production and thus energy generation).
When the controller is requested to operate the plant under full energy recovery ($\eta = 1$), the tracking accuracy worsens: 
\begin{itemize}
\item[$\leadsto$] This results can be understood from an undesired effect of the aforementioned control strategy when the constraints lead to low levels of oxygen in all reactors.
\end{itemize}

\begin{figure}[htb!] \centering
    \includegraphics[width=\columnwidth]{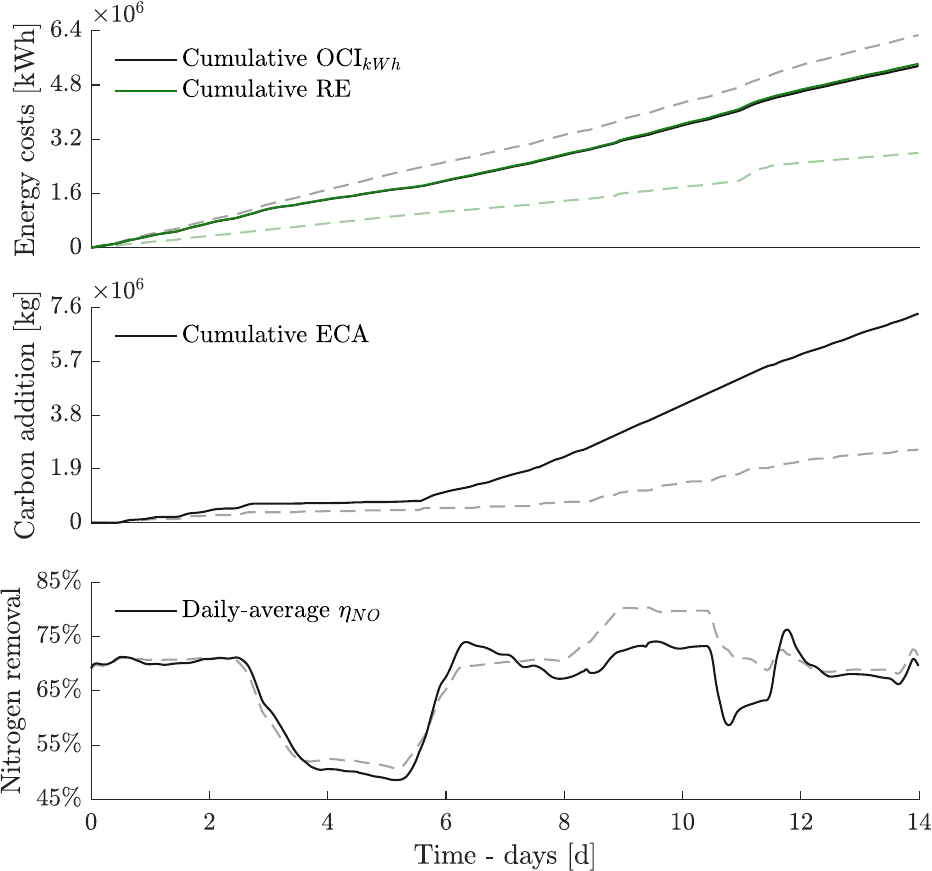}

    \caption{Case-study 2: Control performance in terms of operational cost index, OCI$_{\text{kWh}}$, external carbon addition, ECA, and daily-averaged nitrogen removal efficiency, $\eta_{NO}$, under full energy recovery ($\eta = 1$). Dashed lines refer to the performance obtained when no energy recovery is enforced ($\eta = 0$).} 
    \label{fig: OutMPC_Sustainable_Control_Metrics_Case02} 
\end{figure}

The cumulative operational cost index (OCI$_{kWh}$) and external carbon ($ECA$) needed to enforce a full energy recovery control strategy ($\eta = 1$) and the nitrogen conversion efficiency $\eta_{NO}$ (as daily averages) are shown in Figure \ref{fig: OutMPC_Sustainable_Control_Metrics_Case02}. 
As the electricity production (RE) matches the operating costs, we fully exploited the margins for achieving energetic self-sustenance. 
When compared to the performance for the controller without energy-recovery constraints ($\eta = 0$), the results also show an increased need for external carbon and a decrease in the nitrogen conversion efficiency during the nitrogen removal task ($t \in (8.4,11.2]$ days). This reflects how the Output MPC preferred to improve the denitrification-nitrification process by adding carbon rather than increasing aeration in the reactors.

In conclusion, the Output MPC can operate the activated sludge plant to produce distinct nitrogen profiles while ensuring that $60\%$ of its operational costs can be recovered from converting wastage sludge ito electricity. While the controller is also capable to enforce an operation which fully recovers the energy costs, the desired effluent quality profiles may be compromised. Again, we remind that only sludge from the secondary settler is used as a source for biogas production.

\section{Concluding remarks}\label{sec: Outro}
This work presents a general framework for the advanced control of a common class of activated sludge plants. 
The framework is based on a dynamic model of the plant and its sensors ad actuators.
We designed and configured a highly customisable Output Model-Predictive Controller for the operation of ASPs as facilities for conventional treatment of wastewater, as well as the recovery of materials and energy. 

The controller consists of a Moving-Horizon Estimator used to determine the state of the process, from plant data, and of a Model-Predictive Controller used for computing the optimal actions that drive the the plant to attain high-level operational goals.
By design, the Output MPC is configurable to satisfy all the technological constraints relative to the plant equipment. 
After overviewing the foundations of the control framework, we discuss the performance of the controller in tasks of practical relevance, ranging from depuration, to production of nitrogen on-demand and energy recovery.

\section{Acknowledgments}\label{sec: Acknowledgments} 

This study was done within the project Control4Reuse, part of the IC4WATER programme of the Water Challenges for a Changing World Joint Programme Initiative (Water JPI). The authors thank FUNCAP for the support within this initiative.

\printbibliography


%
%
%
%

\clearpage\onecolumn
\setcounter{equation}{0}
\setcounter{section}{0}
\setcounter{figure}{0}
\setcounter{table}{0}
\setcounter{page}{1}
\renewcommand{\theequation}{S\arabic{equation}}
\renewcommand{\thefigure}{S\arabic{figure}}
\allowdisplaybreaks

\begin{center}
    \Huge
    \textsc{SUPPLEMENTARY MATERIAL}\\
    A model-based framework for controlling activated sludge plants
\end{center}
\vskip1.5cm

\begin{abstract}%
    \noindent This sections provides supplementary material for the article `A model-based framework for controlling activated sludge plants'. The dynamic equations, model parameters, and complementary information for the Benchmark Simulation Model no. 1 (BSM1) are provided. Moreover, the document provides supplementary discussion on the experimental results presented in the main text.
\end{abstract}

\section{The BSM No. 1: Dynamics, parameters, equilibrium point, and smoothification}\label{sec: BSM1_SS}

The dynamic equations, alongside model parameters and complementary information, are provided in the following.

\subsection{Biological reactors: Activated Sludge Model no. 1}

From [1], each concentration $Z^{A(r)}$ in the bio-reactors $A(r)$ with $r=1,\ldots,5$ has dynamics in the form 
\begin{equation} \label{eq: ASM1_Dynamics}
    \dot{Z}^{A(r)} = \cfrac{Q_{IN}^{A(r)}}{V^{A(r)}} \left[ Z^{A(r,IN)} - Z^{A(r)} \right] + R\big( Z^{A(r)} \big),
\end{equation}
where $Q_{IN}^{A(r)}$ is the influent flow-rate to $A(r)$, $Z^{A(r,IN)}$ is the concentration of compound $Z$ in the influent to reactor $A(r)$, and $R\big( Z^{A(r)} \big)$ denotes the net change in concentration of component $Z$ in reactor $A(r)$ due to reactions.
We have, 
{\small
\begin{align}
    Q_{IN}^{A(r)} & = \begin{cases}
        Q_{IN} + Q_A + Q_R + Q_{EC}^{(1)}  & \text{ }(r = 1) \vspace*{0.5em}\\
        Q_{IN}^{A(r-1)} + Q_{EC}^{(r)}     & \text{ otherwise}
    \end{cases}\\
    Z^{A(r,IN)} & = \begin{cases}
        \cfrac{Q_{IN} Z^{IN} + Q_A Z^{A(5)} + Q_R Z^{S(1)} + Q_{EC}^{(1)} Z^{(EC)}}{Q_{IN}^{A(1)}}	& \text{ }(r = 1) \vspace*{0.5em}\\
        \cfrac{Q_{IN}^{A(r-1)} Z^{A(r-1)} + Q_{EC}^{(r)}Z^{(EC)} }{Q_{IN}^{A(r)}}					& \text{ otherwise}
    \end{cases}
\end{align}
}%
where $Z^{EC} = S_S^{EC}$ if $Z^{A(r)} = S_S^{A(r)}$, otherwise we have $Z^{EC} = 0$, with $Z^{IN}$ being the concentration in the influent. 

Explicitly, the dynamics in each reactor $A(r)$ with $r=1,\ldots,5$ are described by the set ordinary differential equations
\begin{subequations}\begin{align}
\dot{S}_I^{A(r)} 
	& = \cfrac{\clU{Q_{IN}^{A(r)}}}{\clP{V^{A(r)}}}\Big[\clX{S_{I}^{A(r,IN)}}-\clX{S_I^{A(r)}}\Big] \\
\dot{S}_S^{A(r)} 
	& = \cfrac{\clU{ Q_{IN}^{A(r)}}}{\clP{ V^{A(r)}}}\Big[\clX{S_{S}^{A(r,IN)}}-\clX{S_S^{A(r)}}\Big] \\
	&\quad - \cfrac{\clP{ \mu_H}}{\clP{ Y_H}} \cfrac{\clX{S_S^{A(r)}}}{\clP{ K_S}+\clX{S_S^{A(r)}}} \Big[ \cfrac{\clX{S_O^{A(r)}}}{\clP{ K_{OH}}+\clX{S_O^{A(r)}}} + \clP{ \eta_g} \cfrac{\clP{ K_{OH}}}{\clP{ K_{OH}}+\clX{S_O^{A(r)}}} \cfrac{\clX{S_{NO}^{A(r)}}}{\clP{ K_{NO}}+\clX{S_{NO}^{A(r)}}}\Big]\clX{X_{BH}^{A(r)}} \nonumber\\
	&\quad + \clP{ k_h}\cfrac{\clX{X_{S}^{A(r)}}}{\clP{ K_X}\clX{X_{BH}^{A(r)}}+\clX{X_S^{A(r)}}} \Big[\cfrac{\clX{S_O^{A(r)}}}{\clP{ K_{OH}}+\clX{S_O^{A(r)}}} + \clP{ \eta_h}\cfrac{\clP{ K_{OH}}}{\clP{ K_{OH}}+\clX{S_O^{A(r)}}} \cfrac{\clX{S_{NO}^{A(r)}}}{\clP{ K_{NO}}+\clX{S_{NO}^{A(r)}}}\Big]\clX{X_{BH}^{A(r)}} \nonumber\\
\dot{X}_I^{A(r)} & =\cfrac{\clU{ Q_{IN}^{A(r)}}}{\clP{ V^{A(r)}}}\Big[\clX{X_{I}^{A(r,IN)}}-\clX{X_I^{A(r)}}\Big] \\
\dot{X}_S^{A(r)} & = \cfrac{\clU{ Q_{IN}^{A(r)}}}{\clP{ V^{A(r)}}}\Big[\clX{X_{S}^{A(r,IN)}}-\clX{X_S^{A(r)}}\Big] \\
    &\quad - \clP{ k_h}\cfrac{\clX{X_{S}^{A(r)}}}{\clP{ K_X}\clX{X_{BH}^{A(r)}}+\clX{X_S^{A(r)}}} \Big[\cfrac{\clX{S_O^{A(r)}}}{\clP{ K_{OH}}+\clX{S_O^{A(r)}}} + \clP{ \eta_h}\cfrac{\clP{ K_{OH}}}{\clP{ K_{OH}}+\clX{S_O^{A(r)}}} \cfrac{\clX{S_{NO}^{A(r)}}}{\clP{ K_{NO}}+\clX{S_{NO}^{A(r)}}}\Big]\clX{X_{BH}^{A(r)}} \nonumber\\ 
    &\quad + \Big[1-\clP{ f_p}\Big] \clP{ b_H}\clX{X_{BH}^{A(r)}}+\Big[1-\clP{ f_p}\Big] \clP{ b_A}\clX{X_{BA}^{A(r)}}  \nonumber\\
\dot{X}_{BH}^{A(r)} & = \cfrac{\clU{ Q_{IN}^{A(r)}}}{\clP{ V^{A(r)}}}\Big[\clX{X_{BH}^{A(r,IN)}}-\clX{X_{BH}^{A(r)}}\Big] \\ 
	&\quad + \clP{ \mu_H} \cfrac{\clX{S_S^{A(r)}}}{\clP{ K_S}+\clX{S_S^{A(r)}}} \Big[ \cfrac{\clX{S_O^{A(r)}}}{\clP{ K_{OH}}+\clX{S_O^{A(r)}}} + \clP{ \eta_g} \cfrac{\clP{ K_{OH}}}{\clP{ K_{OH}}+\clX{S_O^{A(r)}}} \cfrac{\clX{S_{NO}^{A(r)}}}{\clP{ K_{NO}}+\clX{S_{NO}^{A(r)}}}\Big]\clX{X_{BH}^{A(r)}} - \clP{ b_H}\clX{X_{BH}^{A(r)}} \nonumber\\
\dot{X}_{BA}^{A(r)} & = \cfrac{\clU{ Q_{IN}^{A(r)}}}{\clP{ V^{A(r)}}}\Big[\clX{X_{BA}^{A(r,IN)}}-\clX{X_{BA}^{A(r)}}\Big] 
    + \clP{ \mu_A}\cfrac{\clX{S_{NH}^{A(r)}}}{\clP{ K_{NH}}+\clX{S_{NH}^{A(r)}}}\cfrac{\clX{S_O^{A(r)}}}{\clP{ K_{OA}}+\clX{S_O^{A(r)}}}\clX{X_{BA}^{A(r)}}-\clP{ b_A}\clX{X_{BA}^{A(r)}} \\ 
\dot{X}_P^{A(r)} & = \cfrac{\clU{ Q_{IN}^{A(r)}}}{\clP{ V^{A(r)}}}\Big[\clX{X_{P}^{A(r,IN)}}-\clX{X_{P}^{A(r)}}\Big] 
    + \clP{ f_P} \Big[ \clP{ b_H}\clX{X_{BH}^{A(r)}}+\clP{ b_A}\clX{X_{BA}^{A(r)}} \Big]\\
\dot{S}_O^{A(r)} & = \cfrac{\clU{ Q_{IN}^{A(r)}}}{\clP{ V^{A(r)}}} \Big[\clX{S_{O}^{A(r,IN)}}-\clX{S_{O}^{A(r)}}\Big]  + \clU{ K_La^{(r)}}\Big[\clP{ S_O^{sat}}-\clX{S_O^{A(r)}}\Big]  \\
    &\quad - \cfrac{1-\clP{ Y_H}}{\clP{ Y_H}}\clP{ \mu_H} \cfrac{\clX{S_S^{A(r)}}}{\clP{ K_S}+\clX{S_S^{A(r)}}}\cfrac{\clX{S_O^{A(r)}}}{\clP{ K_{OH}}+\clX{S_O^{A(r)}}} \clX{X_{BH}^{A(r)}}  - \cfrac{4.57-\clP{ Y_A}}{\clP{ Y_A}}\clP{ \mu_A}\cfrac{\clX{S_{NH}^{A(r)}}}{\clP{ K_{NH}}+\clX{S_{NH}^{A(r)}}}\cfrac{\clX{S_O^{A(r)}}}{\clP{ K_{OA}}+\clX{S_O^{A(r)}}}\clX{X_{BA}^{A(r)}} \nonumber\\
\dot{S}_{NO}^{A(r)} & = \cfrac{\clU{ Q_{IN}^{A(r)}}}{\clP{ V^{A(r)}}}\Big[\clX{S_{NO}^{A(r,IN)}}-\clX{S_{NO}^{A(r)}}\Big]  \\
    &\quad -\cfrac{1-\clP{ Y_H}}{2.86\clP{ Y_H}}\clP{ \mu_H}\cfrac{\clX{S_S^{A(r)}}}{\clP{ K_S}+\clX{S_S^{A(r)}}}\cfrac{\clP{ K_{OH}}}{\clP{ K_{OH}}+\clX{S_O^{A(r)}}}\cfrac{\clX{S_{NO}^{A(r)}}}{\clP{ K_{NO}}+\clX{S_{NO}^{A(r)}}}\clP{ \eta_g}\clX{X_{BH}^{A(r)}}  + \cfrac{\clP{ \mu_A}}{\clP{ Y_A}}\cfrac{\clX{S_{NH}^{A(r)}}}{\clP{ K_{NH}}+\clX{S_{NH}^{A(r)}}}\cfrac{\clX{S_O}}{\clP{ K_{OA}}+\clX{S_O^{A(r)}}}\clX{X_{BA}^{A(r)}} \nonumber\\
\dot{S}_{NH}^{A(r)} & = \cfrac{\clU{ Q_{IN}^{A(r)}}}{\clP{ V^{A(r)}}}\Big[\clX{S_{NH}^{A(r,IN)}}-\clX{S_{NH}^{A(r)}}\Big]  \\
    &\quad - \clP{ \mu_H}\cfrac{\clX{S_S}}{\clP{ K_S}+\clX{S_S^{A(r)}}}\Big[ \cfrac{\clX{S_O^{A(r)}}}{\clP{ K_{OH}}+\clX{S_O^{A(r)}}} + \clP{ \eta_g} \cfrac{\clP{ K_{OH}}}{\clP{ K_{OH}}+\clX{S_O^{A(r)}}} \cfrac{\clX{S_{NO}^{A(r)}}}{\clP{ K_{NO}}+\clX{S_{NO}^{A(r)}}}\Big] \clP{ i_{XB}}\clX{X_{BH}^{A(r)}} \nonumber\\
    &\quad - \clP{ \mu_A}\Big(\clP{ i_{XB}}+\cfrac{1}{\clP{ Y_A}}\Big)\cfrac{\clX{S_{NH}^{A(r)}}}{\clP{ K_{NH}}+\clX{S_{NH}^{A(r)}}}\cfrac{\clX{S_O^{A(r)}}}{\clP{ K_{OA}}+\clX{S_O^{A(r)}}}\clX{X_{BA}^{A(r)}}+\clP{ k_a}\clX{S_{ND}^{A(r)}}\clX{X_{BH}^{A(r)}} \nonumber\\
\dot{S}_{ND}^{A(r)} & = \cfrac{\clU{ Q_{IN}^{A(r)}}}{\clP{ V^{A(r)}}}\Big[\clX{S_{ND}^{A(r,IN)}}-\clX{S_{ND}^{A(r)}}\Big]  \\
    &\quad - \clP{ k_a}\clX{S_{ND}^{A(r)}}\clX{X_{BH}^{A(r)}} + \clP{ k_h}\cfrac{\clX{X_{ND}^{A(r)}} \clX{X_{BH}^{A(r)}}}{\clP{ K_X}\clX{X_{BH}^{A(r)}}+\clX{X_S^{A(r)}}}\Big[\cfrac{\clX{S_O^{A(r)}}}{\clP{ K_{OH}}+\clX{S_O^{A(r)}}} + \clP{ \eta_h}\cfrac{\clP{ K_{OH}}}{\clP{ K_{OH}}+\clX{S_O^{A(r)}}}\cfrac{\clX{S_{NO}^{A(r)}}}{\clP{ K_{NO}}+\clX{S_{NO}^{A(r)}}}\Big] \nonumber\\
\dot{X}_{ND}^{A(r)} & = \cfrac{\clU{ Q_{IN}^{A(r)}}}{\clP{ V^{A(r)}}}\Big[\clX{X_{ND}^{A(r,IN)}}-\clX{X_{ND}^{A(r)}}\Big]  \\
    &\quad - \clP{ k_h}\cfrac{\clX{X_{ND}^{A(r)}}}{\clP{ K_X}\clX{X_{BH}^{A(r)}}+\clX{X_S^{A(r)}}}\Big[\cfrac{\clX{S_O^{A(r)}}}{\clP{ K_{OH}}+\clX{S_O^{A(r)}}} + \clP{ \eta_h}\cfrac{\clP{ K_{OH}}}{\clP{ K_{OH}}+\clX{S_O^{A(r)}}}\cfrac{\clX{S_{NO}^{A(r)}}}{\clP{ K_{NO}}+\clX{S_{NO}^{A(r)}}}\Big]\clX{X_{BH}^{A(r)}} \nonumber\\
    &\quad + \clP{ b_H} \Big(\clP{ i_{XB}}-\clP{ f_P}\clP{ i_{XP}}\Big)\clX{X_{BH}^{A(r)}} +\clP{ b_A} \Big(\clP{ i_{XB}}-\clP{ f_P}\clP{ i_{XP}}\Big)\clX{X_{BA}^{A(r)}}  \nonumber\\
\dot{S}_{ALK}^{A(r)} & = \cfrac{\clU{ Q_{IN}^{A(r)}}}{\clP{ V^{A(r)}}}\Big[\clX{S_{ALK}^{A(r,IN)}}-\clX{S_{ALK}^{A(r)}}\Big]  \\
    &\quad - \cfrac{\clP{ i_{XB}}}{14}\clP{ \mu_H}\cfrac{\clX{S_S^{A(r)}}}{\clP{ K_S}+\clX{S_S^{A(r)}}}\cfrac{\clX{S_O^{A(r)}}}{\clP{ K_{OH}}+\clX{S_O^{A(r)}}} \clX{X_{BH}^{A(r)}} +\cfrac{1}{14}\clP{ k_a}\clX{S_{ND}^{A(r)}}\clX{X_{BH}^{A(r)}} \nonumber\\
    &\quad + \Big(\cfrac{1-\clP{ Y_H}}{14 \times 2.86 \clP{ Y_H}} - \cfrac{\clP{ i_{XB}}}{14} \Big)\clP{ \mu_H} \cfrac{\clX{S_S^{A(r)}}}{\clP{ K_S}+\clX{S_S^{A(r)}}}\cfrac{\clP{ K_{OH}}}{\clP{ K_{OH}}+\clX{S_O^{A(r)}}}\cfrac{\clX{S_{NO}^{A(r)}}}{\clP{ K_{NO}}+\clX{S_{NO}^{A(r)}}} \clP{ \eta_g} \clX{X_{BH}^{A(r)}} \nonumber\\ 
    &\quad  - \Big(\cfrac{\clP{ i_{XB}}}{14}+\cfrac{1}{7\clP{ Y_A}}\Big)\clP{ \mu_A}\cfrac{\clX{S_{NH}^{A(r)}}}{\clP{ K_{NH}}+\clX{S_{NH}^{A(r)}}}\cfrac{\clX{S_O^{A(r)}}}{\clP{ K_{OA}}+\clX{S_O^{A(r)}}}\clX{X_{BA}^{A(r)}} \nonumber
\end{align}\end{subequations}

\subsection{Secondary settler}

From [2], the dynamics for suspended solids , \clX{$X_{SS}^{S(l)}$}, within each $l$-th settler's layer, are described by
\begin{align} 
\dot{X}_{SS}^{S(l)} & = 
    \begin{cases} 
    \cfrac{\clU{ Q_e}}{\clP{ V^{S(l)}}} \Big[ \clX{X_{SS}^{S(l-1)}} - \clX{X_{SS}^{S(l)}} \Big] - \cfrac{1}{\clP{ h^{S(l)}}} J_{cla}\Big(\clX{X_{SS}^{S(l)}}, \clX{X_{SS}^{S(l-1)}}\Big) 
    & (l = 10) \\
    \cfrac{\clU{ Q_e}}{\clP{ V^{S(l)}}} \Big[ \clX{X_{SS}^{S(l-1)}} - \clX{X_{SS}^{S(l)}} \Big] + \cfrac{1}{\clP{ h^{S(l)}}} \Big[ J_{cla}\Big(\clX{X_{SS}^{S(l+1)}}, \clX{X_{SS}^{S(l)}}\Big) - J_{cla}\Big(\clX{X_{SS}^{S(l)}}, \clX{X_{SS}^{S(l-1)}}\Big) \Big] 
    & (l = 7, \dots, 9) \\
    \cfrac{\clU{ Q_f}}{\clP{ V^{S(l)}}}  \Big[ \clX{X_f} - \clX{X_{SS}^{S(l)}} \Big] + \cfrac{1}{\clP{ h^{S(l)}}} \Big[ J_{cla}\Big(\clX{X_{SS}^{S(l+1)}}, \clX{X_{SS}^{S(l)}}\Big) - J_{st}\Big(\clX{X_{SS}^{S(l)}}, \clX{X_{SS}^{S(l-1)}}\Big) \Big] 
    & (l = 6) \\
    \cfrac{\clU{ Q_u}}{\clP{ V^{S(l)}}} \Big[ \clX{X_{SS}^{S(l+1)}} - \clX{X_{SS}^{S(l)}} \Big] + \cfrac{1}{\clP{ h^{S(l)}}} \Big[ J_{st}\Big(\clX{X_{SS}^{S(l+1)}}, \clX{X_{SS}^{S(l)}}\Big) - J_{st}\Big(\clX{X_{SS}^{S(l)}}, \clX{X_{SS}^{S(l-1)}}\Big) \Big] 
    & (l = 2, \dots, 5) \\
    \cfrac{\clU{ Q_u}}{\clP{ V^{S(l)}}} \Big[ \clX{X_{SS}^{S(l+1)}} - \clX{X_{SS}^{S(l)}} \Big] + \cfrac{1}{\clP{ h^{S(l)}}} J_{st}\Big(\clX{X_{SS}^{S(l+1)}}, \clX{X_{SS}^{S(l)}}\Big) 
    & (l = 1)
    \end{cases}.
\end{align}%
The dynamics of soluble matter $S_{(\cdot)}^{S(l)}$ within each $l$-th layer are described by
\begin{align}
\dot{S}_{(\cdot)}^{S(l)} 
    & = \begin{cases} 
    \cfrac{\clU{ Q_e}}{\clP{ V^{S(l)}}} \Big[ \clX{S_{(\cdot)}^{S(l-1)}} - \clX{S_{(\cdot)}^{S(l)}} \Big] & (l = 7, \dots, 10); \\
    \cfrac{\clU{ Q_f}}{\clP{ V^{S(l)}}} \Big[ \clX{S_{(\cdot)}^{A(5)}} - \clX{S_{(\cdot)}^{S(l)}} \Big] & (l = 6); \\
    \cfrac{\clU{ Q_u}}{\clP{ V^{S(l)}}} \Big[ \clX{S_{(\cdot)}^{S(l+1)}} - \clX{S_{(\cdot)}^{S(l)}} \Big] & (l = 1, \dots, 5), 
    \end{cases}.
\end{align}%
The quantities $(Q_f,X_f)$ denote the flow-rate and concentration of solids entering the settler, at layer $S(l=6)$, with $\clU{ Q_f} = Q_{IN}^{A(5)} - Q_A$ and $X_f = 0.75\big( \clX{X_I^{A(5)}}+\clX{X_S^{A(5)}}+\clX{X_{BH}^{A(5)}}+\clX{X_{BA}^{A(5)}}+\clX{X_P^{A(5)}} \big)$. The (under)flow-rate is $\clU{ Q_u} = (\clU{Q_R}+\clU{ Q_W})$, and the plant's effluent flow-rate is $\clU{ Q_e} = (\clU{Q_{f}}-\clU{Q_u})$. The downward and upward flux of solids are respectively given by 
\begin{align} 
J_{st}\left(\clX{X_{SS}^{S(l)}}, \clX{X_{SS}^{S(l-1)}}\right) 
    & = \min\left[ v_s\big(\clX{X_{SS}^{S(l-1)}}\big)\clX{X_{SS}^{S(l-1)}}, v_s\big(\clX{X_{SS}^{S(l)}}\big)\clX{X_{SS}^{S(l)}} \right]; \label{eq: Settler_solid_flux_upward} \\
J_{cla}\left(\clX{X_{SS}^{S(l)}}, \clX{X_{SS}^{S(l-1)}}\right) 
    & = \begin{cases} 
    \min\left[ v_s\big(\clX{X_{SS}^{S(l-1)}}\big)\clX{X_{SS}^{S(l-1)}}, v_s\big(\clX{X_{SS}^{S(l)}}\big)\clX{X_{SS}^{S(l)}} \right] & \text{if } \clX{X_{SS}^{S(l-1)}} > \clP{ X_t}; \\
    v_s\big(\clX{X_{SS}^{S(l)}}\big)\clX{X_{SS}^{S(l)}} & \text{otherwise},
    \end{cases} \label{eq: Settler_solid_flux_downward}
\end{align}%
in which
\begin{equation} 
    v_s\Big(\clX{X_{SS}^{S(l)}}\Big) = \max\Big\{ 0,\ \min\Big[ \clP{ v_0}\Big( e^{-\clP{ r_h}(\clX{X_{SS}^{S(l)}} - \clP{ f_{ns}} X_f)} - e^{-\clP{ r_p} (\clX{X_{SS}^{S(l)}} - \clP{ f_{ns}} X_f)} \Big),\ \clP{ v_0^{max}} \Big] \Big\}.
\end{equation}

\subsection{Smoothification of discontinuities}
  
Jacobian linearisations of the dynamics require that functions $f(\cdot)$ and $g(\cdot)$ are differentiable with respect to state and input variables. 
Due to the discontinuities in the model of the settler, a smooth approximation of the BSM1 was obtained by replacing the terms corresponding to minimum and maximum functions between two terms by a log-sum-exp or soft-max function, whereas a hyperbolic tangent function was used for approximating conditional statements. 

We rewrite below the condition for the downward flux of solids, $J_{cla}(\cdot)$ given in Eq. \eqref{eq: Settler_solid_flux_downward},
\begin{align}
    J_{cla}(\cdot)
        = {\varphi(\clX{X_{SS}^{S(l-1)}})} \min\left[ v_s\big(\clX{X_{SS}^{S(l-1)}}\big)\clX{X_{SS}^{S(l-1)}}, v_s\big(\clX{X_{SS}^{S(l)}}\big)\clX{X_{SS}^{S(l)}} \right] 
        + \big[ 1 - \varphi(\clX{X_{SS}^{S(l-1)}}) \big] v_s\big(\clX{X_{SS}^{S(l)}}\big)\clX{X_{SS}^{S(l)}}
\end{align}
with $\varphi(\clX{X_{SS}^{S(l-1)}}) = 1$ when $\clX{X_{SS}^{S(l-1)}} - \clP{X_t} > 0$ and $\varphi(\clX{X_{SS}^{S(l-1)}}) = 0$ otherwise. 
Then, we approximate the step function 
\begin{equation}
    \varphi(\clX{X_{SS}^{S(l-1)}}) \approx 0.5 + 0.5 \tanh\Big( 50 \big(\clX{ X_{SS}^{S(l-1)}} - \clP{X_t} \big) \Big).
\end{equation}

\subsection{Performance metrics}

The effluent concentrations of biochemical oxygen demand ($BOD_5$), chemical oxygen demand ($COD$) and total nitrogen ($N_{TOT}$) are defined from the concentrations at the top layer ($S^{(10)}$) of the settler as
\begin{align}
    BOD_5^{S(10)}	& = \big( (1 - {f_P}) (X_{BH}^{S(10)} + X_{BA}^{S(10)}) + S_S^{S(10)} + X_{S}^{S(10)} \big)/4; \\
    COD^{S(10)}		& = S_S^{S(10)} + S_I^{S(10)} + X_S^{S(10)} + X_{I}^{S(10)} + X_{BH}^{S(10)} + X_{BA}^{S(10)} + X_{P}^{S(10)};	\\
    N_{TOT}^{S(10)}	& = S_{NO}^{S(10)} + S_{NH}^{S(10)} + S_{ND}^{S(10)} + X_{ND}^{S(10)} + {i_{XB}} \big(X_ {BH}^{S(10)} + X_{BA}^{S(10)} \big) + {i_{XP}} \big( X_{P}^{S(10)} + X_{I}^{S(10)} \big).
\end{align}
The effluent concentrations of particle compounds are given as $X_{(\cdot)}^{S(10)} = \big( {X_{SS}^{S(10)}}/{X_f} \big) X_{(\cdot)}^{A(5)}$, with feed concentration $X_f = 0.75\big( \clX{X_I^{A(5)}}+\clX{X_S^{A(5)}}+\clX{X_{BH}^{A(5)}}+\clX{X_{BA}^{A(5)}}+\clX{X_P^{A(5)}} \big)$. 

Based on these quantities, the treatment performance can be quantified by the effluent quality index (EQI, in kg d$^{-1}$), 
\begin{equation}
	\text{EQI} = Q_{e}\big( 2 X_{SS}^{S(10)} + COD^{S(10)} + 2 BOD_5^{S(10)} \\+ 30 N_{TKN}^{S(10)} + 10 S_{NO}^{S(10)} \big),
\end{equation}
with effluent flow-rate $Q_{e} = Q_{IN}^{A(5)} - Q_{A}$ and Kjeldahl nitrogen $N_{TKN}^{S(10)} = N_{TOT}^{S(10)} - S_{NO}^{S(10)}$. 
As the EQI defines the total quantity of pollutants in the effluent, the performance of an ASP can be based on measuring and minimising this metric.

The energy to implement a control strategy can be quantified by the operational cost index (OCI$_{\text{kWh}}$, in kWh d$^{-1}$), 
\begin{equation} \label{eq: OCI}
    \text{OCI}_{\text{kWh}}(t) = \text{AE}(t) + \text{PE}(t) + \text{ME}(t),
\end{equation}
where the aeration (AE, in kWh d$^{-1}$), pumping (PE, in kWh d$^{-1}$), and mixing (ME, in kWh d$^{-1}$) energies are
\begin{subequations}
\begin{align}
    \text{AE}(t) &= \cfrac{S_O^{sat}}{1800} { \sum_{r=1}^5 V^{A(r)} K_La^{(r)}(t)};         \label{eq: Energies_AE}\\
    \text{PE}(t) &= \cfrac{1}{1000} {\big( 4Q_A(t) + 8Q_R(t) + 50Q_W(t) \big) };   \label{eq: Energies_PE}\\
    \text{ME}(t) &= \cfrac{24}{1000} { \sum_{r=1}^5 5V^{A(r)} H\big(20 - K_La^{(r)}(t)\big) }.       \label{eq: Energies_ME}
\end{align} \label{eq: Energies}%
\end{subequations}
The step function is $H(x) = 1$ if $x \geq 0$, otherwise we have $H(x) = 0$. 

This  metric differs from the standard definition, as it uses only terms corresponding to actual energy (in kWh d$^{-1}$ units).

\newpage\subsection{Model parameters}

The model equations depend on the set of stoichiometric, kinetic and general parameters reported below in Table \ref{tab: ASP_Parameters}. 

\begin{table}[htb!] \centering
    \caption{Benchmark Simulation Model No. 1: Model constant parameters.}
    \begin{tabular} {c | l | c | l} 
                    & Stoichiometric parameter 			        	& Value 	& Units		                                                \\ 
    \doublehline
    $\clP{Y_A}$	    & Autotrophic yield						        & 0.24		& g X$_{BA}$ COD formed $\cdot$ (g N oxidised)$^{-1}$	    \\
    $\clP{Y_H}$	    & Heterotrophic yield 					        & 0.67		& g X$_{BH}$ COD formed $\cdot$ (g COD utilised)$^{-1}$	    \\
    $\clP{f_P}$	    & Fraction of biomass to particulate products 	& 0.08		& g X$_{P}$ COD formed $\cdot$ (g X$_{BH}$ decayed)$^{-1}$	\\
    $\clP{i_{XB}}$	& Fraction nitrogen in biomass 				    & 0.08		& g N (g COD)$^{-1}$ in biomass                             \\
    $\clP{i_{XP}}$	& Fraction nitrogen in particulate products 	& 0.06		& g N (g COD)$^{-1}$ in X$_P$                               \\[0.5ex]
    \hline			&							                    &			&					                                        \\
    \multicolumn{4}{c}{}\\
                    & Kinetic parameter					        & Value 	& Units						        \\ 
    \doublehline
    $\clP{\mu_{H}}$	& Maximum heterotrophic growth rate 	    & 4.00		& d$^{-1}$						    \\
    $\clP{K_S}$		& Half-saturation (heterotrophic growth) 	& 10.0		& g COD m$^{-3}$				    \\
    $\clP{K_{OH}}$	& Half-saturation (heterotrophic oxygen) 	& 0.20		& g O$_2$ m$^{-3}$				    \\
    $\clP{K_{NO}}$	& Half-saturation (nitrate)				    & 0.50		& g NO$_3$-N m$^{-3}$			    \\
    $\clP{b_H}$		& Heterotrophic decay rate 			        & 0.30		& d$^{-1}$						    \\
    $\clP{\eta_g}$	& Anoxic growth rate correction factor 	    & 0.80		& dimensionless					    \\
    $\clP{\eta_h}$	& Anoxic hydrolysis rate correction factor 	& 0.80		& dimensionless					    \\
    $\clP{k_h}$		& Maximum specific hydrolysis rate		    & 3.00		& g X$_S$ (g X$_{BH}$ COD d)$^{-1}$	\\
    $\clP{K_X}$		& Half-saturation (hydrolysis)			    & 0.10		& g X$_S$ (g X$_{BH}$ COD)$^{-1}$	\\
    $\clP{\mu_A}$	& Maximum autotrophic growth rate		    & 0.50		& d$^{-1}$						    \\	
    $\clP{K_{NH}}$	& Half-saturation (autotrophic growth)	    & 1.00		& g NH$_4$-N m$^{-3}$			    \\
    $\clP{b_{A}}$	& Autotrophic decay rate				    & 0.05		& d$^{-1}$						    \\
    $\clP{K_{OA}}$	& Half-saturation (autotrophic oxygen) 	    & 0.40		& g O$_2$ m$^{-3}$				    \\		
    $\clP{k_a}$		& Ammonification rate 				        & 0.05		& m$^3$ (g COD d)$^{-1}$			\\[0.5ex]
    \hline		    &								            &			&							        \\
    \multicolumn{4}{c}{}\\  	
                    & Secondary settler parameter			& Value 	& Units				    \\ 
    \doublehline
    $\clP{v_0^{max}}$	& Maximum settling velocity			& 250.0 	& m d$^{-1}$		    \\
    $\clP{v_0}$		& Maximum Vesilind settling velocity 	& 474.0		& m d$^{-1}$		    \\
    $\clP{r_h}$		& Hindered zone settling parameter 		& 0.000576	& m$^{3}$ (g SS)$^{-1}$	\\
    $\clP{r_p}$		& Flocculant zone settling parameter 	& 0.00286	& m$^{3}$ (g SS)$^{-1}$	\\
    $\clP{f_{ns}}$	& Non-settleable fraction				& 0.00228	& dimensionless		    \\[0.5ex]
    \hline		    &								        &			&						\\
    \multicolumn{4}{c}{}\\  	
                                                & General parameter			            & Value 		& Units				\\ 
    \doublehline
    $\clP{V^{A(1\leadsto 2)}}$			        & Reactor volume (anoxic section)		& 1000 		    & m$^{3}$		    \\
    $\clP{V^{A(3\leadsto 5)}}$	                & Reactor volume (aerobic section) 		& 1333		    & m$^{3}$		    \\
    $\clP{V^{S(l)}}$	                        & Settler layer volume 				    & 600		    & m$^{3}$	        \\
    $\clP{h^{S(l)}}$	                        & Settler layer height 				    & 0.4			& m		            \\
    $\clP{S_S^{EC}}$							& External carbon source concentration	& $4\cdot 10^5$	& g COD m$^{-3}$    \\
    $\clP{S_O^{sat}}$							& Oxygen saturation concentration		& 8.0			& g O$_2$ m$^{-3}$	\\
    $\clP{X_t}$									& Settling threshold concentration 		& 3000		    & g m$^{-3}$		\\[0.5ex]
    \hline
\end{tabular}\label{tab: ASP_Parameters} 
\end{table}

\newpage\subsection{Common equilibrium point for linearisation}

The conventional operation of the BMS1 corresponds to the \textit{steady-state} point $P \coloneqq (\clX{\bar{x}}, \clU{\bar{u}}, \clW{\bar{w}}, \clY{\bar{y}})$ presented in Table \ref{tab: Linearisation}.

\begin{table}[htb!] \centering
\caption{Benchmark Simulation Model No. 1: Fixed point $P \coloneqq (\bar{x}, \bar{u}, \bar{w}, \bar{y})$.}
\vskip0.15cm
\begin{tabular} {c | c | c | c | c | c | c | l} 
			\cline{2-7}
				& Influent  & \multicolumn{5}{c|}{Reactor}\\
			\cline{2-7}
				& IN        & A(1)   & A(2)      & A(3)  & A(4)  & A(5)   	& Units\\ 
	\doublehline 
	$S_{I}$	    & 30        & 30     & 30        & 30    & 30    & 30 		& g COD m$^{-3}$ 	\\
	$S_{S}$	    & 69.5      & 2.81   & 1.46      & 1.15  & 0.995 & 0.889 	& g COD m$^{-3}$ 	\\
	$X_{I}$	    & 51.2      & 1149   & 1149      & 1149  & 1149  & 1149 	& g COD m$^{-3}$ 	\\
	$X_{S}$	    & 202.32    & 82.1   & 76.4      & 64.9  & 55.7  & 49.3 	& g COD m$^{-3}$ 	\\
	$X_{BH}$    & 28.17     & 2552   & 2553      & 2557  & 2559  & 2559 	& g COD m$^{-3}$ 	\\
	$X_{BA}$    & 0         & 148    & 148       & 149   & 150   & 150 		& g COD m$^{-3}$ 	\\
	$X_{P}$	    & 0         & 449    & 450       & 450   & 451   & 452 		& g COD m$^{-3}$ 	\\
	$S_{O}$	    & 0         & 0.0043 & 6.31E-5 & 1.72  & 2.43  & 0.491    & g O$_2$ m$^{-3}$ 	\\
	$S_{NO}$    & 0         & 5.37   & 3.66      & 6.54  & 9.30  & 10.4 	& g N m$^{-3}$ 	    \\
	$S_{NH}$    & 31.56     & 7.92   & 8.34      & 5.55  & 2.97  & 1.73 	& g N m$^{-3}$ 	    \\
	$S_{ND}$    & 6.95      & 1.22   & 0.882     & 0.829 & 0.767 & 0.688 	& g N m$^{-3}$ 	    \\
	$X_{ND}$    & 10.59     & 5.28   & 5.03      & 4.39  & 3.88  & 3.53 	& g N m$^{-3}$ 	    \\
	$S_{ALK}$	& 7         & 4.93   & 5.08      & 4.67  & 4.29  & 4.13 	& mol HCO$_3^-$ m$^{-3}$	\\[0.5ex]
\hline
\end{tabular} \vskip0.35cm
\begin{tabular} {c | c | c | c | c | c | c | c | c | c | c | l} 
                \cline{2-11}
                &  \multicolumn{10}{c|}{Settler Layer}	\\
                \cline{2-11}
                & S(1) & S(2) & S(3) & S(4) & S(5) & S(6) & S(7) & S(8) & S(9) & S(10) 					& Units \\ 
    \doublehline 
    $X_{SS}$	& 6394  & 356.07 & 356.07 & 356.07 & 356.07 & 356.07 & 68.98 & 29.54 & 18.11 & 12.5 	& g COD m$^{-3}$ 	\\
    $S_{I}$	    & 30    & 30     & 30 & 30 & 30  & 30 & 30 & 30 & 30 & 30							    & g COD m$^{-3}$ 	\\
    $S_{S}$	    & 0.89  & 0.89   & 0.89 & 0.89 & 0.89 & 0.89 & 0.89 & 0.89 & 0.89 & 0.89 				& g COD m$^{-3}$ 	\\
    $S_{O}$	    & 0.49  & 0.49   & 0.49 & 0.49 & 0.49 & 0.49 & 0.49 & 0.49 & 0.49 & 0.49 				& g O$_2$ m$^{-3}$ 	\\
    $S_{NO}$	& 10.42 & 10.42  & 10.42 & 10.42 & 10.42 & 10.42 & 10.42 & 10.42 & 10.42 & 10.42 		& g N m$^{-3}$	\\
    $S_{NH}$	& 1.73  & 1.73   & 1.73 & 1.73 & 1.73 & 1.73 & 1.73 & 1.73 & 1.73 & 1.73 				& g N m$^{-3}$	\\	
    $S_{ND}$	& 0.69  & 0.69   & 0.69 & 0.69 & 0.69 & 0.69 & 0.69 & 0.69 & 0.69 & 0.69 				& g N m$^{-3}$ 	\\		
    $S_{ALK}$	& 4.13  & 4.13   & 4.13 & 4.13 & 4.13 & 4.13 & 4.13 & 4.13 & 4.13 & 4.13 				& mol HCO$_3^-$ m$^{-3}$	\\[0.5ex]
    \hline
\end{tabular}\vskip0.35cm
\begin{tabular} {c | c | l} 
            & Input & Units \\ 
    \doublehline 
    $Q_{IN}$	    & 18846 & m$^{3}$ d$^{-1}$	\\
    $Q_{A}$	        & 55338 & m$^{3}$ d$^{-1}$	\\
    $Q_{R}$	        & 18446 & m$^{3}$ d$^{-1}$	\\
    $Q_{W}$	        & 385   & m$^{3}$	\\
    $K_La^{(1)}$	& 0     & d$^{-1}$	\\
    $K_La^{(2)}$	& 0     & d$^{-1}$	\\
    $K_La^{(3)}$	& 240   & d$^{-1}$	\\
    $K_La^{(4)}$	& 240   & d$^{-1}$	\\
    $K_La^{(5)}$	& 84    & d$^{-1}$	\\
    $Q_{EC}^{(1)}$	& 0     & m$^{3}$ d$^{-1}$	\\
    $Q_{EC}^{(2)}$	& 0     & m$^{3}$ d$^{-1}$	\\
    $Q_{EC}^{(3)}$	& 0     & m$^{3}$ d$^{-1}$	\\
    $Q_{EC}^{(4)}$	& 0     & m$^{3}$ d$^{-1}$	\\
    $Q_{EC}^{(5)}$	& 0     & m$^{3}$ d$^{-1}$	\\[0.5ex]
    \hline
\end{tabular} \label{tab: Linearisation}%
\end{table}	

\clearpage\section{Case-studies: Controller parameters and supplementary discussions} \label{sec: Results_Full}

In the following, we provide the tuning parameters of the model predictive controllers and moving horizon estimators used by the Output MPC. 
Further discussion of the results from Section V and extended visualisations are also provided.

\subsection{MPC }
The MPC used in the Output MPC is designed to minimise the quadratic stage and terminal cost functions below
$$
    L^c_{k+n_c\kappa_c}(\cdot) = \| x_{k+n_c\kappa_c} - x_{k+n_c\kappa_c}^{\text{ref}} \|_{Q_c}^{2} + \| u_{k+n_c\kappa_c} - u_{k+n_c\kappa_c}^{\text{ref}} \|_{R_{c}}^{2} 
    \quad\text{ and }\quad
    L^f_k(\cdot) = \| x_{k+N_c\kappa_c} - x_{k+N_c\kappa_c}^{\text{ref}} \|_{Q_f}^2,
$$
with weighting matrices $Q_{c|k+n_c\kappa_c} = C_{\Delta t | k+n_c\kappa_c}^{\tran} Q_{y|k+n_c\kappa_c} C_{\Delta t | k+n_c\kappa_c}$ and $Q_{f|k+N_c\kappa_c} = Q_{k+N_c\kappa_c}$ given the matrix
\begin{equation} \label{eq: Qn_Tuning}
    Q_{y|k+n_c\kappa_c} = \begin{cases}
        \texttt{diag}\big(0.01, \ldots, 0.01, 0.2, 3, 1 \big)                           	& \text{if } N_{TOT}^{SP}\big( t_{k+n_c\kappa_c} \big) = \frac{2}{3} 14 \\[1.5ex]
        \texttt{diag}\big(0.01, \ldots, 0.01, \cfrac{0.41}{3}, \frac{6.01}{3}, 1 \big)  	& \text{if } N_{TOT}^{SP}\big( t_{k+n_c\kappa_c} \big) = 14 \\[1.5ex]
        \texttt{diag}\big(0.01, \ldots, 0.01, 0.01, 0.01, 1 \big)                       	& \text{if } N_{TOT}^{SP}\big( t_{k+n_c\kappa_c} \big) = \frac{5}{3} 14
    \end{cases},
\end{equation}
and $R_{c|k+n_c\kappa_c} = \texttt{diag}\big( 10^{-10},\  10^{-6},\  10^{-5},\  10^{-4} I_5,\  10^{-1} I_5 \big)$ for all $n_c \in \mathbb{N}$. 
The matrix in Eq. \eqref{eq: Qn_Tuning} can also be written
\begin{equation}
    Q_{y|k+n_c\kappa_c} = \alpha(t_{k+n_c\kappa_c})~\texttt{diag}\big( 0.01, \ldots, 0.01, 0.01, 0.01, 1 \big)  + \big( 1 - \alpha(t_{k+n_c\kappa_c}) \big)~\texttt{diag}\big( 0.01, \ldots, 0.01, 0.2, 3, 1 \big)
\end{equation}
with $\alpha(t) = \big( \frac{1}{14} N_{TOT}^{SP}( t ) - \frac{2}{3} \big)$. 

This design choice of the objective penalises state deviations from reference (here: stricter nitrogen removal, standard treatment, and producing reuse water), whereas control effort are penalised relatively less to allow for flexible adjustments. 

\subsubsection{Steady-state optimiser}
Given output references $\big( X_{SS}^{S(10)},S_{NH}^{S(10)},N_{TOT}^{S(10)} \big)$, references $\big( x_{k+n_c\kappa_c}^{\text{ref}}, u_{k+n_c\kappa_c}^{\text{ref}} \big)$ are given by the steady-state optimiser 
$$
    h\big( g(x_{k+n_c\kappa_c}^{\text{ref}}) \big) = H g(x_{k+n_c\kappa_c}^{\text{ref}}), \quad \text{with } H = (0,\ldots,0,I_3).
$$
We consider weights $W_y = \texttt{diag}(1,\  10,\  20)$ and $W_u = R_0 = \texttt{diag}(10^{-10},\  10^{-6}, \  10^{-5}, \  10^{-4} I_5, \  10^{-1} I_5)$, and assume the fixed-point disturbances $w_{k+n_c\kappa_c}^{\text{ref}} = \bar{w}$ for all $n_c$. Finally, steady-states are constrained to be positive ($\mathcal{X}_{k+n_c\kappa_c}^{\text{ref}} = \mathbb{R}^{N_x}_{\geq 0}$), whereas the steady-inputs are constrained by the same actuator limits considered for the MPC ($\mathcal{U}_{k+n_c\kappa_c}^{\text{ref}} = \mathcal{U}_{k+n_c\kappa_c}$).

\subsection{MHE}
The MHE used in the Output MPC is designed to determine state- and disturbance-vectors that minimise the costs
$$
    L^i_k(\cdot) = \| {x}_{k-N_e\kappa_e} - \overline{x}_{k-N_e\kappa_e} \|_{Q_{x_0}^{-1}}^2,
    \quad\text{ and }\quad
    L^e_{k-n_e\kappa_e}(\cdot) = \| g_t({x}_{k-n_e\kappa_e}) - y^{\text{data}}_{k-n_e\kappa_e} \|_{Q_v^{-1}}^{2} + \| {w}_{k-n_e\kappa_e} - \overline{w}_{k-n_e\kappa_e} \|_{R_w^{-1}}^{2} 
$$
arising from the assumption that initial state, disturbances and measurements are normally distributed variables 
$$
    x_{k - N_e\kappa_e} \sim	\mathcal{N}(\overline{x}_{k - N_e\kappa_e}, Q_{x_0}); \quad
    w_{k - n_e\kappa_e} \sim	\mathcal{N}(\overline{w}_{k - n_e\kappa_e}, R_{w}), \quad n_e=0,\ldots,N_e; \quad
    v_{k - n_e\kappa_e} \sim	\mathcal{N}(0, Q_{v}),
$$
with covariance matrices $Q_{x_0|k-N_e\kappa_e} = \texttt{diag}(0.01 \bar{x})^2$, $R_{w|k-n_e\kappa_e}  = \texttt{diag}(2,100,250,350,16,15,1,2)$, and $Q_{v|k-n_e\kappa_e}$ $=\texttt{diag}(0.005 I_5, 0.05, I_5)$, for all $n_e \in \mathbb{N}$.
We assumed that noise in the measurements of $\{ S_O^{A(r)} \}_{r=1}^5$ is small while measurements of $\{ S_{NO}^{A(r)} \}_{r=1}^5$ and $\{ X_{SS}^{S(10)}, S_{NH}^{S(10)}, N_{TOT}^{S(10)} \}$ are relatively more noisy. 
The means of the initial state and disturbance are set recursively to be equal to the previous estimates ($\overline{x}_{k-N_e\kappa_e} = \widehat{x}^*_{k-N_e\kappa_e}$ and $(\overline{w}_{k-n_e\kappa_e} = \widehat{w}^*_{k - n_e\kappa_e})_{n_e=0}^{N_e}$). 

\clearpage
\subsection{Case-study 1} \label{sec: Treatment_Appendix}

\subsubsection{Wastewater treatment}
\begin{figure}[htb!] \centering
    \includegraphics[width=\textwidth]{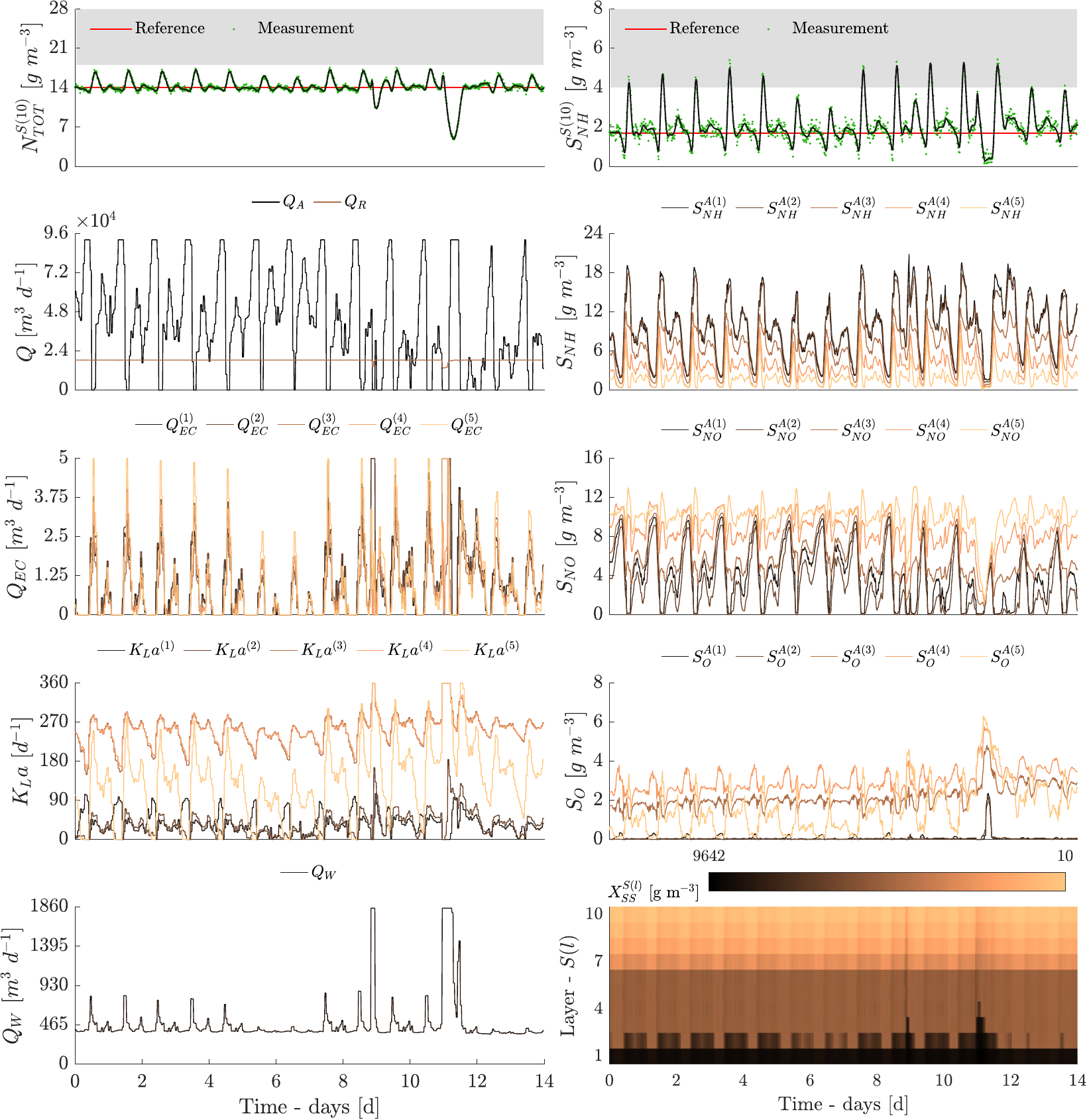}

    \caption{Case-study I, $\eta = 0$: Effluent total $N_{TOT}^{S(10)}$ and ammonium $S_{NH}^{S(10)}$ nitrogen (topmost panels) control actions for flow-rates $(Q_A, Q_R, Q_W)$, oxygen transfer coefficients $K_La^{(1\leadsto 5)}$, and extra carbon flow-rates $Q_{EC}^{(1\leadsto 5)}$ (left-column panels) and system responses for nitrogen forms $S_{NH}^{A(1\leadsto 5)}$ and $S_{NO}^{A(1\leadsto 5)}$, soluble oxygen $S_O^{A(1\leadsto 5)}$, and suspended solids $X_{SS}^{S(1\leadsto 10)}$ (right-column panels). 
    The shaded regions in the topmost panels denote concentration values above the quality limits.}
    \label{fig: OutMPC_Treatment_Eta0_Summary} 
\end{figure}

In this section, we present additional visualisations for the results obtained by our control framework when operating the activated sludge plant to satisfy conventional regulatory constraints. 
Firstly, all the control actions and a selection of system responses are provided in Figure \ref{fig: OutMPC_Treatment_Eta0_Summary} for the controller without energy recovery constraints ($\eta = 0$). 
\begin{itemize}
\item This visualisation shows how the controller operates the  plant mainly by enforcing certain desired of oxygen in all reactors $A(1{\leadsto} 5)$, by manipulating the aeration strength through $K_La^{(1\leadsto 5)}$, specially for the fifth reactor $A(5)$. 
\item The control strategy results in consistent patterns in the daily concentrations $S_{O}^{A(1\leadsto 5)}$, with mild variation during weekends. As expected, the dominant patterns are disrupted by the storm events occurring in the second week. 
\item The control actions commanded to the flow-rates $Q_A$ and $Q_{EC}^{(1\leadsto 5)}$ respond similarly to the influent variations, with internal recirculation and addition of external carbon being increased when influent carbon matter is diluted. 
\item To remove sludge $X_{SS}^{S(1)}$ when particle matter accumulates in the settler bottom layers, the controller commands an increase in wastage flow-rate $Q_W$. This automatic decision is enforced especially after the two storm events. 
\item The profile of wastage concentrations seems to result from external recirculation $Q_{R}$ being always almost constant.  
\end{itemize}

\clearpage
\subsubsection{Wastewater treatment, with energy recovery}   

\begin{figure}[b!] \centering
    \includegraphics[width=\textwidth]{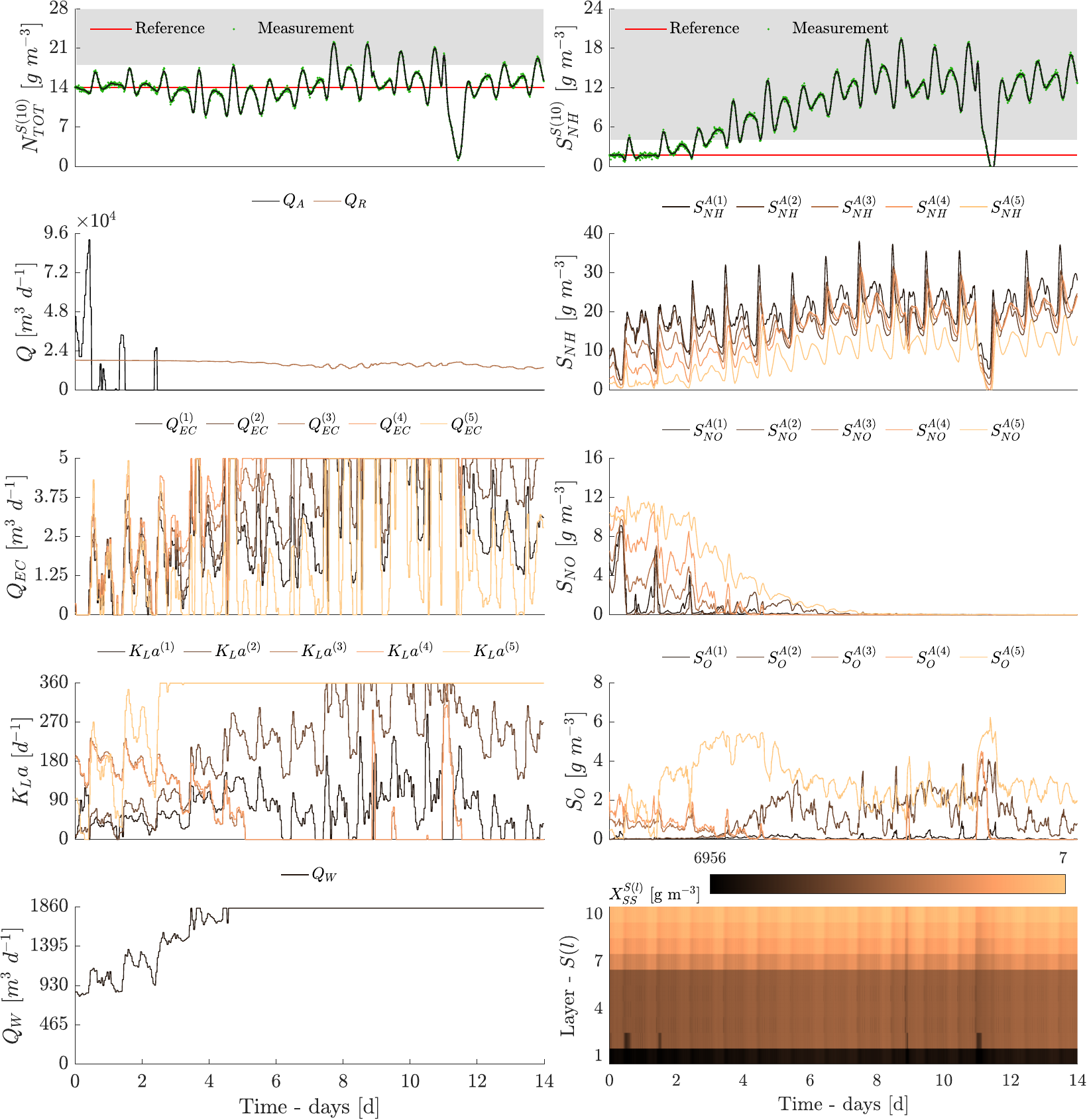}
    
    \caption{Case-study I, $\eta = 1$: Effluent total $N_{TOT}^{S(10)}$ and ammonium $S_{NH}^{S(10)}$ nitrogen (topmost panels) control actions for flow-rates $(Q_A, Q_R, Q_W)$, oxygen transfer coefficients $K_La^{(1\leadsto 5)}$, and extra carbon flow-rates $Q_{EC}^{(1\leadsto 5)}$ (left-column panels) and system responses for nitrogen forms $S_{NH}^{A(1\leadsto 5)}$ and $S_{NO}^{A(1\leadsto 5)}$, soluble oxygen $S_O^{A(1\leadsto 5)}$, and suspended solids $X_{SS}^{S(1\leadsto 10)}$ (right-column panels). The shaded regions in the topmost panels denote concentration values above the quality limits.}
    \label{fig: OutMPC_Treatment_Eta1_Summary}
\end{figure}

In this section, we consider the Output MPC when tracking the effluent profile, while being restricted to recover all of the energy needs from disposed sludge (when $\eta = 1$). The control actions and a selection of responses are in Figure \ref{fig: OutMPC_Treatment_Eta1_Summary}. 

\begin{itemize}
\item The control strategy starts similarly to the previous case, albeit with decreased aeration to reactors $A(3{\leadsto} 5)$ and increased wastage flow-rate $Q_W$. Due to this strategy, denitrification is favoured throughout the entire process. 
\item As a result, the concentrations of ammonium $S_{NH}^{A(1\leadsto 5)}$ increase as nitrogen $S_{NO}^{A(1\leadsto 5)}$ is removed from all reactors. 
\item The controller tries to recover the original levels of nitrogen firstly by increasing the wastage flow-rate $Q_W$, to increase the production of biogas, and secondly by decreasing $Q_A$, to contain energy costs, in such a way that aeration to reactors $A(2,5)$ can be increased through $K_La^{(2,5)}$ without violating the energy-recovery constraints. 
\item External carbon addition is also increased in all reactors to compensate for the interruption of the internal recirculation. However, these control actions fail to recover the concentrations of nitrogen to the desired levels. 
\end{itemize}

\clearpage
\subsection{Case-study 2} \label{sec: Tracking_Appendix}

\subsubsection{Nitrogen on-demand}

In this section, we present additional visualisations for the results obtained by the Output MPC framework when configured to operate the activated sludge plants for water resource recovery tasks (Section V-B). 
Again, we start with the control actions and a selection of system responses for the controller without energy recovery constraints ($\eta = 0$). 

\begin{figure}[htb!] \centering
    \includegraphics[width=\textwidth]{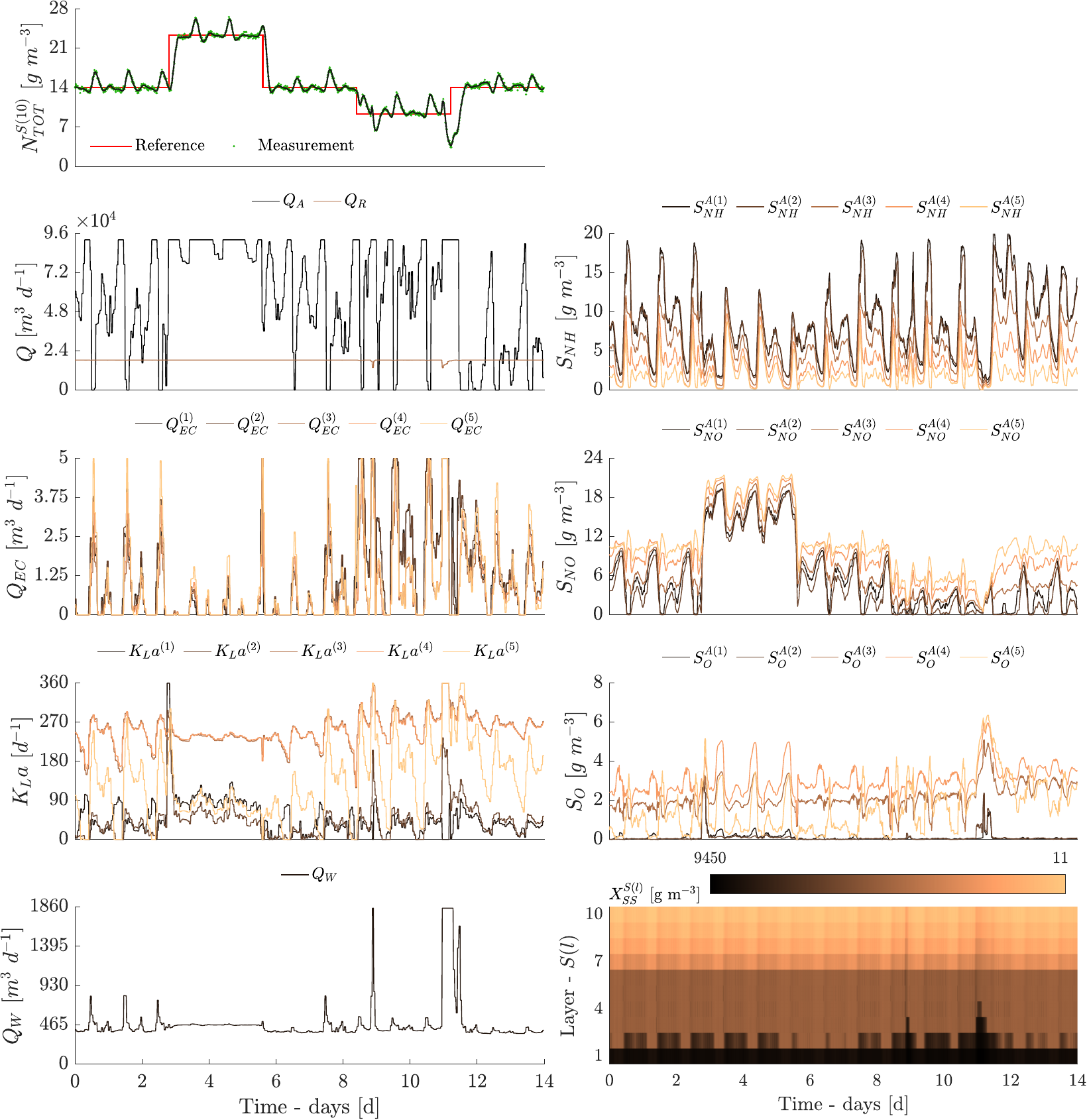}
    
    \caption{Case-study II, $\eta = 0$: Effluent total $N_{TOT}^{S(10)}$ and ammonium $S_{NH}^{S(10)}$ nitrogen (topmost panels) control actions for flow-rates $(Q_A, Q_R, Q_W)$, oxygen transfer coefficients $K_La^{(1\leadsto 5)}$, and extra carbon flow-rates $Q_{EC}^{(1\leadsto 5)}$ (left-column) and system responses for nitrogen forms $S_{NH}^{A(1\leadsto 5)}$ and $S_{NO}^{A(1\leadsto 5)}$, soluble oxygen $S_O^{A(1\leadsto 5)}$, and suspended solids $X_{SS}^{S(1\leadsto 10)}$ (right-column).}
    \label{fig: OutMPC_Track_Eta0_Summary}
\end{figure}

\begin{itemize}
\item The results, in Figure \ref{fig: OutMPC_Track_Eta0_Summary}, highlight how the controller acts to adapt operations to the effluent nitrogen references. 
\item During the \textit{water reuse} task (for $t \in [2.8,5.6)$ days), in order to satisfy the higher of nitrogen, the controller increases the internal recirculation $Q_A$, while also increasing the aeration to all reactors $A(1{\leadsto} 5)$ through $K_La^{(1\leadsto 5)}$. 
\item As showcased in Section V-B, this strategy leads to an increase on effluent $N_{TOT}^S(10)$ to match the demand. 
\item During the \textit{stricter nitrogen removal} task (for $t \in [8.4, 11.2)$ days), the controller applies a control strategy which is similar to the one adopted for conventional treatment (when $N_{TOT}^{SP} \approx 14$ g m$^{-3}$), with the exception that the addition of external carbon $Q_{EC}^{(1\leadsto 5)}$ is increased in all five reactors $A(1{\leadsto} 5)$. 
\item For the other tasks, the controller operates with a strategy similar to what used for the case-study in Section \ref{sec: Treatment_Appendix}.
\end{itemize}

\clearpage
\subsubsection{Nitrogen on-demand, with energy recovery}

\begin{figure}[h!] \centering
    \includegraphics[width=\textwidth]{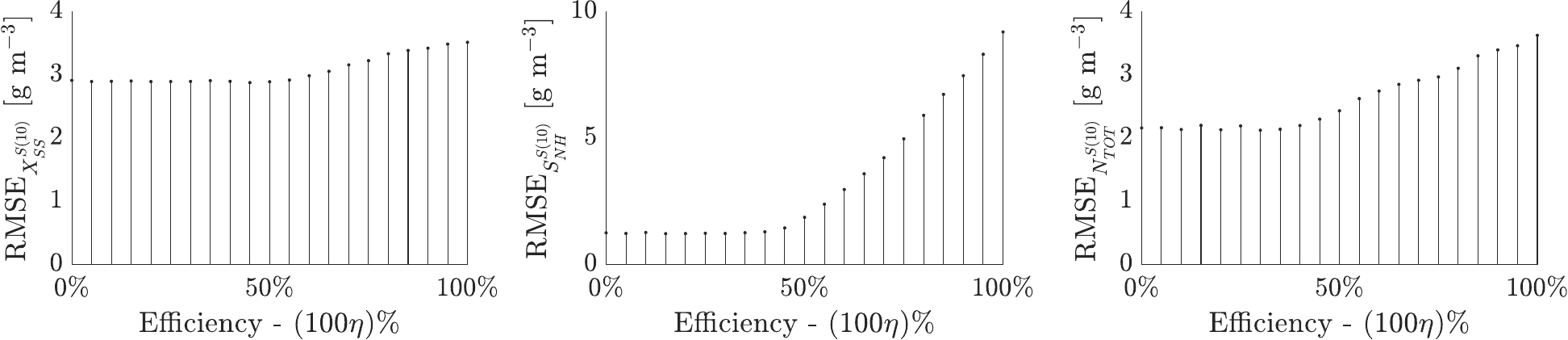}
    
    \caption{Case-study II: Reference tracking accuracy (RMSE) for the different key outputs $\{X_{SS}^{S(10)}, S_{NH}^{S(10)}, N_{TOT}^{S(10)}\}$ for different recovery levels $\eta \in [0,1] $.}
    \label{fig: OutMPC_Track_Eta0_AllRMSE}
\end{figure}

We consider the results from the Output MPC when configured to satisfy the effluent requirements, while restricted to recover the energy demand from disposed sludge, that is, $\eta \in (0,1]$. Figure \ref{fig: OutMPC_Track_Eta0_AllRMSE} shows how requested recovery degrees directly affect on the achieved tracking accuracy, in terms of root-mean-squared-error (RMSE),
\begin{equation}
    \text{RMSE} = \sqrt{ \frac{1}{14} \int_{0}^{14}\hspace*{-0.4em} \left( Z^{SP}(t) - Z(t) \right)^2 dt} \qquad \text{with } Z \in \{ X_{SS}^{S(10)}, S_{NH}^{S(10)}, N_{TOT}^{S(10)} \} .
\end{equation}

The Output MPC is capable to achieve satisfactory tracking accuracies when recovering up to $60\%$ of the process energetic demands. 
The controller performance degrades when higher recovery levels are requested, especially with respect to ammonium nitrogen $S_{NH}^{S(10)}$ references. 
Conversely, the accuracy in tracking $X_{SS}^{S(10)}$ remains consistently good for all recovery levels. 
This indicates that tracking this variable is only lightly affected by energy recovery procedures.

We show the control actions and a selection of responses obtained by the controller restricted to full-recovery ($\eta = 1$). 

\begin{figure}[htb!] \centering
    \includegraphics[width=\textwidth]{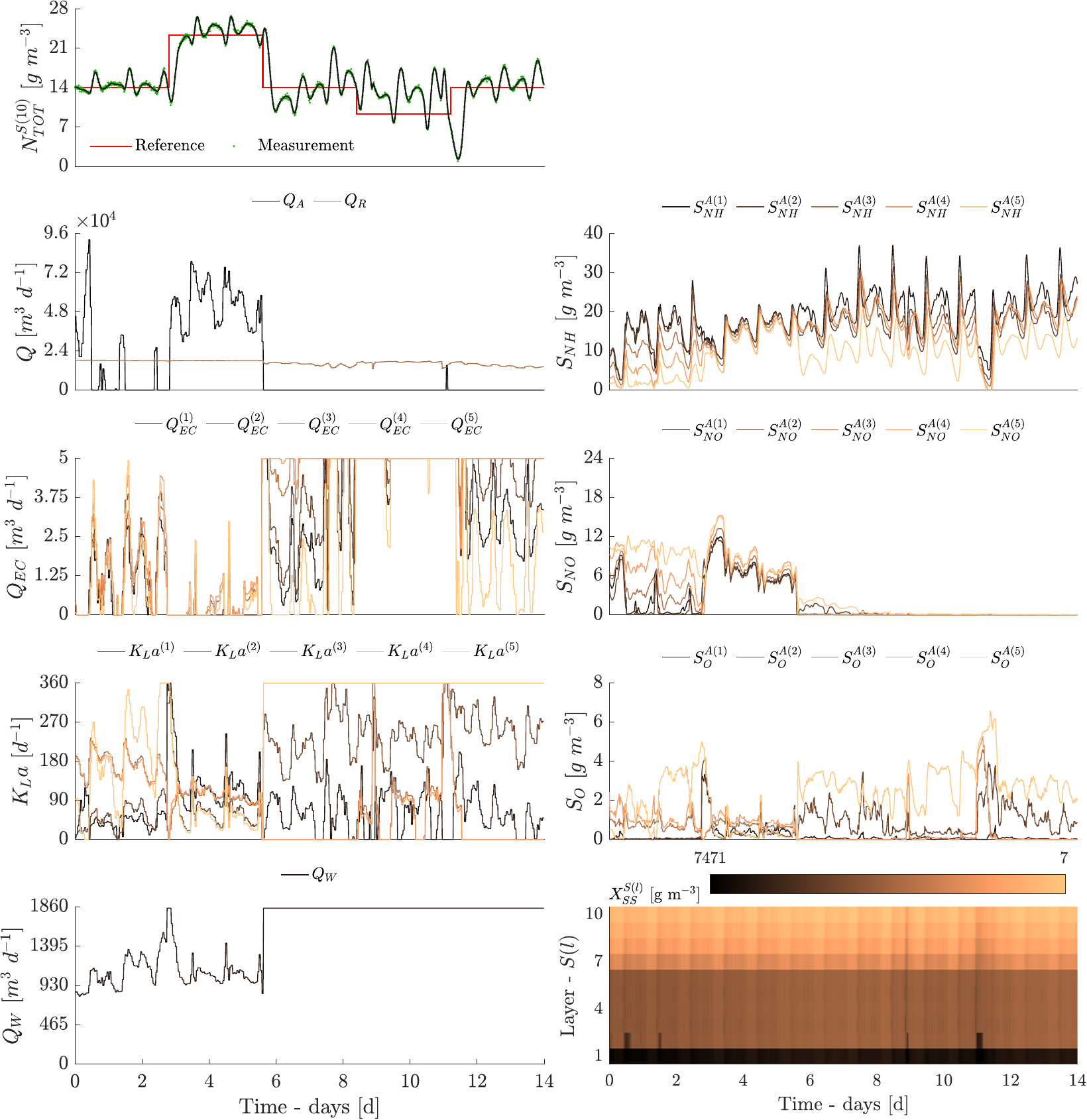}
    
    \caption{Case-study II, $\eta = 1$: Effluent total $N_{TOT}^{S(10)}$ and ammonium $S_{NH}^{S(10)}$ nitrogen (topmost panels) control actions for flow-rates $(Q_A, Q_R, Q_W)$, oxygen transfer coefficients $K_La^{(1\leadsto 5)}$, and extra carbon flow-rates $Q_{EC}^{(1\leadsto 5)}$ (left-column) and system responses for nitrogen forms $S_{NH}^{A(1\leadsto 5)}$ and $S_{NO}^{A(1\leadsto 5)}$, soluble oxygen $S_O^{A(1\leadsto 5)}$, and suspended solids $X_{SS}^{S(1\leadsto 10)}$ (right-column).}
    \label{fig: OutMPC_Track_Eta1_Summary}
\end{figure}

\begin{itemize}
\item The results, i Figure \ref{fig: OutMPC_Track_Eta1_Summary}, show that the controller is unable to satisfy the effluent requirements due to an undesired complete removal of nitrogen $S_{NO}^{A(1\leadsto 5)}$. This is observed also for standard wastewater treatment (Section \ref{sec: Treatment_Appendix}). 
\item The controller is still capable to manipulate the plant towards satisfying the reuse-related nitrogen reference (during $t \in [2.8,5.6)$ days), while satisfying the constraint to recover the entire energetic demand from sludge. 
\item This specific results is achieved by strengthening the aeration to only three reactors ($A(1,2,5)$), by increasing $K_La^{(1,2,5)}$, when the reference change occurs, then by reducing the aeration strength to all the five reactors. 
\end{itemize}

\clearpage\subsection{Performance of the moving horizon estimator} 

For completeness, we provide visualisations on the estimates obtained by the moving horizon estimator. 
As the estimator has similar performance in all cases, we restrict ourselves to show the results relative to the case-study with no energy-recovery (when $\eta = 0$). 
We focus on the estimates determined in the second week, when the storm events occur. 
\begin{itemize}
\item The estimates with respect to the disturbance variables are shown below in Figure \ref{fig: MHE_Treatment_Eta0_Disturbances_Summary}. The results show that the MHE is capable to accurately reconstruct the influent concentration profile, for most of the involved components. 
\item The worst estimation performances are observed for the concentrations $\{ S_{I}^{IN}, X_I^{IN}, X_{BH}^{IN} \}$, whose estimates are kept constant around their mean value, indicating that their evolution does not affect the changes in measurements. 
\end{itemize}

\begin{figure}[htb!] \centering 
    \includegraphics[width=\textwidth]{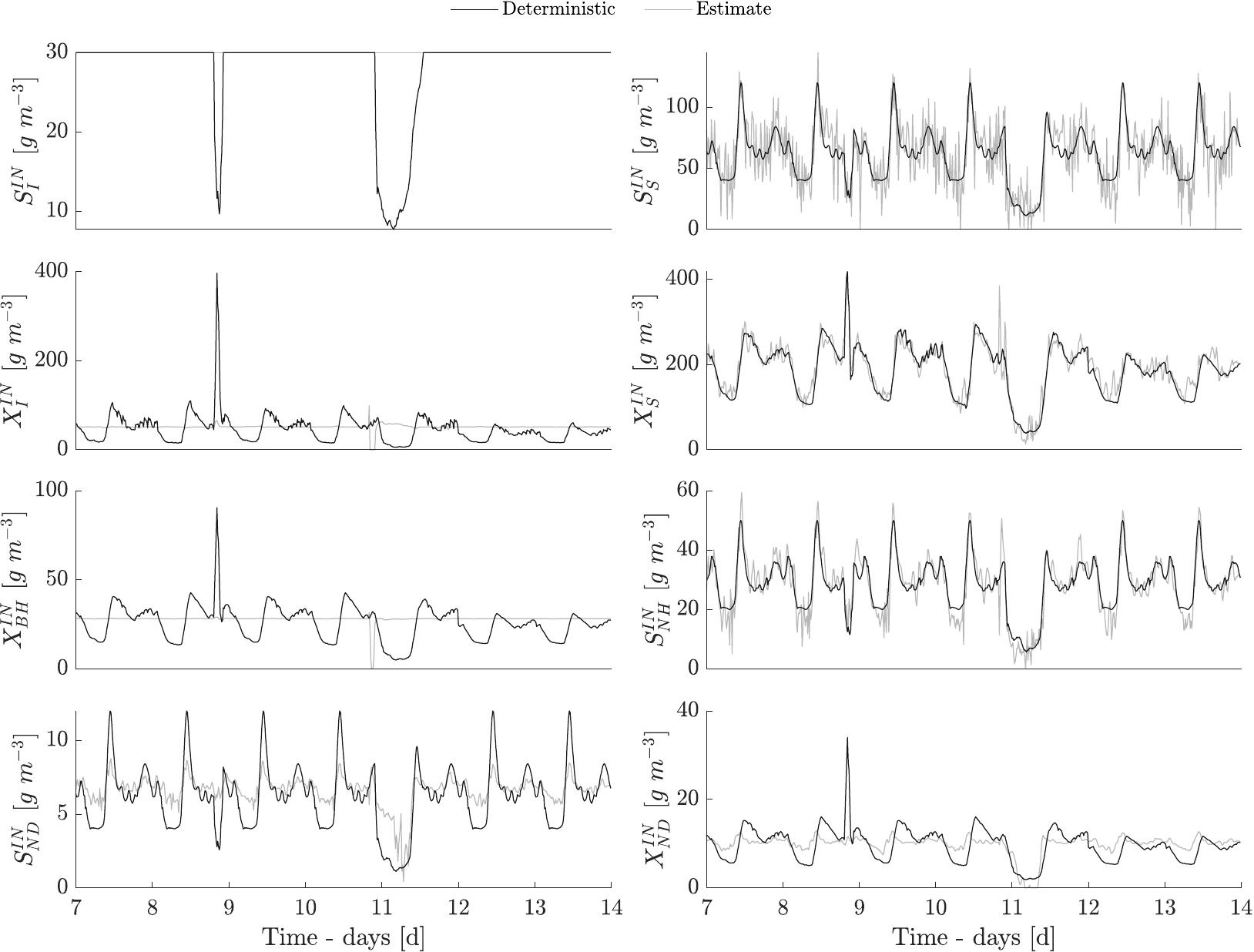}
    
    \caption{Case-study I, $t \in [7,14]$, $\eta = 0$: True values (black) and estimates (grey) for the influent concentrations $x^{A(IN)}$. Concentrations $X_{BA}^{IN} = X_P^{IN} = S_O^{IN} = S_{NO}^{IN} = 0$ g m$^{-3}$ and $S_{ALK}^{IN} = 7$ mol HCO$_3^-$ m$^{-3}$ are omitted.}
    \label{fig: MHE_Treatment_Eta0_Disturbances_Summary}
\end{figure}

\begin{figure}[htb!] \centering
    \includegraphics[width=\textwidth]{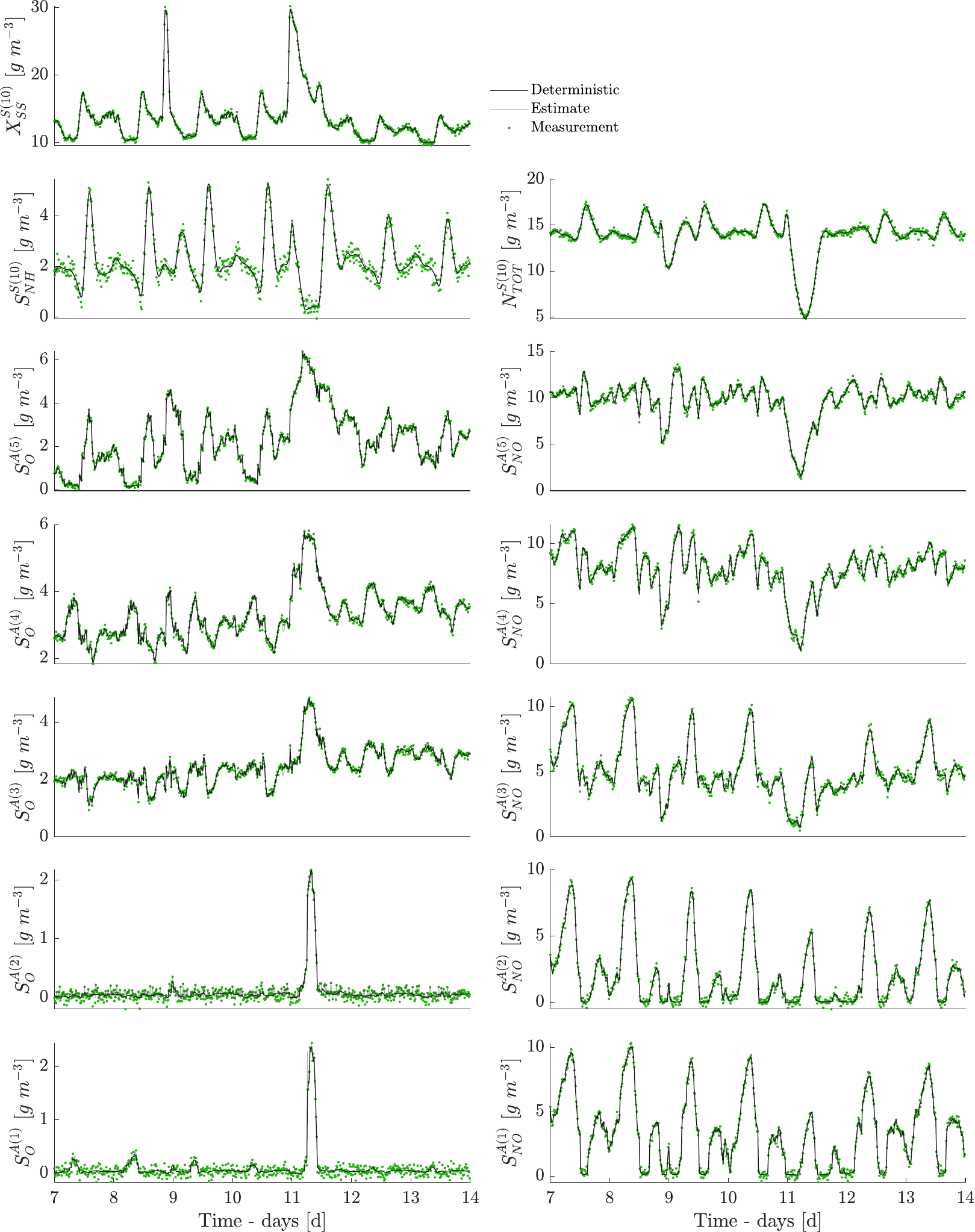}
    
    \caption{Case-study I, $t \in [7,14]$, $\eta = 0$: True values (black) and estimates (grey) for the measured concentrations $y(t)$.}
    \label{fig: MHE_Treatment_Eta0_Outputs_Summary}
\end{figure}

\begin{itemize}
\item The estimates with respect to the output variables are in Figure \ref{fig: MHE_Treatment_Eta0_Outputs_Summary}. The results show that the MHE is capable to accurately reconstruct the true values of all outputs, with inferior performances for $S_O^{A(2)}$ relatable to data overfit.
\end{itemize}

\begin{figure}[htb!] \centering
    \includegraphics[width=\textwidth]{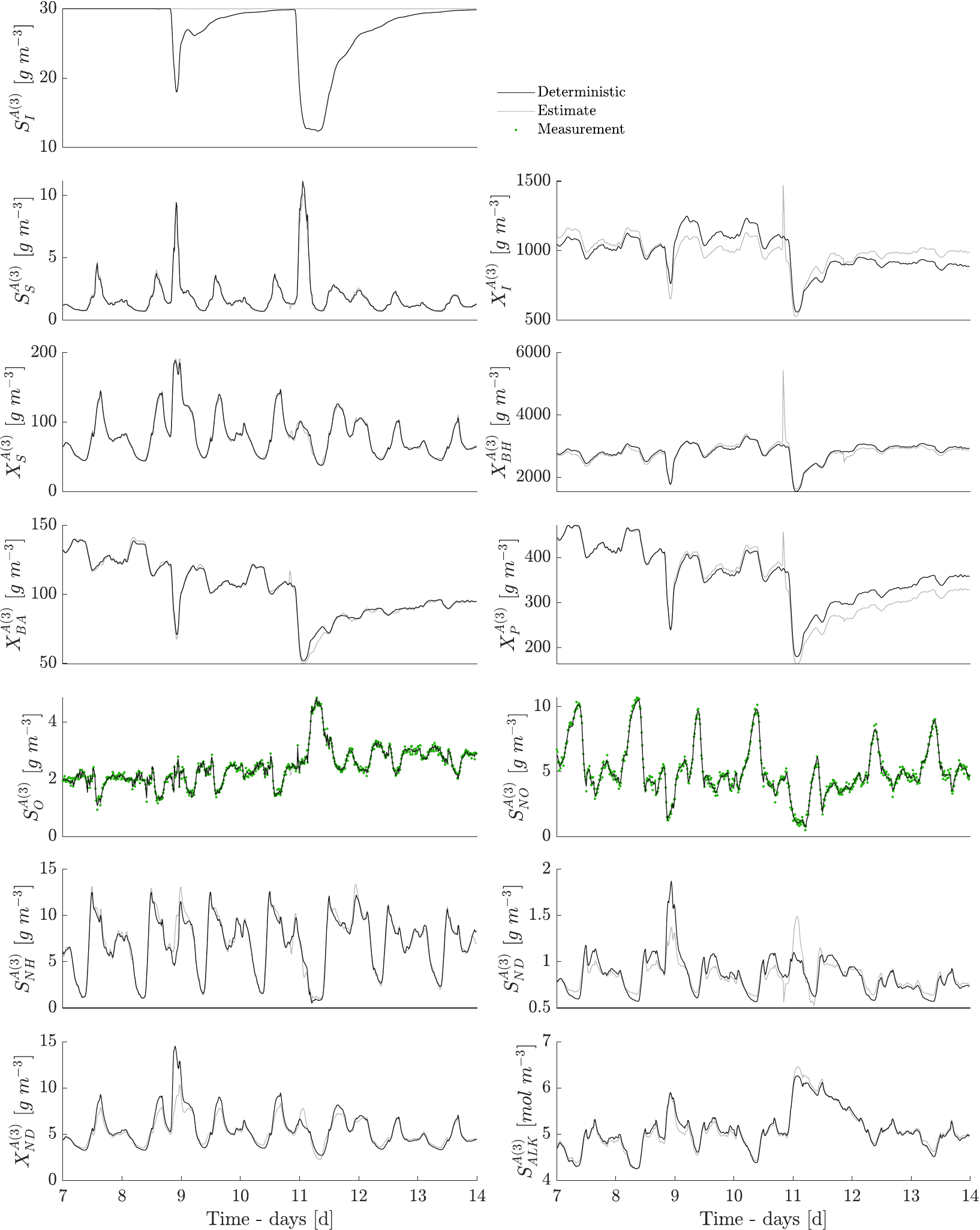}
    
    \caption{Case-study I, $t \in [7,14]$, $\eta = 0$: True values (black) and estimates (grey) for the concentrations $x^{A(3)}$ in the third reactor, $A(3)$. As $S_O^{A(3)}$ and $S_{NO}^{A(3)}$ are measured, the measurement values for these concentrations are also shown.}
    \label{fig: MHE_Treatment_Eta0_State_Summary}
\end{figure}

The estimates for the concentrations in the first aerated reactor, $A(3)$, in Figure \ref{fig: MHE_Treatment_Eta0_State_Summary} show the good performance of the MHE. 
The results indicate that $\{ X_I^{A(3)}, X_{BH}^{A(3)}, X_P^{A(3)} \}$ are amongst the concentrations which are the hardest to estimate. 
\begin{itemize} 
\item Our results thus agree with those obtained by [3] for the moving-horizon estimation of this class of activated sludge plants, although  different sensor configurations and some model simplifications were used in that case.
\end{itemize}

\clearpage

\clearpage


\end{document}